\documentclass[a4paper,11pt]{article}
\pdfoutput=1 

\usepackage{jheppub} 

\usepackage[T1]{fontenc} 
\usepackage{feyn}
\usepackage{feynmf} 

\usepackage{float}
\usepackage{multirow}

\usepackage{subfig}
\usepackage{amsmath}
\usepackage[dvipsnames]{xcolor}
\usepackage{hyperref}
\usepackage[normalem]{ulem}
\usepackage{xspace}

\newcommand{\new}[1]{{\color{red}#1}}

\newcommand{\gambit}{\textsf{GAMBIT}\xspace}

\newcommand{\RD}{$R(D)$}
\newcommand{\RDs}{$R(D^{*})$}

\title{\boldmath Likelihood analysis of the flavour anomalies and $g-2$ in the general
two Higgs doublet model}

\author[a,b]{Peter Athron,}
\author[b]{Csaba Balazs,}
\author[c]{Tom\'as E. Gonzalo,}
\author[b]{Douglas Jacob,}
\author[d,e]{Farvah Mahmoudi,}
\author[b]{Cristian Sierra}

\affiliation[a]{Department of Physics and Institute of Theoretical Physics, Nanjing Normal University, Wenyuan Road, Nanjing, Jiangsu, 210023, China}

\affiliation[b]{School of Physics and Astronomy, Monash University,
Wellington Road, Clayton, VIC 3800, Australia}

\affiliation[c]{Institute for Theoretical Particle Physics and Cosmology (TTK), RWTH Aachen,
Sommerfeldstrasse 12, 52074 Aachen, Germany}

\affiliation[d]{Universit\'e de Lyon, Universit\'e Claude Bernard Lyon 1, CNRS/IN2P3, Institut de Physique des 2 Infinis de Lyon, UMR 5822, 69622 Villeurbanne, France}
\affiliation[e]{Theoretical Physics Department, CERN, CH-1211 Geneva 23, Switzerland}

\emailAdd{cristian.sierra@monash.edu}


\preprint{\hfill {\tt CERN-TH-2021-194, TTK-21-47}}

\abstract{We present a likelihood analysis of the general two Higgs doublet model, using the most important currently measured flavour observables, in view of the anomalies in charged current tree-level and neutral current one-loop rare decays of $B$ mesons in $b\to c l \overline{\nu}$ and $b\to s\mu^{+}\mu^{-}$ transitions, respectively. 
We corroborate that the model explains the latter and it is able to simultaneously fit the experimental values of the $R(D)$ charged current ratio at $1\sigma$, but it can not accommodate the $D^{*}$ charmed meson observables $R(D^{*})$ and $F_{L}(D^{*})$. We find that the fitted values for the angular observables in $b\to s\mu^{+}\mu^{-}$ transitions exhibit better agreement with the general two Higgs double model in comparison to the SM. We also make predictions for future collider observables $\mathrm{BR}(t\to ch)$, $\mathrm{BR}(h\to bs)$, $\mathrm{BR}(h\to \tau\mu)$, $\mathrm{BR}(B_{s}\rightarrow\tau^{+}\tau^{-})$, $\mathrm{BR}(B^{+}\rightarrow K^{+}\tau^{+}\tau^{-})$ and the flavour violating decays of the $\tau$ lepton,  $\mathrm{BR}(\tau\rightarrow3\mu)$ and $\mathrm{BR}(\tau\to\mu\gamma)$. The model predicts values of $\mathrm{BR}(t\to ch)$, $\mathrm{BR}(B_{s}\rightarrow\tau^{+}\tau^{-})$ and $\mathrm{BR}(B^{+}\rightarrow K^{+}\tau^{+}\tau^{-})$ that are out of reach of future experiments, but its predictions for $\mathrm{BR}(h\to bs)$ and $\mathrm{BR}(h\to \tau\mu)$ are within the future sensitivity of the HL-LHC or the ILC.  We also find that the predictions for the $\tau\rightarrow3\mu$ and $\tau\to\mu\gamma$ decays are well within the projected limits of the Belle II experiment.  Finally, using the latest measurement of the Fermilab Muon $g-2$ Collaboration, we performed a simultaneous fit to $\Delta a_{\mu}$ constrained by the charged anomalies, finding solutions at the $1\sigma$ level. Once the neutral anomalies are included, however, a simultaneous explanation is unfeasible. }

\begin{document} 
\maketitle
\flushbottom

\section{Introduction}
\label{sec:intro}

The Standard Model (SM) of particle physics contains three fermion families which acquire mass by means of the interaction with the Higgs boson. The two Higgs doublet model  (2HDM) is one of the simplest ways to extend the Higgs sector, which is the least constrained sector of the Standard Model.  Two Higgs doublets also appear in many more elaborate extensions of the SM that are based on fundamental principles, such as supersymmetry (see e.g. \cite{Martin:1997ns}), the Peccei-Quinn symmetry~\cite{Peccei:1977hh,Peccei:1977ur} or grand unified theories (see \cite{Croon:2019kpe} for a recent review). Two Higgs doublet models are also motivated from electroweak baryogenesis studies, where it has been shown that contributions coming from the new physical Higgs bosons to the effective Higgs potential can strengthen the phase transition and in addition introduce new sources of charge-parity (CP) violation, from both fermion and scalar sectors \cite{Carena:1997gx, Cline:1997vk, Konstandin:2005cd, Cirigliano:2006wh, Buchmuller:2012tv, Morrissey:2012db, Konstandin:2013caa, Basler:2016obg, Fuyuto:2017ewj}.  As a result, the 2HDM is one of the most popular SM extensions and has been frequently used as a benchmark for phenomenological studies (see e.g. \cite{Branco:2011iw} for a review of 2HDM studies).  Furthermore, the presence of another doublet can contribute to resolving anomalies in lepton flavour universality observables \cite{Iguro:2018qzf, Martinez:2018ynq} and muon g-2 \cite{Broggio:2014mna,Wang:2014sda,Abe:2015oca,Chun:2015hsa,Chun:2015xfx,Chun:2016hzs,Wang:2018hnw,Chun:2019oix,Chun:2019sjo,Keung:2021rps,Ferreira:2021gke,Han:2021gfu,Eung:2021bef,Jueid:2021avn,Dey:2021pyn,Ilisie:2015tra,Han:2015yys,Cherchiglia:2016eui,Cherchiglia:2017uwv,Li:2020dbg,Athron:2021iuf,Omura:2015nja,Crivellin:2015hha,Iguro:2019sly,Jana:2020pxx,Ghosh:2020tfq,Hou:2021sfl,Hou:2021qmf,Atkinson:2021eox,Hou:2021wjj}, while scenarios where the extra doublet is "inert" can also explain dark matter \cite{LopezHonorez:2006gr, Gustafsson:2007pc, Dolle:2009fn, Honorez:2010re, LopezHonorez:2010tb, Chao:2012pt, Goudelis:2013uca, Arhrib:2013ela, Bonilla:2014xba, Queiroz:2015utg, Arcadi:2018pfo, Tsai:2019eqi, Camargo:2019ukv}.

The new interactions between the SM fermions and the physical states arising from the introduction of a second Higgs doublet imply a richer phenomenology than the SM. This is further enhanced by the new free parameters and couplings in the general two Higgs doublet model (GTHDM), also known as type-III 2HDM \cite{Hou:1991un}. Physical effects such as CP violation, scalar mixing and flavour changing transitions are expected \cite{Mahmoudi:2009zx},
allowing for signatures to be observed in particle colliders. One of the most interesting experimental consequences of the flavour changing currents present in the GTHDM is lepton flavour universality (LFU) violation. Experimental measurements of LFU violation come from flavour changing charged currents (FCCCs), such as those in $B$ meson decays, and flavour changing neutral currents (FCNCs), for instance in kaon decays. The observed deviations from the SM in the measurements of FCCCs (around $3.1\sigma$ from the SM~\cite{Amhis:2019ckw}) and FCNCs (close to a combined $6\sigma$ deviation, see for example~\cite{Alguero:2021anc,Hurth:2021nsi,Bhom:2020lmk}), hint at the existence of new physics (NP) contributions and thus serve as a clear motivation for the study of NP models capable of explaining the anomalies.

It has indeed been shown that the GTHDM is able to explain the charged anomalies at $2\sigma$ \cite{Cline:2015lqp,Iguro:2018qzf, Martinez:2018ynq,Cardozo:2020uol}.  Similar analyses for the neutral anomalies have also been presented previously \cite{Arnan:2017lxi,Arhrib2017, Iguro:2018qzf, Crivellin:2019dun}, finding solutions at the $2\sigma$ level and up to the $1\sigma$ level including right-handed neutrinos \cite{Crivellin:2019dun}. Nevertheless, the majority of these studies have only explored solutions in restricted regions of the parameter space, with a lack of discussion of the role of (marginally) statistically preferred regions, and often considering the $b\to sll$ observables from model independent global fits \cite{Iguro:2018qzf,Crivellin:2019dun}. Statistically rigorous explorations of the parameter space of the model contrasted directly to experimental constraints have rarely been performed, and even those were focused exclusively on interactions in the quark sector \cite{Herrero-Garcia:2019mcy}.

Furthermore, the longstanding discrepancy between the experimentally measured and SM predicted values of the anomalous magnetic moment of the muon $a_\mu$ has recently been brought back to the spotlight with the new measurement by the Muon g-2 experiment at Fermilab~\cite{PhysRevLett.126.141801}. The latest experimental value, taking into account the measurements at both Brookhaven National Laboratory and Fermilab, is $a^{\textrm{Exp}}_\mu = 116592061\pm41\times10^{-11}$. Compared to the theoretical prediction in the SM from the recent Muon $g-2$ Theory Initiative White Paper, $a^{\textrm{SM}}_\mu = 116591810\pm43\times10^{-11}$ \cite{Aoyama:2020ynm}, building on the extensive work examining the various SM contributions in  \cite{davier:2017zfy,keshavarzi:2018mgv,colangelo:2018mtw,hoferichter:2019gzf,davier:2019can,keshavarzi:2019abf,kurz:2014wya,melnikov:2003xd,masjuan:2017tvw,Colangelo:2017fiz,hoferichter:2018kwz,gerardin:2019vio,bijnens:2019ghy,colangelo:2019uex,colangelo:2014qya,Blum:2019ugy,aoyama:2012wk,Aoyama:2019ryr,czarnecki:2002nt,gnendiger:2013pva}, the measured value differs from the SM prediction by $\Delta a_\mu = 2.51\pm59\times10^{-9}$, corresponding to a discrepancy of $4.2\sigma$. Models with a second Higgs doublet have been studied extensively in the literature as sources to explain this deviation~\cite{Broggio:2014mna,Wang:2014sda,Abe:2015oca,Chun:2015hsa,Chun:2015xfx,Chun:2016hzs,Wang:2018hnw,Chun:2019oix,Chun:2019sjo,Keung:2021rps,Ferreira:2021gke,Han:2021gfu,Eung:2021bef,Jueid:2021avn,Dey:2021pyn,Ilisie:2015tra,Han:2015yys,Cherchiglia:2016eui,Cherchiglia:2017uwv,Li:2020dbg,Athron:2021iuf,Omura:2015nja,Crivellin:2015hha,Iguro:2019sly,Jana:2020pxx,Ghosh:2020tfq,Hou:2021sfl,Hou:2021qmf,Atkinson:2021eox,Hou:2021wjj}. However, no simultaneous global fit of the flavour anomalies and $a_\mu$ in the GTHDM has been attempted giving a proper statistical insight into the whole parameter space.


Therefore, in this paper we present a frequentist inspired likelihood analysis for the GTHDM, simultaneously including the  FCCC observables, both $b\to s\mu^{+}\mu^{-}$ transitions and the muon anomalous magnetic moment, along with other flavour observables. We perform a global fit of all constraints using the inference package \gambit, the Global And Modular Beyond-the-Standard-Model
Inference Tool \cite{Athron:2017ard, grev}. \gambit is a powerful software framework capable of performing statistical inference studies using constraints from collider~\cite{ColliderBit}, dark matter~\cite{DarkBit}, flavour~\cite{Workgroup:2017myk} and neutrino~\cite{RHN} physics, as well as cosmology~\cite{CosmoBit}. It has already been used for detailed statistical analyses of a variety of beyond the Standard Model (BSM) models, including supersymmetry~\cite{CMSSM, MSSM, EWMSSM}, scalar singlet DM~\cite{SSDM,SSDM2,HP,GUM,DMEFT}, axion and axion-like particles~\cite{Axions,XENON1T}, and neutrinos~\cite{RHN,CosmoBit_numass}, as well as an initial analysis of the 2HDM \cite{Rajec:2020orn}.  Our work enhances the \textsf{FlavBit} \cite{Workgroup:2017myk} and \textsf{PrecisionBit} \cite{GAMBITModelsWorkgroup:2017ilg} modules of \textsf{GAMBIT} to support the GTHDM.  We also make use of various external codes: \textsf{SuperIso 4.1} \cite{Mahmoudi:2007vz,Mahmoudi:2008tp,Mahmoudi:2009zz,Neshatpour:2021nbn} for computing flavour observables, the \textsf{2HDMC 1.8} package \cite{Eriksson:2009ws} for precision electroweak constraints, the \textsf{HEPLike} package \cite{Bhom:2020bfe} which provides likelihoods for the neutral anomaly related observables, and the differential evolution sampler \textsf{Diver 1.0.4} \cite{Workgroup:2017htr}.

The paper is organised as follows. In section \ref{sec:GTHDM} we present the Higgs and Yukawa sectors along the theoretical bounds for their parameters. In section \ref{sec:Hamiltonian} we define the effective Hamiltonian and the Wilson coefficients (WCs) for $b\to s\mu^{+}\mu^{-}$ transitions. Then, in section \ref{sec:Observables} we list the observables to be used in our scans.  Following this, our results from the global fit and predictions for future experiments in colliders are discussed in section \ref{sec:Results}.  Finally, we summarise our conclusions in section \ref{sec:Conclusions}.

\section{GTHDM}
\label{sec:GTHDM}
The GTHDM has been actively investigated in both its scalar and Yukawa sectors. These can be written in three different ways, namely in the generic, Higgs and physical bases, all of them related via basis transformations  \cite{Davidson:2005cw}. Particularly, with respect to the Yukawa sector, in the past theorists imposed discrete symmetries to avoid flavour changing transitions,
the most popular being the $\mathbb{Z}_{2}$ symmetry in the type-II 2HDM \cite{Glashow:1976nt,Gunion:1989we}.  However, it has been shown that there is no fundamental reason for forbidding flavour changing couplings \cite{Hou2019}: if the mixing angle is small, the non-observation of several tree level flavour changing transitions can be explained by the alignment phenomenon.  This, and a suppression inversely proportional to the mass of the heavy Higgses in the tree level amplitudes, could suppress the effects coming from the off-diagonal Yukawa couplings, without invoking the so called natural flavour conservation (NFC) condition \cite{Glashow:1976nt}. 

Here we review the Higgs potential and the Yukawa Lagrangian of the model as well as the relevant theoretical constraints coming from stability, unitarity and perturbativity at leading order (LO). We also make use of the precision electroweak constraints from the oblique parameters. For a more comprehensive review of the model the reader is referred to \cite{Branco:2011iw,Haber:2010bw,HernandezSanchez:2012eg,Crivellin2013}.

\subsection{Higgs potential}

The most general renormalizable scalar potential in the GTHDM is commonly written as \cite{Branco:2011iw,Gunion:2002zf}
\begin{alignat}{1}
V(\Phi_{1},\Phi_{2})=\: & m_{11}^{2}(\Phi_{1}^{\dag}\Phi_{1})+m_{22}^{2}(\Phi_{2}^{\dag}\Phi_{2})-m_{12}^{2}(\Phi_{1}^{\dag}\Phi_{2}+\Phi_{2}^{\dag}\Phi_{1})\nonumber \\
 & +\frac{1}{2}\lambda_{1}(\Phi_{1}^{\dag}\Phi_{1})^{2}+\frac{1}{2}\lambda_{2}(\Phi_{2}^{\dag}\Phi_{2})^{2}+\lambda_{3}(\Phi_{1}^{\dag}\Phi_{1})(\Phi_{2}^{\dag}\,\Phi_{2})+\lambda_{4}(\Phi_{1}^{\dag}\Phi_{2})(\Phi_{2}^{\dag}\Phi_{1})\nonumber \\
 & +\left(\frac{1}{2}\lambda_{5}(\Phi_{1}^{\dag}\Phi_{2})^{2}+\left(\lambda_{6}(\Phi_{1}^{\dag}\Phi_{1})+\lambda_{7}(\Phi_{2}^{\dag}\Phi_{2})\right)(\Phi_{1}^{\dag}\Phi_{2})+{\rm ~h.c.}\right) ,
 \label{HiggsPotential}
\end{alignat}
where the two scalar doublets are given by
\begin{equation}
\ensuremath{\Phi_{i}=\left(\begin{array}{c}
\phi_{i}^{+}\\
\frac{1}{\sqrt{2}}(\upsilon_{i}+\rho_{i}+i\eta_{i})
\end{array}\right)},\quad i=1,2,
\end{equation}
\noindent with $\upsilon_{i}$ the vacuum expectation values (VEV) of the fields, while linear combinations of the fields $\rho_i$, $\eta_i$ and $\phi_{i}^{\pm}$ form mass eigenstates:
\begin{equation}\label{eqn:HiggsMixingMatrix}
\begin{pmatrix}G_{Z}\\
A
\end{pmatrix}=R_{\beta}\begin{pmatrix}\eta_{1}\\
\eta_{2}
\end{pmatrix},\quad\begin{pmatrix}G_{W^{\pm}}\\
H^{\pm}
\end{pmatrix}=R_{\beta}\begin{pmatrix}\phi_{1}^{\pm}\\
\phi_{2}^{\pm}
\end{pmatrix},\quad\begin{pmatrix}H\\
h
\end{pmatrix}=R_{\alpha}\begin{pmatrix}\rho_{1}\\
\rho_{2}
\end{pmatrix},
\end{equation}
where the fields $\phi_{i}^{+}$ are charged complex scalars. From the eight degrees of freedom, three of them ($G_{W^{\pm}}$ and $G_{Z}$) get absorbed
by the longitudinal components of the vector bosons. The remaining
five make up the new particle spectrum of the model, namely, $h$
and $H$ are physical CP-even states, $A$ is a CP-odd state and $H^{\pm}$
are two charged Higgs bosons. The rotation matrices are defined
according to
\begin{equation} \label{eqn:GenericMixingMatrix}
R_{\theta}=\left(\begin{array}{cc}
\cos\theta & \sin\theta\\
-\sin\theta & \cos\theta
\end{array}\right).
\end{equation}

In this work, we assume a CP conserving
scalar sector, which implies all the parameters in Eq.~\eqref{HiggsPotential} to be real \cite{Gunion:2002zf}. Additionally, for simplicity, we set $\lambda_{6}=\lambda_{7}=0$. In particular, for this choice of the quartic couplings, the necessary and sufficient conditions to ensure positivity of the potential along all directions are given by \cite{Gunion:2002zf,Branco:2011iw}
\begin{align}
\lambda_1 \geq 0 & , &  \lambda_2 \geq 0 \; ,\label{eq:l1_l2} \\
\lambda_3 \geq -\sqrt{\lambda_1 \lambda_2} & , &
\lambda_3 + \lambda_4 - |\lambda_5| \geq -\sqrt{\lambda_1 \lambda_2} \;,
\label{eq:stability_cond}
\end{align}
\noindent whereas the tree level unitarity of the couplings imposes \cite{Branco:2011iw}
\begin{equation}
\left| a_\pm \right|,\
\left| b_\pm \right|,\
\left| c_\pm \right|,\
\left| d_\pm \right|,\
\left| e_\pm \right|,\
\left| f_\pm \right|
< 8 \pi,
\end{equation}
where
\begin{align}a_{\pm} & =\frac{3}{2}\left(\lambda_{1}+\lambda_{2}\right)\pm\sqrt{\frac{9}{4}\left(\lambda_{1}-\lambda_{2}\right)^{2}+\left(2\lambda_{3}+\lambda_{4}\right)^{2}},\\
b_{\pm} & =\frac{1}{2}\left(\lambda_{1}+\lambda_{2}\right)\pm\frac{1}{2},\,\sqrt{\left(\lambda_{1}-\lambda_{2}\right)^{2}+4\lambda_{4}^{2}},\\
c_{\pm} & =\frac{1}{2}\left(\lambda_{1}+\lambda_{2}\right)\pm\frac{1}{2}\,\sqrt{\left(\lambda_{1}-\lambda_{2}\right)^{2}+4\lambda_{5}^{2}},\\
d_{\pm} & =\lambda_{3}+2\lambda_{4}\pm3\lambda_{5},\\
e_{\pm} & =\lambda_{3}\pm\lambda_{5},\\
f_{\pm} & =\lambda_{3}\pm\lambda_{4}.
\end{align}

Following \cite{Haber:2010bw,Herrero-Garcia:2019mcy} we also include the oblique parameters $S$, $T$ and $U$, which parametrise radiative corrections to electroweak gauge boson propagators. In this study we computed these oblique parameters with the \textsf{2HDMC} package and these are contrasted with the most probable values inferred from experimental data, as found by the \textsf{Gfitter} group~\cite{Baak:2014ora}
\begin{equation}
\begin{aligned}S=0.05\pm0.11,\,\,\,T=0.09\pm0.13,\,\,\,U=0.01\pm0.11,\end{aligned}
\end{equation}
with correlations given by
\begin{equation}
\Sigma=\left(\begin{array}{ccc}
1.0 & 0.9 & -0.59\\
0.9 & 1.0 & -0.83\\
-0.59 & -0.83 & 1.0
\end{array}\right)\,.
\end{equation}

\subsection{Yukawa Lagrangian} \label{sec:Yukawas}

The most general Yukawa Lagrangian in the generic scalar basis $\{\Phi_{1},\Phi_{2}\}$
reads \cite{Herrero-Garcia:2019mcy}:
\begin{equation}
    -\mathcal{L}_{Yukawa}=\bar{Q}^{0}\,(Y_{1}^{u}\tilde{\Phi}_{1}+Y_{2}^{u}\tilde{\Phi}_{2})u_{{\rm R}}^{0}+\bar{Q}^{0}\,(Y_{1}^{d}\Phi_{1}+Y_{2}^{d}\Phi_{2})d_{{\rm R}}^{0}+\bar{L}^{0}\,(Y_{1}^{l}\Phi_{1}+Y_{2}^{l}\Phi_{2})l_{{\rm R}}^{0}+{\rm ~h.c.}\,\label{eq:yuk2d},
\end{equation}
where the superscript "0" notation refers to the flavour eigenstates, and $\tilde{\Phi}_j = i \sigma_2 \Phi_j^\dagger$. The fermion mass matrices are given by
\begin{equation}
    M_{f}=\frac{1}{\sqrt{2}}(v_{1}Y_{1}^{f}+v_{2}Y_{2}^{f}),\qquad f=u,d,l.\label{masa-fermiones}  
\end{equation}
Notice that this matrices need to be diagonalized. This can be done through a bi-unitary transformation
\begin{equation}
    \bar{M}_{f}=V_{fL}^{\dagger}M_{f}V_{fR},\label{masa-diagonal}
\end{equation}
where the fact that $M_{f}$ is Hermitian implies that
$V_{fL}=V_{fR}$, and the mass eigenstates for the fermions are given by
\begin{equation}
    u=V_{u}^{\dagger}u^{0},\qquad d=V_{d}^{\dagger}d^{0},\qquad l=V_{l}^{\dagger}l^{0}.\label{redfields}
\end{equation}
Then, Eq.~(\ref{masa-fermiones}) takes the form
\begin{equation}
    \bar{M}_{f}=\frac{1}{\sqrt{2}}(v_{1}\tilde{Y}_{1}^{f}+v_{2}\tilde{Y}_{2}^{f}),\label{diag-Mf}
\end{equation}
where $\tilde{Y}_{i}^{f}=V_{fL}^{\dagger}Y_{i}^{f}V_{fR}$, though
each Yukawa matrix is not diagonalized by this transformation. For this reason we shall drop the tilde from now on. Solving for $Y_{1}^{f}$ we have
\begin{equation}
Y_{1,ba}^{f}=\frac{\sqrt{2}}{v\cos\beta}\bar{M}_{f,ba}-\tan\beta Y_{2,ba}^{f}.
\end{equation}
Using the expressions above we can write the Yukawa Lagrangian in the mass basis as\footnote{This Yukawa Lagrangian differs from the one defined in Eq.(2.3) in \cite{Crivellin:2019dun} by an overall factor of $\sqrt{2}$.} 
\begin{equation}
\begin{aligned}-\mathcal{L}_{Yukawa}\, =& \bar{u}_{b} \left(V_{bc}\xi_{ca}^{d}P_{R} - V_{ca}\xi_{cb}^{u*}P_{L}\right) d_{a}\,H^{+} + \bar{\nu}_{b}\xi_{ba}^{l}P_{R}l_{a}\,H^{+} + \mathrm{h.c.}\\
 & +\sum_{f=u,d,e}\sum_{\phi=h,H,A}\bar{f}_{b} \Gamma_{\phi ba}^{f}P_{R}f_{a}\phi+\mathrm{h.c.},
\end{aligned}
\label{eq:YukawaLagran}
\end{equation}
where $a,b=1,2,3$ and
\begin{equation}
\xi_{ba}^{f}\equiv\dfrac{Y_{2,ba}^{f}}{\cos\beta}-\dfrac{\sqrt{2}\tan\beta\bar{M}_{f,ba}}{v},\label{eq:Xis}
\end{equation}
\begin{align}
\Gamma_{hba}^{f} & \equiv\dfrac{\bar{M}_{f,ba}}{v}s_{\beta-\alpha}+\dfrac{1}{\sqrt{2}}\xi_{ba}^{f}c_{\beta-\alpha},\label{eq:Gammafhba}\\
\Gamma_{Hba}^{f} & \equiv\dfrac{\bar{M}_{f,ba}}{v}c_{\beta-\alpha}-\dfrac{1}{\sqrt{2}}\xi_{ba}^{f}s_{\beta-\alpha},\label{eq:GammafHba}\\
\Gamma_{Aba}^{f} & \equiv\begin{cases}
-\dfrac{i}{\sqrt{2}}\xi_{ba}^{f} & \textrm{if }f=u,\\
\dfrac{i}{\sqrt{2}}\xi_{ba}^{f} & \textrm{if }f=d,l.
\end{cases}\label{eq:GammafAba}
\end{align}

At first, the total number of new complex Yukawa couplings
to consider is 54. Considering only their real parts and the ansatz
\begin{equation} 
\xi^{u}=\left(\begin{array}{ccc}
0 & 0 & 0\\
0 & \xi_{cc}^{u} & \xi_{ct}^{u}\\
0 & \xi_{tc}^{u} & \xi_{tt}^{u}
\end{array}\right),\qquad\xi^{d}=\left(\begin{array}{ccc}
0 & 0 & 0\\
0 &  \xi_{ss}^{d} & \xi_{sb}^{d}\\
0 &  \xi_{bs}^{d} & \xi_{bb}^{d}
\end{array}\right),\qquad\xi^{l}=\left(\begin{array}{ccc}
0 & 0 & 0\\
0 & \xi_{\mu\mu}^{l} & \xi_{\mu\tau}^{l}\\
0 & \xi_{\tau\mu}^{l} & \xi_{\tau\tau}^{l}
\end{array}\right),\label{eq:Textures}
\end{equation}
we get only 12 Yukawa parameters (i.e., ignoring $3\to1$ and $2\to1$ generation transitions). Here, the $\xi^{u}$ matrix has been previously  considered to be asymmetric from $B_{s}-\overline{B}_{s}$ oscillations constraints at one loop level and for heavy Higgs masses of order $\lesssim 700\,\mathrm{GeV}$ \cite{Altunkaynak:2015twa,Hou:2020chc}. However, since we are approaching the dominant contribution process at LO and we are exploring masses in the range  $[0.5,\,4.0]\,\mathrm{TeV}$ as in \cite{Herrero-Garcia:2019mcy}, we will consider only the symmetric case, i.e., $\xi_{tc}^{u}=\xi_{ct}^{u}$. Hence, assuming the remaining $\xi^{d}$ and $\xi^{l}$ matrices to be symmetric as well, the total number of parameters to scan over is reduced by 3. 

\section{Effective Hamiltonians for flavour changing transitions}
\label{sec:Hamiltonian}

Most of the relevant flavour observables that we consider in this work arise from processes with either suppressed or negligible contributions from SM particles. Hence, these processes are often dominated by BSM contributions, which can be generated by a large variety of UV complete theories. It is often convenient to study these transitions using the, model-agnostic, effective Hamiltonian approach, where transition operators are decomposed using the Operator Product Expansion (OPE) into a collection of simple, low-energy, operators. Associated with each of these operators comes a WC, which encodes the knowledge of the high-energy theory. Even for complete high-energy theories, as it is our case, it is extremely useful to work with the effective Hamiltonian, since one can easily compute most observables of interest in terms of a small set of WCs. In fact, there are only two independent flavour changing transitions that give rise to the majority of the studied observables, and these are the neutral  $b\to s\ell^+\ell^-$ transition and the charged $b\to c\ell\bar{\nu}$ transition. In this section we write down the effective Hamiltonian for both of these transitions and provide expressions for the BSM contributions to the WCs that arise in our model.\footnote{These BSM new contributions for $b\to s\ell^+\ell^-$ and $b\to c\ell\bar{\nu}$ transitions were included in our local version of \textsf{FlavBit} and might appear in a future release.}

\subsection{\texorpdfstring{$b\to s\ell^+\ell^-$ transitions}{b to smumu}}

\begin{figure}[h]
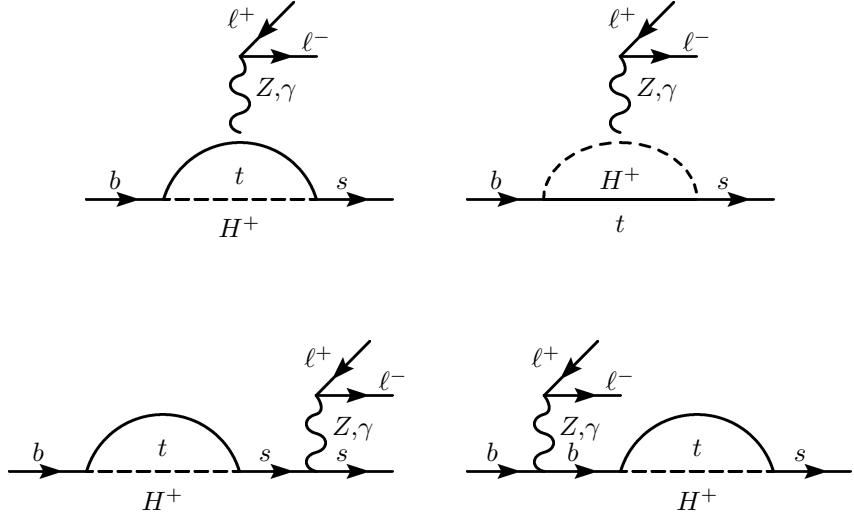

{\Large{}
\[
\Diagram{ &  & \momentum[top]{fuV}{\ell^{+}\quad}\\
 &  & \momentum{fA}{}\vertexlabel^{\ell^{-}}\\
 &  & \momentum[urt]{gv}{\:Z,\gamma}\\
\momentum{fA}{b} & h\vertexlabel_{H^{+}} & \momentum[bot]{fl}{t}h\momentum{fA}{s\;}
}
\quad\quad\Diagram{ &  & \momentum[top]{fuV}{\ell^{+}\quad}\\
 &  & \momentum{fA}{}\vertexlabel^{\ell^{-}}\\
 &  & \momentum[urt]{gv}{\:Z,\gamma}\\
\momentum{fA}{b} & f\vertexlabel_{t} & \momentum[bot]{hl}{H^{+}}f\momentum{fA}{s\;}
}
\]
}{\Large\par}

{\Large{}$ $}{\Large\par}

{\Large{}$ $}{\Large\par}

{\Large{}
\[
\Diagram{\\
 &  &  & \momentum[top]{fuV}{\ell^{+}\quad}\\
 &  &  & \momentum{fA}{}\vertexlabel^{\ell^{-}}\\
\momentum{fA}{b} & h\vertexlabel_{H^{+}} & \momentum[bot]{fl}{t}h\momentum{fA}{s\;} & \momentum[urt]{gv}{\:Z,\gamma}\\
 &  &  & \momentum{fA}{s\;}
}
\quad\quad\Diagram{ & \momentum[top]{fuV}{\ell^{+}\quad}\\
 & \momentum{fA}{}\vertexlabel^{\ell^{-}}\\
 & \momentum[urt]{gv}{\:Z,\gamma} & h\vertexlabel_{H^{+}} & \momentum[bot]{fl}{t}h\momentum{fA}{s\;}\\
\momentum{fA}{b\;} & \momentum{fA}{b}
}
\]
}{\Large\par}

{\Large{}\caption{\emph{Penguin diagrams for $b\to s\ell^{+}\ell^{-}$ transitions.
\label{fig:Penguin-diagrams.}}}
}{\Large\par}
\end{figure}

\begin{figure}[h]
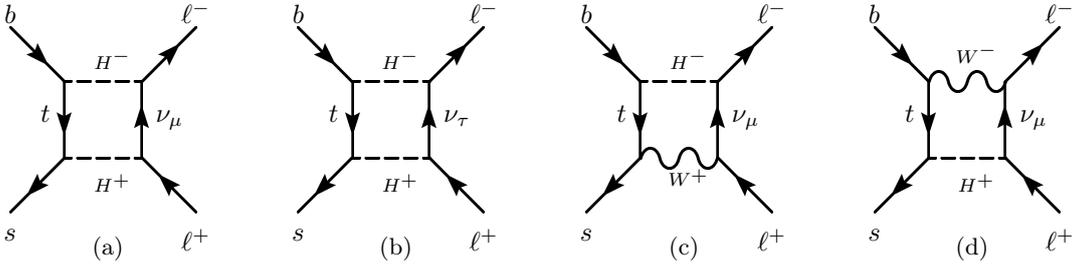

\begin{centering}
{\Large{}}\subfloat[\label{fig:a}]{\begin{centering}
{\Large{}$\Diagram{\vertexlabel^{b} &  &  &  & \vertexlabel^{\ell^{-}}\\
\momentum{fdA}{} &  &  & \momentum{fuA}{}\\
 &  & \momentum{h}{\quad\overset{H^{-}}{}}\\
 & \momentum[ulft]{fvV}{t\,} & \momentum[bot]{h}{\quad\underset{H^{+}}{}} & \momentum{fvA}{\,\nu_{\mu}}\\
\vertexlabel_{s}\momentum{fuV}{} &  &  & \momentum{fdV}{}\vertexlabel_{\ell^{+}}
}
$}{\Large\par}
\par\end{centering}
{\Large{}}{\Large\par}}{\Large{}$\quad\quad$}\subfloat[\label{fig:b}]{\begin{centering}
{\Large{}$\Diagram{\vertexlabel^{b} &  &  &  & \vertexlabel^{\ell^{-}}\\
\momentum{fdA}{} &  &  & \momentum{fuA}{}\\
 &  & \momentum{h}{\quad\overset{H^{-}}{}}\\
 & \momentum[ulft]{fvV}{t\,} & \momentum[bot]{h}{\quad\underset{H^{+}}{}} & \momentum{fvA}{\,\nu_{\tau}}\\
\vertexlabel_{s}\momentum{fuV}{} &  &  & \momentum{fdV}{}\vertexlabel_{\ell^{+}}
}
$}{\Large\par}
\par\end{centering}
{\Large{}}{\Large\par}}{\Large{}$\quad\quad$}\subfloat[\label{fig:c}]{\begin{centering}
{\Large{}$\Diagram{\vertexlabel^{b} &  &  &  & \vertexlabel^{\ell^{-}}\\
\momentum{fdA}{} &  &  & \momentum{fuA}{}\\
 &  & \momentum{h}{\quad\overset{H^{-}}{}}\\
 & \momentum[ulft]{fvV}{t\,} & \momentum[bot]{g}{\quad\underset{W^{+}}{}} & \momentum{fvA}{\,\nu_{\mu}}\\
\vertexlabel_{s}\momentum{fuV}{} &  &  & \momentum{fdV}{}\vertexlabel_{\ell^{+}}
}
$}{\Large\par}
\par\end{centering}
{\Large{}}{\Large\par}}{\Large{}$\quad\quad$}\subfloat[\label{fig:d}]{\begin{centering}
{\Large{}$\Diagram{\vertexlabel^{b} &  &  &  & \vertexlabel^{\ell^{-}}\\
\momentum{fdA}{} &  &  & \momentum{fuA}{}\\
 &  & \momentum{g}{\quad\overset{W^{-}}{}}\\
 & \momentum[ulft]{fvV}{t\,} & \momentum[bot]{h}{\quad\underset{H^{+}}{}} & \momentum{fvA}{\,\nu_{\mu}}\\
\vertexlabel_{s}\momentum{fuV}{} &  &  & \momentum{fdV}{}\vertexlabel_{\ell^{+}}
}
$}{\Large\par}
\par\end{centering}
{\Large{}}{\Large\par}}{\Large\par}
\par\end{centering}
{\Large{}\caption{\emph{Box diagrams for $b\to s\ell^{+}\ell^{-}$ transitions.\label{fig:Box-diagrams.}}}
}{\Large\par}
\end{figure}

The effective Hamiltonian responsible for $b\to s\ell^+\ell^-$ transitions can be written as
\begin{align}
{\cal H}_{\mathrm{eff}} & =-\frac{4G_{F}}{\sqrt{2}}V_{tb}V_{ts}^{*}\left[\sum_{i=S,P}C_{i}(\mu)\mathcal{O}_{i}+C_{i}^{\prime}(\mu)\mathcal{O}_{i}^{\prime}+\sum_{i=7}^{10} C_{i}(\mu)\mathcal{O}_{i}+C_{i}^{\prime}(\mu)\mathcal{O}_{i}^{\prime}\right],\label{eq:heff}
\end{align}
where $\mu$ is the energy scale at which the WCs are defined, and
\begin{alignat}{3}
\mathcal{O}_{9} & =\frac{e^{2}}{16\pi^{2}}(\bar{s}\gamma_{\mu}P_{L}b)(\bar{\ell}\gamma^{\mu}\ell),\qquad\qquad &  & \mathcal{O}_{10} &  & =\frac{e^{2}}{16\pi^{2}}(\bar{s}\gamma_{\mu}P_{L}b)(\bar{\ell}\gamma^{\mu}\gamma_{5}\ell),\label{eq:basisA}\\
\mathcal{O}_{S} & =\frac{e^{2}}{16\pi^{2}}m_{b}(\bar{s}P_{R}b)(\bar{\ell}\ell),\qquad\qquad &  & \mathcal{O}_{P} &  & =\frac{e^{2}}{16\pi^{2}}m_{b}(\bar{s}P_{R}b)(\bar{\ell}\gamma_{5}\ell),\\
\mathcal{O}_{7} & =\frac{e}{16\pi^{2}}m_{b}(\bar{s}\sigma^{\mu\nu}P_{R}b)F_{\mu\nu},\qquad\qquad &  & \mathcal{O}_{8} &  & =\frac{g}{16\pi^{2}}m_{b}\bar{s}\sigma^{\mu\nu}T^{a}P_{R}bG_{\mu\nu}^{a},
\end{alignat}
are the FCNC local operators encoding the low-energy description of the high energy physics that has been integrated out. The prime operators are obtained by the replacement $P_{R(L)}\rightarrow P_{L(R)}$.  The WCs can be written as
\begin{align}
C_{i} & =C_{i}^{\mathrm{SM}}+\Delta C_{i}, 
\end{align}
\noindent where $C_{i}^{\mathrm{SM}}$ is the SM contribution to the $i$th WC and $\Delta C_{i}$ is the NP contribution, a prediction of the GTHDM model. The SM contribution to the scalar WCs, $C_{S,P}^{(\prime)}$, is negligible, whereas for $C_{7-10}$ we have
\begin{align}
\mathrm{Re}(C_{7,8,9,10}^{\mathrm{SM}})=-0.297,\,-0.16,\,4.22,\,-4.06,
\end{align}
as computed with \textsf{SuperIso}. We evaluate the NP scalar and pseudoscalar coefficients $\Delta C_{S,P}^{(\prime)}$ at tree level, which is the LO contribution from the GTHDM \cite{Crivellin:2019dun}. Henceforth we will use the scalar and pseudoscalar coefficients in the basis defined in \textsf{SuperIso}, i.e.,  $C_{Q_{1},Q_{2}}^{(')}=m_{b(s)}C_{S,P}^{(')}$. The remaining coefficients, $\Delta C_{7,8,9,10}$ first appear at one loop level and we therefore include the one-loop BSM contributions to these in our analysis.  These one-loop corrections can be split by contribution as follows,
\begin{align}
\Delta C_{7,8} & =C_{7,8}^{\gamma,\,g},\label{eq:DeltaC78}\\
\Delta C_{9} & =C_{9}^{\gamma}+C_{9}^{Z}+C_{9}^{\textrm{box}},\label{eq:DeltaC9}\\
\Delta C_{10} & =C_{10}^{Z}+C_{10}^{\textrm{box}}.\label{eq:DeltaC10}
\end{align}
where $C_{9,10}^{Z}$ and $C_{7,9}^{\gamma}$ come from the $Z$ and $\gamma$ penguins, respectively (figure\ \ref{fig:Penguin-diagrams.}), and $C_{9,10}^{\textrm{box}}$ are contributions from box diagrams, (figure\ \ref{fig:Box-diagrams.}). At this level, the $\Delta C_{9}^{'}$ and $\Delta C_{10}^{'}$ coefficients are suppressed as $m_{b}/m_{t}$ with respect to their non-prime counterparts.  However, for studying the effects of flavour-changing Yukawa couplings we include these coefficients for completeness. $C_{8}^{g}$ is the WC related to the chromomagnetic operator coming from gluon penguins and the NP contributions $\Delta C_{7,8}^{'}$ are computed in \cite{Crivellin:2019dun}.

{\Large{}}
{\Large\par}

{\Large{}}
{\Large\par}

\subsubsection{Penguins and boxes computation}

We review the computation of the WCs in Eqs. (\ref{eq:DeltaC78}-\ref{eq:DeltaC10}) which have been obtained already for both the flavour conserving general THDM in \textsf{SuperIso} and for the GTHDM itself in \cite{Iguro:2017ysu,Iguro:2018qzf,Crivellin:2019dun}. In these latter works, the Yukawa couplings related to $\xi^{d}$ were assumed to be zero or negligibly small from the beginning, avoiding the appearance of possible mixed terms between the down and up couplings that, at first, might not be as suppressed as those involving only down quarks. This computation is performed assuming $\ell=\mu$ in the final state, as inspired by our choice of Yukawa textures in Eq.~\eqref{eq:Textures}, but it can be easily generalised for all flavours when required.

Using the model files provided by \texttt{FeynRules} from \cite{Degrande:2014vpa}, we generate in \texttt{FeynArts} the one loop level Feynman diagrams for $b\to s\mu^{+}\mu^{-}$ transitions.
After this, the amplitudes are tensor decomposed in\texttt{ FeynCalc} \cite{Shtabovenko:2016sxi} and then, the resulting Passarino-Veltman functions are Taylor expanded in the external momenta up to second
order. Finally, the functions are integrated with \texttt{Package X} \cite{Patel:2015tea}. In this way, with $x_{tH^{\pm}}=m_{t}^{2}(\mu_{W})/m_{H^{\pm}}^{2}$ for $\mu_W=\mathcal{O}(m_{W})$
we obtain \new{\footnote{We additionally computed the WCs using the Modern ARtificial Theoretical phYsicist (\texttt{MARTY-1.4}) C++ package \cite{Uhlrich:2020ltd}, obtaining a very good numerical agreement compared to the resultant expressions from \texttt{Package X}.}}
\begin{flalign}
&\begin{aligned}
C_{9}^{\gamma}=\frac{-\Gamma_{tb}^{L}\Gamma_{ts}^{L}}{\sqrt{2}G_{F}V_{tb}V_{ts}^{*}m_{t}^{2}\lambda_{tt}^{2}}\mathcal{D}^{H(0)}(x_{tH^{\pm}}),
\end{aligned}&&
\end{flalign}
\begin{flalign}
&\begin{aligned}
C_{9}^{Z} & =\frac{\Gamma_{tb}^{L}\Gamma_{ts}^{L}}{\sqrt{2}G_{F}V_{tb}V_{ts}^{*}m_{t}^{2}\lambda_{tt}^{2}}\frac{\left(1-4s_{W}^{2}\right)}{s_{W}^{2}}\mathcal{C}^{H(0)}(x_{tH^{\pm}})+\frac{m_{b}}{m_{t}}\frac{\Gamma_{tb}^{R}\Gamma_{ts}^{L}}{\sqrt{2}G_{F}V_{tb}V_{ts}^{*}}\mathcal{C}_{\textrm{mix}}^{H(0)}(x_{tH^{\pm}}),\label{eq:C9Z}
\end{aligned}&&
\end{flalign}
\begin{flalign}
&\begin{aligned}
C_{9}^{\textrm{box}}=C_{10}^{\textrm{box}} & =\frac{\Gamma_{tb}^{L}\Gamma_{ts}^{L}}{32G_{F}^{2}V_{tb}V_{ts}^{*}m_{t}^{2}}\left|\Gamma_{\nu_{i}\mu}^{R}\right|^{2}\mathcal{B}^{H(0)}(x_{tH^{\pm}})+\frac{m_{\mu}\,\xi_{\mu\mu}^{l}}{8\sqrt{2}G_{F}m_{W}^{3}s_{W}^{2}\text{\ensuremath{V_{tb}}}V_{ts}^{*}}\mathcal{B}^{H(0)}_{\textrm{mix}}(x_{tH^{\pm}},\,H),\label{eq:C9box}
\end{aligned}&&
\end{flalign}
\begin{flalign}
&\begin{aligned}
C_{10}^{Z} & =\frac{1}{\left(4s_{W}^{2}-1\right)}C_{9}^{Z},
\end{aligned}&&
\end{flalign}
\begin{flalign}
&\begin{aligned}
C_{7,8}^{\gamma,g} & =\frac{\Gamma_{tb}^{L}\Gamma_{ts}^{L}}{3\sqrt{2}G_{F}V_{tb}V_{ts}^{*}m_{t}^{2}\lambda_{tt}^{2}}F_{7,8}^{(1)}(x_{tH^{\pm}})-\frac{\Gamma_{tb}^{R}\Gamma_{ts}^{L}}{\sqrt{2}G_{F}V_{tb}V_{ts}^{*}m_{b}m_{t}\lambda_{tt}\lambda_{bb}}F_{7,8}^{(2)}(x_{tH^{\pm}}),
\end{aligned}&&
\end{flalign}
where
\begin{equation}
\Gamma_{ts}^{L}=\frac{1}{\sqrt{2}}\sum_{l=1}^{3}\xi_{l3}^{u}V_{l2}^{*},\quad\Gamma_{tb}^{L}=\frac{1}{\sqrt{2}}\sum_{k=1}^{3}V_{kt}\xi_{k3}^{u*},\label{eq:GammaLts}
\end{equation}
\begin{equation}
\Gamma_{tb}^{R}=\frac{1}{\sqrt{2}}\sum_{k=1}^{3}V_{kt}\xi_{k3}^{d*},\quad\left|\Gamma_{\nu_{i}\mu}^{R}\right|^{2}=\frac{1}{2}\left(\left|\xi_{\mu\mu}^{l}\right|^{2}+\left|\xi_{\tau\mu}^{l}\right|^{2}\right),\label{eq:GammaRtb}
\end{equation}

with the Green functions $\mathcal{D}^{H(0)},\,\mathcal{C}^{H(0)},\,F_{7,8}^{(1)}$ and $F_{7,8}^{(2)}$  defined in appendices C1 and C2 in \cite{SuperIso4.1}. Here, $\lambda_{ii}$ are the diagonal Yukawa couplings defined in \textsf{SuperIso}, $G_{F}$ is the Fermi constant and $s_{W}$ is the sine of the Weinberg angle. The Green function $\mathcal{B}^{H(0)}$ for the  box diagram contribution in $C_{9,10}^{\textrm{box}}$ coming from the new lepton flavour violating (LFV) couplings is given by
\begin{equation}
\mathcal{B}^{H(0)}(t)=\frac{t\left(t-t\log t-1\right)}{m_{W}^{2}s_{W}^{2}(t-1)^{2}}.\label{eq:Greenf_Box}
\end{equation}

Our computation shows two new terms absent in both the \textsf{SuperIso} manual and in \cite{Crivellin:2019dun}, namely the mixed term in the $C_{9}^{Z}$ expression where 
\begin{equation}
\mathcal{C}_{\textrm{mix}}^{H(0)}(t)=-\frac{\left(1-4s_{W}^{2}\right)t\left(t^{2}-2\,t\log t-1\right)}{16m_{W}^{2}s_{W}^{2}(t-1)^{3}},
\end{equation}
and a gauge dependent contribution to $C_{9}^{\textrm{box}}$ coming from the box diagrams in figures \ref{fig:c} and \ref{fig:d} proportional to $\mathcal{B}^{H(0)}_{\textrm{mix}}(t,\,H)$ with $H=m_{H^{\pm}}^{2}/m_{W}^{2}$ (see appendix \ref{sec:gauge-term}).

For all remaining terms, we obtained full agreement with \cite{Crivellin:2019dun} once the overall
$\sqrt{2}$ factor in their Yukawa Lagrangian is taken into account
compared to our Eq.~(\ref{eq:YukawaLagran}). It is important to mention here that once the full quantum field theory matches with the effective theory at a scale $\mu_W=\mathcal{O}(m_{W})$, the evolution of the WC $C_{7}$ (and $C_{7}^{'}$) from $\mu=\mu_W$ down to $\mu=\mu_b$, where $\mu_b$ is of the order of $m_b$, is given at LO by \cite{Buras:1998raa}
\begin{equation}
C_{7}^{\textrm{eff}}(\mu_b)=\eta^{\frac{16}{23}}C_{7}+\frac{8}{3}\left(\eta^{\frac{14}{23}}-\eta^{\frac{16}{23}}\right)C_{8}+\sum_{i=1}^{8}h_{i}\eta^{a_{i}}\,C_{2}\,,
\label{C7RGE}
\end{equation}
where $\eta=\alpha_{S}(\mu_{W})/\alpha_{S}(\mu_{b})$ and the renormalisation group evolution of the QCD coupling is
\begin{equation}
\alpha_{S}(\mu_{b})=\frac{\alpha_{S}(m_{Z})}{1-\beta_{0}\frac{\alpha_{S}(m_{Z})}{2\pi}\log(m_{Z}/\mu_{b})},
\end{equation}
with $\beta_{0}=23/3$. The $\sum_{i=1}^{8}h_{i}\eta^{a_{i}}$ factor in Eq.~\eqref{C7RGE} is given in Eq.(12.23) of  \cite{Buras:1998raa} and references therein. The $C_{2}$ coefficient
comes from four-quark operators generated by $W$ boson exchange in the SM and contributes importantly when computing the branching ratio $\mathrm{BR}(\overline{B}\rightarrow X_{s}\gamma)$.
In the GTHDM, as shown in \cite{Crivellin:2019dun}, an analogous  contribution comes from charged Higgs exchange at tree level. In this way, following \cite{Buchalla:1995vs} with $\alpha_{S}(m_{Z})=0.117$, we use the following parametric expression at LO: 

\begin{equation}
C_{7}^{\textrm{eff}}(\mu_b)=0.698\,C_{7}+0.086\,C_{8}-0.158\,C_{2},
\end{equation}
where $C_{2}=C_{2}^{\mathrm{SM}}+\Delta C_{2}$ for $C_{2}^{\mathrm{SM}}=1$ and
\begin{equation}
    \Delta C_{2}=  -\dfrac{7}{18}\dfrac{m_{W}^{2}}{m_{H^{\pm}}^{2}}\dfrac{V_{k2}^{*}\xi_{k2}^{u}\xi_{n2}^{u*}V_{n3}}{g_{2}^{2}V_{tb}V_{ts}^{*}}-\dfrac{1}{3}\dfrac{m_{c}}{m_{b}}\dfrac{m_{W}^{2}}{m_{H^{\pm}}^{2}}\dfrac{V_{k2}^{*}\xi_{k2}^{u}V_{2n}\xi_{n3}^{d}}{g_{2}^{2}V_{tb}V_{ts}^{*}}\left(3+4\log\left(\dfrac{\mu_{b}^{2}}{m_{H^{+}}^{2}}\right)\right).\,
\end{equation}
with $g_{2}$ the weak coupling constant. Similarly, there will be a contribution to the $C_{9}$ (and $C_{9}^{'}$) WC coming from those four-quark operators given by \cite{Crivellin:2019dun} 
\begin{equation}
C_{9}^{4-\mathrm{quark}}(\mu_b)=\dfrac{2}{{27}}\dfrac{{V_{k2}^{*}\xi_{k2}^{u}\xi_{n2}^{u*}{V_{n3}}}}{{g_{2}^{2}{V_{tb}}V_{ts}^{*}}}\dfrac{m_{W}^{2}}{m_{H^{\pm}}^{2}}\Bigg(19+12\log\!\left(\!\dfrac{\mu_{b}^{2}}{m_{H^{\pm}}^{2}}\!\right)\Bigg),
\end{equation}
 which can be added at LO to both the penguins and boxes contributions, obtaining 
 
 \begin{equation}
C_{9}^{\mathrm{eff}}(\mu_b)=C_{9}+C_{9}^{4-\mathrm{quark}}(\mu_b).
\end{equation}

\subsubsection{Summary of contributions}

\begin{figure}[h]
\begin{centering}
\includegraphics[scale=0.4]{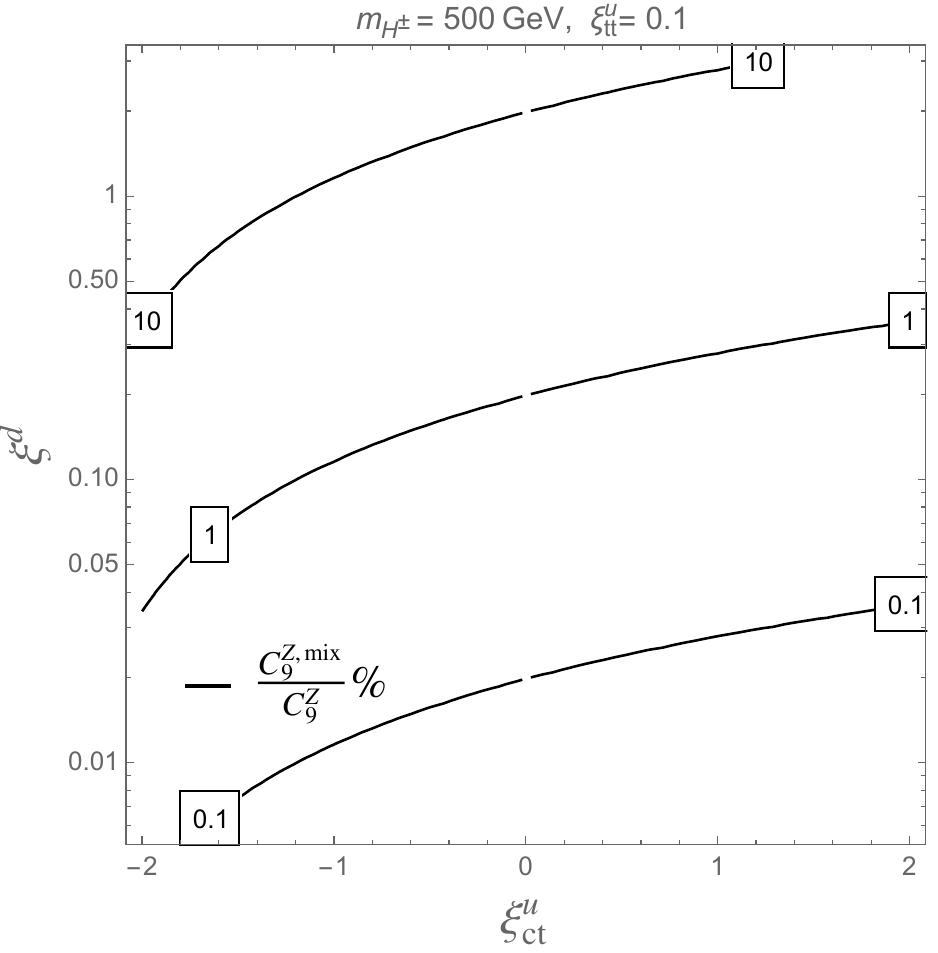} $\qquad$\includegraphics[scale=0.4]{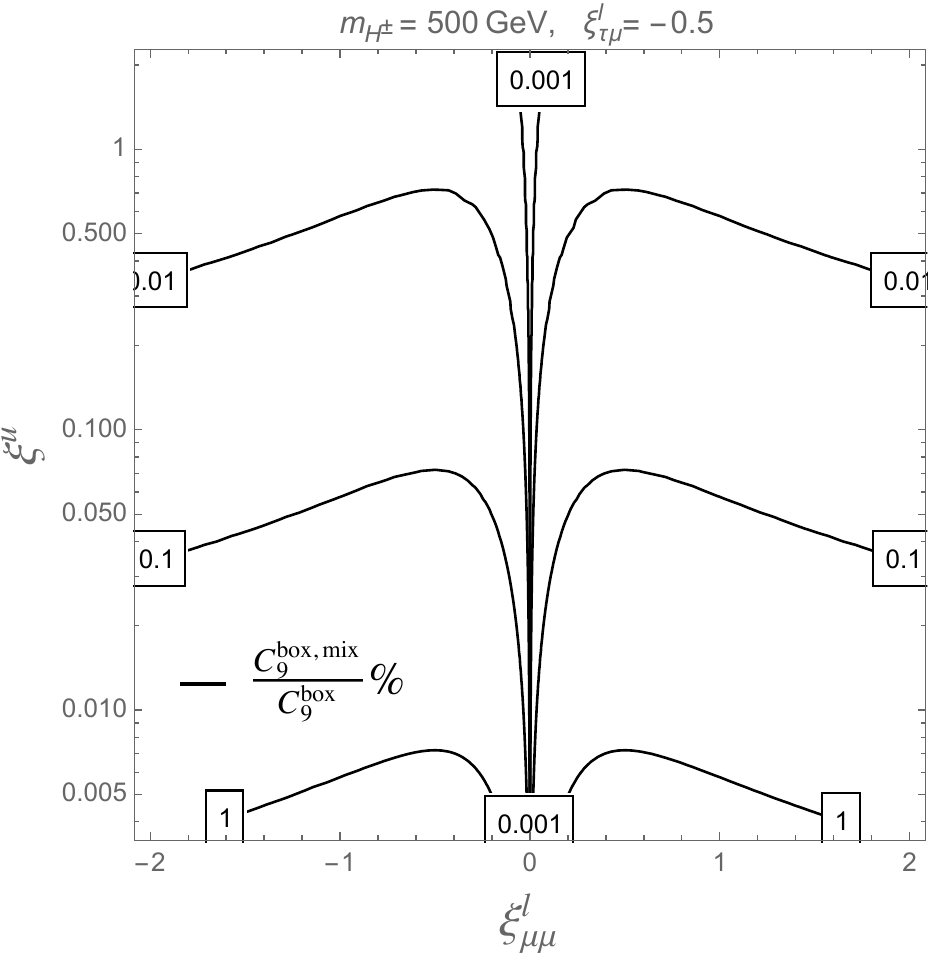} 
\par\end{centering}
\caption{\emph{Left: $C_{9}^{Z,mix}/C_{9}^{Z}$ contour levels for $\xi_{sb}^{d}=\xi_{bb}^{d}=\xi^{d}$.
Here $C_{9}^{Z}$ and $C_{9}^{Z,mix}$ refers to the first and second
terms in Eq.~(\ref{eq:C9Z}) respectively. Right: $C_{9}^{box,mix}/C_{9}^{box}$
contour levels for $\xi_{tt}^{u}=\xi_{ct}^{u}=\xi^{u}$. $C_{9}^{box}$ and $C_{9}^{box,mix}$ refers to the first and second
terms in Eq.~(\ref{eq:C9box}) respectively.\label{fig:comparison}}}
\end{figure}

As already mentioned, in view of the flavour changing couplings in the GTHDM, there are two new contributions compared to the ones present in \textsf{SuperIso}. These contributions come from the box diagrams in figures\ \ref{fig:a}-\ref{fig:b} and from the $Z$ penguin in figure\ \ref{fig:Penguin-diagrams.}. The $\Gamma_{tb}^{L}\Gamma_{ts}^{L}$ contribution is the largest and dominates the amplitude for most of the parameter space, with a strong dependence on $\tan\beta$, $m_{H^{\pm}},$ $Y_{2,ct/tc}^{u},\,Y_{2,tt}^{u},\,Y_{2,\mu\mu}^{l},\,Y_{2,\mu\tau}^{l}$. There are also two subdominant contributions, the first one coming from the part proportional to $\Gamma_{tb}^{R}\Gamma_{ts}^{L}$ in the $Z$ penguin diagram in figure\ \ref{fig:Penguin-diagrams.}. When comparing its contribution relative to the $\Gamma_{tb}^{L}\Gamma_{ts}^{L}$ term, we find regions of the parameter space in which it can make up to 10$\%$ of the total contribution (see figure\ \ref{fig:comparison} \textit{left}). The second subdominant contribution is the already mentioned gauge dependent part of the boxes diagrams (figures\ \ref{fig:c}-\ref{fig:d}) which is suppressed by the muon mass (see figure\ \ref{fig:comparison} \textit{right}). Additionally we verified that when varying the mass of the charged Higgs from 500 GeV to 4000 GeV these ratios were essentially unaffected. In this way, we keep in our calculations the  $\Gamma_{tb}^{R}\Gamma_{ts}^{L}$ term from the $Z$ penguin and neglect the gauge dependent part of the boxes diagrams.

\subsection{\texorpdfstring{$b\to c \ell \overline{\nu}$}{b to clnu} semileptonic transitions}

As a consequence of the new interactions between the fermions and the charged Higgs, semileptonic tree level flavour changing transitions appear in the GTHDM (figure\ \ref{fig:RD-tree-level}) which have been extensively studied in the literature \cite{Celis:2012dk,Crivellin:2012ye,Crivellin2013,Alonso:2016oyd,Iguro:2018qzf,Martinez:2018ynq}. Therefore we include tree-level calculations of the Wilson coefficients related to these in our analysis. The effective Hamiltonian responsible for the $b\to c \ell \overline{\nu}$ 
transitions for the semileptonic decays of $B$-mesons, including the SM and tree level GTHDM contributions can be
written in terms of scalar operators in the form
\begin{equation}
\begin{array}{l}
{\cal H}_{{\rm eff}}=C_{SM}^{cb}{\cal O}_{SM}^{cb}+C_{R}^{cb}{\cal O}_{R}^{cb}+C_{L}^{cb}{\cal O}_{L}^{cb},
\end{array}
\label{eq:Heffective}
\end{equation}
where $C_{SM}^{cb}=4G_{F}V_{cb}/\sqrt{2}$ and the operators are given by 
\begin{equation}
\begin{array}{l}
{\cal O}_{SM}^{cb}=\left(\bar{c}\gamma_{\mu}P_{L}b\right)\left(\bar{\ell}\gamma_{\mu}P_{L}\nu\right),\\
{\cal O}_{R}^{cb}=\left(\bar{c}P_{R}b\right)\left(\bar{\ell}P_{L}\nu\right),\\
{\cal O}_{L}^{cb}=\left(\bar{c}P_{L}b\right)\left(\bar{\ell}P_{L}\nu\right).
\end{array}\label{eq:Oeffective}
\end{equation}
\begin{figure}[h]
\begin{centering}
{\Large{}$\Diagram{\vertexlabel^{c} &  &  &  & \vertexlabel^{\ell}\\
 & fdV &  & fuA\\
 &  & \momentum[bot]{h}{\quad\,H^{-}}\\
 & fuA &  & fdV\vertexlabel_{\overline{\nu}}\\
\vertexlabel_{b}
}
$}{\Large\par}
\par\end{centering}
\caption{\emph{Tree level contribution to $b\to c \ell \overline{\nu}$.\label{fig:RD-tree-level}}}
\end{figure}

Given that the flavour of the neutrino in the final state can not be discerned by experiments, one has to add (incoherently) to the SM the NP contributions associated with the LFV couplings $\xi_{ij}^{l}$. As the existing constraints will apply separately to the scalar and
the pseudoscalar couplings, it is convenient to define 
\begin{eqnarray}
g_{S}^{\ell\ell^{\prime}}\equiv\frac{C_{R}^{cb}+C_{L}^{cb}}{C_{SM}^{cb}},\ g_{P}^{\ell\ell^{\prime}}\equiv\frac{C_{R}^{cb}-C_{L}^{cb}}{C_{SM}^{cb}},
\end{eqnarray}
where in our analysis we evaluate the WCs $C_{R}^{cb}$ and $C_{L}^{cb}$ at tree-level,  with the expressions,
\begin{equation}
C_{R}^{cb}=-2\frac{(V_{cb}\xi_{bb}^{d}+V_{cs}\xi_{sb}^{d})\xi_{\ell\ell^{\prime}}^{l*}}{m_{H^{\pm}}^{2}},\quad C_{L}^{cb}=2\frac{V_{tb}\xi_{tc}^{u*}\xi_{\ell\ell^{\prime}}^{l*}}{m_{H^{\pm}}^{2}}.
\label{semileptonicWCs}
\end{equation}

\section{Observables}
\label{sec:Observables}

In this section we present the observables to be included in the fit. We divide them in four sets: The first one for FCNCs in $b\to s$ transitions and $B$ meson rare decay observables, both of them affected by the new WC contributions. The second set is associated with FCCCs observables that arise from semileptonic $b\to c \ell \overline{\nu}$ decays and the mass difference $\Delta M_{s}$ from $B_{s}-\overline{B}_{s}$ oscillations. Various leptonic decays of mesons form the third set. Finally, the fourth set contains leptonic observables associated with $\tau$ and $\mu$ decays, among them the anomalous magnetic moment of the muon in particular.

\subsection{FCNCs and \texorpdfstring{$B$}{B} rare decays}
\label{sec:FCNCObservables}

Lepton flavour universality in the SM means that all couplings between leptons and gauge bosons are the same (up to mass differences). This implies that any departure from this identity could be a clear sign of NP. The most interesting tests of LFU violation with FCNC are given by the ratios of $b\rightarrow sll$ transitions
\begin{equation}
R(K^{(*)})=\frac{\Gamma(B\rightarrow K^{(*)}\mu^{+}\mu^{-})}{\Gamma(B\rightarrow K^{(*)}e^{+}e^{-})},
\end{equation}
with $\Gamma$ representing the decay width and $K^{(*)}$ are kaons. As per our choice of Yukawa textures in Eq.~\eqref{eq:Textures}, here we only consider NP effects coming from the muon specific WCs, i.e., electronic WCs are SM-like. Aside from this $R(K^{(*)})$ ratios, hints for LFU violation are found in many branching fractions and angular observables related
to $B\rightarrow K^{(*)}\mu^{+}\mu^{-}$ decays as a function of the dimuon mass squared $q^2$. In this work we use the same observables as in \cite{Bhom:2020lmk}, with the predicted values obtained with  \textsf{SuperIso} and with likelihoods provided via \textsf{HEPLike}. In particular, among the observables included are the optimised angular observables $P_{i}^{(\prime)}$ which have been constructed in order to minimise the hadronic uncertainties emerging from form factor contributions to the $B^{0}\rightarrow K^{*0}\mu^{+}\mu^{-}$ decay at leading order \cite{Descotes-Genon:2013vna}. In view of that, experimentally these observables are obtained by fitting $q^{2}$-binned
angular distributions and they are defined in the theory as CP-averages integrated in the $q^{2}$ bins:
\begin{align}
\left\langle P_{1}\right\rangle_{{\rm bin}} & =\frac{1}{2}\frac{\int_{{\rm bin}}dq^{2}[J_{3}+\bar{J}_{3}]}{\int_{{\rm bin}}dq^{2}[J_{2s}+\bar{J}_{2s}]}\ , & \left\langle P_{2}\right\rangle_{{\rm bin}} & =\frac{1}{8}\frac{\int_{{\rm bin}}dq^{2}[J_{6s}+\bar{J}_{6s}]}{\int_{{\rm bin}}dq^{2}[J_{2s}+\bar{J}_{2s}]}\ ,
\end{align}
\begin{equation}
\left\langle P_{5}^{\prime}\right\rangle_{{\rm bin}}=\frac{1}{2\,{\cal N}_{{\rm bin}}^{\prime}}\int_{{\rm bin}}dq^{2}[J_{5}+\bar{J}_{5}]\ ,
\end{equation}
\textbf{}where the $J_{i}$ functions and the normalisation
constant $\mathcal{N}_{\mathrm{bin}}^{\prime}$ are given in  \cite{Bhom:2020lmk}. Additionally, they can be related to the form factor dependent observables $S_{i}$ \cite{Altmannshofer:2008dz} as
\begin{equation}
\begin{aligned}P_{1} & =\frac{2\,S_{3}}{(1-F_{{\rm L}})},\qquad P_{2}=\frac{2}{3}\frac{A_{{\rm FB}}}{(1-F_{{\rm L}})},\label{eq:P1P2}\end{aligned}
\end{equation}
\begin{equation}
P_{5}^{\prime}=\frac{S_{5}}{\sqrt{F_{{\rm L}}(1-F_{{\rm L}})}},\label{eq:P5p}
\end{equation}
where $A_{\rm FB}$ is the forward-backward asymmetry of the dimuon system and $F_{L}$ is the fraction of longitudinal polarisation of the $K^{*0}$ meson.

The most sensitive observable to scalar operators is the branching ratio $\mathrm{BR}(B_{s}\rightarrow\mu^{+}\mu^{-})$ which also depends on the muon specific $C_{10}$ and $C_{10}^{'}$ WCs \cite{Bhom:2020lmk}:
\begin{align}
\mathrm{BR}(B_{s} & \rightarrow\mu^{+}\mu^{-})=\dfrac{G_{F}^{2}\alpha^{2}}{64\pi^{3}}f_{B_{s}}^{2}\tau_{B_{s}}m_{B_{s}}^{3}\big|V_{tb}V_{ts}^{*}\big|^{2}\sqrt{1-\frac{4m_{\mu}^{2}}{m_{B_{s}}^{2}}}\nonumber\\
 & \times\left[\left(1-\frac{4m_{\mu}^{2}}{m_{B_{s}}^{2}}\right)\left|\dfrac{m_{B_{s}}\left(C_{Q_{1}}-C_{Q_{1}}^{'}\right)}{(m_{b}+m_{s})}\right|^{2} 
 +\left|\dfrac{m_{B_{s}}\left(C_{Q_{2}}-C_{Q_{2}}^{'}\right)}{\left(m_{b}+m_{s}\right)}-2\left(C_{10}-C_{10}^{\prime}\right)\frac{m_{\mu}}{m_{B_{s}}}\right|^{2}\right],
\label{eq:Bsmumu}
\end{align}
where $f_{B_{s}}$ is the decay constant and $\tau_{B_{s}}$ is the
mean lifetime. 

With respect to the inclusive
$\overline{B}\rightarrow X_{s}\gamma$ decay, we use the full expression given in the works of \cite{Czarnecki:1998tn,Misiak:2006zs,Misiak:2006ab,Czakon:2015exa,Misiak:2017bgg,Misiak:2020vlo} and implemented in \textsf{SuperIso}. The WCs $C_{7}$ and $C_{7}'$ are constrained by this decay, given at the quark level
by $b\rightarrow s\gamma$, which at LO is
\begin{equation}
\Gamma(b\rightarrow s\gamma)=\frac{G_{F}^{2}}{32\pi^{4}}\big|V_{tb}V_{ts}^{*}\big|^{2}\alpha_{{\rm em}}\,m_{b}^{5}\,\left(\vert C_{7{\rm eff}}\,(\mu_{b})\vert^{2}+\vert C_{7{\rm eff}}^{\prime}(\mu_{b})\vert^{2}\right).
\end{equation}

We also take into account the rare decays $B_{s}\rightarrow \tau^{+}\tau^{-}$ and $B^{+}\rightarrow K^{+}\tau^{+}\tau^{-}$ as well as the LFV processes $B_{s}\rightarrow\mu^{\pm}\tau^{\mp}$, $B^{+}\rightarrow K^{+}\mu^{\pm}\tau^{\mp}$ and $b\rightarrow s\nu\overline{\nu}$ with theoretical expressions given in \cite{Crivellin:2019dun}. A list of the included FCNC observables\footnote{New measurements of BR$(B_s\to\mu+\mu^-)$ have been performed recently by LHCb~\cite{LHCb:2021trn,LHCb:2021awg}, as well as a combination with previous results~\cite{Hurth:2021nsi}, giving a combined measured value of $2.85^{+0.34}_{-0.31}$. Nevertheless, we do not expect significant deviations from our results with this new measurement.} can be found in Table \ref{tab:neutral-observables}.

\begin{table}[H]
\begin{centering}
\begin{tabular}{|c|c|}
\hline 
Observable  & Experiment \tabularnewline
\hline
\hline 
$R(K^{*})[0.045,\,1.1]\,\mathrm{GeV^{2}}$  & $0.66\pm0.09\pm0.03$ \cite{LHCb:2017avl} \tabularnewline
\hline 
$R(K^{*})[1.1,\,6.0]\,\mathrm{GeV^{2}}$  & $0.69\pm0.09\pm0.05$ \cite{LHCb:2017avl} \tabularnewline
\hline 
$R(K)[1.1,\,6.0]\,\mathrm{GeV^{2}}$  & $0.846\pm0.042\pm0.013$ \cite{LHCb:2021trn} \tabularnewline
\hline 
$\mathrm{BR}(B_{s}\rightarrow\mu^{+}\mu^{-})\times10^{9}$  & $2.69^{+0.37}_{-0.35}$ \cite{LHCb-CONF-2020-002} \tabularnewline
\hline 
$\mathrm{BR}(B\rightarrow X_{s}\gamma)\times10^{4}$  & $3.32\pm0.15$ \cite{Amhis:2019ckw} \tabularnewline
\hline 
$\mathrm{BR}(B_{s}\rightarrow\tau^{+}\tau^{-})\times10^{3}$ & $<6.8$ at 95\% C.L. \cite{Zyla:2020zbs}\tabularnewline
\hline 
$\mathrm{BR}(B^{+}\rightarrow K^{+}\tau^{+}\tau^{-})\times10^{3}$ & $<2.25$ at 90\% C.L. \cite{Zyla:2020zbs}\tabularnewline
\hline 
$\mathrm{BR}(B_{s}\rightarrow\mu^{\pm}\tau^{\mp})\times10^{5}$ & $<4.2$ at 95\% C.L. \cite{Zyla:2020zbs}\tabularnewline
\hline 
$\mathrm{BR}(B^{+}\rightarrow K^{+}\mu^{\pm}\tau^{\mp})\times10^{5}$ & $<4.8$ at 90\% C.L. \cite{Zyla:2020zbs}\tabularnewline
\hline 
$\mathcal{R}_{K}^{\nu\overline{\nu}}$ & $<3.9$ at 90\% C.L. \cite{Grygier:2017tzo}\tabularnewline
\hline 
$\mathcal{R}_{K^{*}}^{\nu\overline{\nu}}$ & $<2.7$ at 90\% C.L. \cite{Grygier:2017tzo}\tabularnewline
\hline 
\end{tabular}
\par\end{centering}
\caption{\emph{Experimental measurements of FCNCs observables and bounds for rare $B$ decays considered in our study. The $\mathcal{R}_{K^{(*)}}^{\nu\overline{\nu}}$ parameters are related to $b\rightarrow s\nu\overline{\nu}$ transitions as introduced in Eq.(4.6) in \cite{Crivellin:2019dun}.  We also include all the angular distributions and branching fractions of $B^{0}\rightarrow K^{*0}\mu^{+}\mu^{-}$ decays, the branching fractions of both $B_{s}\rightarrow \phi\mu^{+}\mu^{-}$ and $B^{+}\rightarrow K^{+}\mu^{+}\mu^{-}$ with measurements provided by the \textsf{HEPLikeData} repository \cite{HEPLikeData}. \label{tab:neutral-observables}}}
\end{table}

\subsection{FCCCs observables}
\label{sec:FCCCObservables}

The most relevant FCCC observables are the ratios of semileptonic $B$ meson decays to $\tau$ and light leptons, that is
\begin{equation}
R(D^{(*)})=\frac{\Gamma(\overline{B}\rightarrow D^{(*)}\tau\overline{\nu})}{\Gamma(\overline{B}\rightarrow D^{(*)}l\overline{\nu})} ,
\end{equation}
where $D^{(*)}$ are charmed mesons and $l$ is either an electron $(e)$ or a muon $(\mu)$. As of the time of writing, the world average for the experimental measurement of the ratios $R(D^{(*)})$ sits at a 3.1$\sigma$ deviation from the SM prediction~\cite{Amhis:2019ckw}.

The GTHDM contributions to \RD$\,$  and  \RDs$\,$   from the effective Hamiltonian in Eq.~(\ref{eq:Heffective}) can be written as, 
\begin{equation}
R(D)=\frac{1+1.5\,\mathrm{Re}(g_{S}^{\tau\tau})+1.0\sum\left|g_{S}^{\tau l}\right|^{2}}{3.34+4.8\sum\left|g_{S}^{\mu l}\right|^{2}},
\end{equation}
\begin{equation}
R(D^{*})=\frac{1+0.12\,\mathrm{Re}(g_{P}^{\tau\tau})+0.05\sum\left|g_{P}^{\tau l}\right|^{2}}{3.89+0.25\sum\left|g_{P}^{\mu l}\right|^{2}}.
\end{equation}

In addition to \RD$\,$   and \RDs, a third ratio has been measured by the Belle collaboration \cite{Belle:2018ezy}, the ratio  $R_{e/\mu}=\mathrm{BR}(\overline{B}\rightarrow D e\overline{\nu})/\mathrm{BR}(\overline{B}\rightarrow D \mu\overline{\nu})$ which is considered to be the  stringent test of LFU in $B$ decays. It can be expressed in the GTHDM as
\begin{equation}
R_{e/\mu}=\frac{1}{0.9964+0.18\,\mathrm{Re}(g_{S}^{\mu\mu})+1.46\sum\left|g_{S}^{\mu l}\right|^{2}},
\end{equation}
where we have obtained the NP leptonic contributions by integrating the heavy quark effective theory (HQET) amplitudes of the scalar type operators from \cite{Murgui:2019czp, Tanaka:2012nw}. 

The $B_{c}$ meson lifetime has contributions from the SM, given by $\tau_{B_{c}}^{\mathrm{SM}}=0.52_{-0.12}^{+0.18}$ ps~\cite{Beneke:1996xe}, and the GTHDM, which can be written as
\begin{align}
1/\tau_{B_c}^\mathrm{GTHDM} = \Gamma_{B_{c}\rightarrow\tau\bar{\nu}}^{\mathrm{GTHDM}}= & \frac{m_{B_{c}}(m_{\tau}f_{B_{c}}G_{F})^{2}\left|V_{cb}\right|^{2}}{8\pi}\left(1-\frac{m_{\tau}^{2}}{m_{B_{c}}^{2}}\right)^{2}\nonumber\\
 & \times\left[\left|1+\frac{m_{B_{c}}^{2}}{m_{\tau}(m_{b}+m_{c})}g_{P}^{\tau \tau}\right|^{2}+\left|\frac{m_{B_{c}}^{2}}{m_{\tau}(m_{b}+m_{c})}g_{P}^{\tau  l}\right|^{2}-1\right],
\end{align}
where the -1 term accounts for the subtraction of the SM contribution. By using the lifetime of the $B_c$ meson as the constraining observable, we can compare it to the current experimental measurement of $\tau_{B_{c}}=0.510\pm0.009$(ps) \cite{Zyla:2020zbs}, instead of using the theoretical limits on the branching ratio $\mathrm{BR}(B_{c}\rightarrow\tau\bar{\nu})$, which are reported to be either 10$\%$~\cite{Akeroyd:2017mhr} and 30$\%$~\cite{Alonso:2016oyd} \footnote{In \cite{Blanke:2018yud} it was found that values even as large as 60$\%$ could not be excluded, in agreement with a recent calculation of the SM prediction \cite{Aebischer:2021eio,Aebischer:2021ilm}.}.

Another related measurement, $B_{c}^{+}\to J/\psi\tau^{+}{\nu}_{\tau}$,
has been reported by LHCb \cite{Aaij:2017tyk} and also hints to
disagreement with the SM. However the errors are too large at present
to reach a definitive conclusion, with $\mathcal{R}(J/\psi)=0.71\pm0.17\pm0.18$. In addition it has been claimed that the hadronic uncertainties are not at the same level as for the observables related to $\overline{B}\rightarrow D^{*}$ transitions \cite{Murgui:2019czp}, so we do not include it in our fit.

In contrast,
a recent measurement of the longitudinal polarization fraction of the $D^{*}$ meson,
defined as
\begin{equation}
F_{L}(D^{*})=\frac{\Gamma_{\lambda_{D^{*}}=0}\left(\overline{B}\rightarrow D^{*}\tau\overline{\nu}\right)}{\Gamma\left(\overline{B}\rightarrow D^{*}\tau\overline{\nu}\right)},
\end{equation}
has been recently announced by the Belle collaboration~\cite{Abdesselam:2019wbt},
\begin{equation}
F_{L}(D^{*})=0.6\pm0.08\,(\textrm{stat})\pm0.04\,(\textrm{syst)},
\end{equation}
deviating from the SM prediction $F_{L}^{\mathrm{SM}}(D^{*})=0.457\pm0.010$~\cite{Bhattacharya:2018kig} by $1.6\sigma$. The $B\to D^{*}\tau\overline{\nu}$ differential decay width into
longitudinally-polarized ($\lambda_{D^{*}}=0$) $D^{*}$ mesons is
given (keeping NP from scalar contributions only) by 
\begin{eqnarray}
\frac{d\Gamma_{\lambda_{D^{*}}=0}^{D^{*}}}{dq^{2}} & = & \frac{G_{F}^{2}|V_{cb}|^{2}}{192\pi^{3}m_{B}^{3}}\,q^{2}\sqrt{\lambda_{D^{*}}(q^{2})}\left(1-\frac{m_{\tau}^{2}}{q^{2}}\right)^{2}\left\{\left[\left(1+\frac{m_{\tau}^{2}}{2q^{2}}\right)H_{V,0}^{2}+\frac{3}{2}\,\frac{m_{\tau}^{2}}{q^{2}}\,H_{V,t}^{2}\right]\right.\nonumber \\
 &  & \left.\hskip1.5cm\mbox{}+\frac{3}{2}\,|C_{R}^{cb}-C_{L}^{cb}|^{2}H_{S}^{2}+3\,\mathrm{Re}(C_{R}^{cb*}-C_{L}^{cb*})\,\frac{m_{\tau}}{\sqrt{q^{2}}}\,H_{S}H_{V,t}\right\}\,, 
\end{eqnarray}
where the helicity amplitudes are defined in appendix B of  \cite{Murgui:2019czp}. In addition, we  also include the normalised distributions $d\Gamma(B\to D\tau\overline{\nu})/(\Gamma dq^{2})$
and $d\Gamma(B\to D^{\star}\tau\overline{\nu})/(\Gamma dq^{2})$, as measured by the BaBar collaboration \cite{Lees:2013uzd}.

Lastly, the mass difference $\Delta M_{s}$ of $B_{s}-\overline{B}_{s}$ oscillations is included in our study and (for $m_A=m_H$) is given by \cite{Herrero-Garcia:2019mcy}
\begin{align}
\Delta M_{s}^{\mathrm{GTHDM}}= & -\frac{f_{B_{s}}^{2}M_{B_{s}}^{3}}{4(m_{b}+m_{s})^{2}}\biggl[c_{\beta\alpha}^{2}\,\biggl(\frac{1}{m_{h}^{2}}-\frac{1}{m_{H}^{2}}\biggr)+\frac{2}{m_{H}^{2}}\biggr]\nonumber \\
 & \times\biggl\{(U_{22}\tilde{\mathcal{B}}_{B_{s}}^{(2)}\,b_{2}+U_{32}\tilde{\mathcal{B}}_{B_{s}}^{(3)}\,b_{3})\,\biggl[(\xi_{bs}^{d*})^{2}+(\xi_{sb}^{d})^{2}\biggr]+2\,(U_{44}\tilde{\mathcal{B}}_{B_{s}}^{(4)}b_{4})\,\xi_{bs}^{d*}\xi_{sb}^{d}\biggr\}\,,
\end{align}
with  $\vec{b}=\{8/3,\;-5/3,\;1/3,\;2,\;2/3\}$, bag factors $\tilde{\mathcal{B}}_{B_{s}}^{(2)}=0.806$, $\tilde{\mathcal{B}}_{B_{s}}^{(3)}=1.1$ and $\tilde{\mathcal{B}}_{B_{s}}^{(4)}=1.022$ \cite{Bazavov:2016nty,Straub:2018kue}, and the $U$ running matrix being defined in \cite{Herrero-Garcia:2019mcy}.  A summary of all FCCC observables included in this study is provided in Table \ref{tab:charged-observables}.

\begin{table}[H]
\begin{centering}
\begin{tabular}{|c|c|}
\hline 
Observable & Experiment\tabularnewline
\hline 
\hline 
$R(D)$ & $0.340\pm0.027\pm0.013$ \cite{Amhis:2019ckw}\tabularnewline
\hline 
$R(D^{*})$ & $0.295\pm0.011\pm0.008$ \cite{Amhis:2019ckw}\tabularnewline
\hline 
$R_{e/\mu}$ & $1.01\pm0.01\pm0.03$ \cite{Belle:2018ezy}\tabularnewline
\hline 
$\tau_{B_{c}}$(ps) & $0.510\pm0.009$ \cite{Zyla:2020zbs}\tabularnewline
\hline 
$F_{L}(D^{*})$ & $0.6\pm0.08\pm0.04$ \cite{Abdesselam:2019wbt}\tabularnewline
\hline
$\Delta M_{s}(\mathrm{ps}^{-1})$ & 17.741 $\pm$ 0.020 \cite{Amhis:2019ckw}\tabularnewline
\hline 
\end{tabular}
\par\end{centering}

\caption{\emph{Observables related to the charged anomalies considered in our study. We also include the normalised distributions $d\Gamma(B\to D\tau\overline{\nu})/(\Gamma dq^{2})$
and $d\Gamma(B\to D^{\star}\tau\overline{\nu})/(\Gamma dq^{2})$ as measured by the BaBar collaboration \cite{Lees:2013uzd}. \label{tab:charged-observables}}}
\end{table}

\subsection{Leptonic decays of mesons}

Beyond those described in Sections~\ref{sec:FCNCObservables} and \ref{sec:FCCCObservables}, there are additional leptonic decays included in this study. The total decay width at LO for the process $M\to{l}\nu$ in the GTHDM is computed as \cite{HernandezSanchez:2012eg,Jung:2010ik,Iguro:2017ysu}
\begin{eqnarray}
\mathrm{BR}(M_{ij}\to l\nu)=G_{F}^{2}m_{l}^{2}f_{M}^{2}\tau_{M}|V_{ij}|^{2}\frac{m_{M}}{8\pi}\left(1-\frac{m_{l}^{2}}{m_{M}^{2}}\right)^{2}\left[|1-\Delta_{ij}^{ll}|^{2}+|\Delta_{ij}^{ll^{\prime}}|^{2}\right],
\end{eqnarray}
where $i$, $j$ are the valence quarks of the meson $M$, $f_{M}$ is
its decay constant and $\Delta_{ij}^{ll^{\prime}}$ is the NP
correction given by 
\begin{eqnarray}
\Delta_{ij}^{ll^{\prime}}=\bigg(\frac{m_{M}}{m_{H^{\pm}}}\bigg)^{2}Z_{ll^{\prime}}\bigg(\frac{Y_{ij}m_{u_{i}}+X_{ij}m_{d_{j}}}{V_{ij}(m_{u_{i}}+m_{d_{j}})}\bigg),\quad\,\,\,l,l^{\prime}=2,3.
\end{eqnarray}
where the relations
\begin{equation}
X_{ij}=\frac{v}{\sqrt{2}m_{d_{j}}}V_{ik}\,\xi_{kj}^{d},\qquad Y_{ij}=-\frac{v}{\sqrt{2}m_{u_{i}}}\xi_{ki}^{u*}\,V_{kj},\qquad Z_{ij}=\frac{v}{\sqrt{2}m_{j}}\xi_{ij}^{l},
\end{equation}
depend on the Yukawa textures. The list of fully leptonic decays of mesons included in this analysis, for various mesons $M$, can be seen in Table~\ref{tab:leptonic-meson-decays}.

\begin{table}[h]
\begin{centering}
\begin{tabular}{|c|c|}
\hline 
Observable & Experiment\tabularnewline
\hline 
\hline 
{\small{}$\mathrm{BR}(B_{u}\rightarrow\tau\nu)\times10^{4}$} & $1.09\pm0.24$ \cite{Barberio:2008fa}\tabularnewline
\hline 
{\small{}$\frac{\mathrm{BR}(K\rightarrow\mu\nu)}{\mathrm{BR}(\pi\rightarrow\mu\nu)}$} & $0.6358\pm0.0011$ \cite{Mahmoudi:2008tp}\tabularnewline
\hline 
{\small{}$\mathrm{BR}(D_{s}\rightarrow\tau\nu)\times10^{2}$} & $5.48\pm0.23$ \cite{Akeroyd:2009tn}\tabularnewline
\hline 
{\small{}$\mathrm{BR}(D_{s}\rightarrow\mu\nu)\times10^{3}$} & $5.49\pm0.16$ \cite{Akeroyd:2009tn}\tabularnewline
\hline 
{\small{}$\mathrm{BR}(D\rightarrow\mu\nu)\times10^{4}$} & $3.74\pm0.17$ \cite{Zyla:2020zbs}\tabularnewline
\hline 
{\small{}$\mathrm{BR}(D\rightarrow\tau\nu)\times10^{3}$} & $1.20\pm0.27$ \cite{ParticleDataGroup:2020ssz}\tabularnewline
\hline 
\end{tabular}
\par\end{centering}
\caption{\emph{Additional leptonic decays of mesons considered in this work. \label{tab:leptonic-meson-decays}}}
\end{table}

\subsection{Leptonic observables} \label{sec:leptonic_observables}

There are a number of leptonic processes that are forbidden or suppressed in the SM but can occur in the GTHDM. These include modifications to the form factors for $\ell\ell^\prime\gamma$, $\ell\ell^\prime Z$ and other interactions, which lead to contributions to the anomalous magnetic moment of the muon, $(g-2)_{\mu}$, and LFV decays such as $\tau\rightarrow\mu\gamma$, $\tau\to3\mu$ and $h\to\tau\mu$. In the SM, the contributions to these LFV observables are suppressed by the GIM mechanism, giving a very low experimental background, but in the GTHDM LFV is allowed at one- and two-loop level through the couplings $\xi^l_{ij}$ in Eqs.\ (\ref{eq:Gammafhba}-\ref{eq:GammafAba},\ref{eq:GammafCba}).\footnote{Note that in this study we will focus solely on the decays involving $\tau$ and $\mu$ leptons due to our choice of including only second and third generations in the $\xi^l_{ij}$ matrix from Eq.~(\ref{eq:Textures}).}\ 

A second Higgs doublet has been examined as a way to explain the muon $g-2$ anomaly.  In the Type-X \cite{Wang:2014sda,Abe:2015oca,Chun:2015hsa,Chun:2015xfx,Chun:2016hzs,Wang:2018hnw,Chun:2019oix,Chun:2019sjo,Keung:2021rps,Ferreira:2021gke,Han:2021gfu,Eung:2021bef,Jueid:2021avn,Dey:2021pyn} and Flavour-Aligned \cite{Ilisie:2015tra,Han:2015yys,Cherchiglia:2016eui,Cherchiglia:2017uwv,Li:2020dbg,Athron:2021iuf} versions of the THDM the contributions from two-loop diagrams are dominant in most of the parameter space thanks to mechanisms also available in the GTHDM.  Additionally, with LFV, the one-loop diagrams can receive a chirality flip enhancement from including the tau lepton in the diagram loop, as was investigated by \cite{Omura:2015nja,Crivellin:2015hha,Iguro:2019sly,Jana:2020pxx,Hou:2021sfl,Hou:2021qmf,Atkinson:2021eox,Hou:2021wjj}, however they only examined muon $g-2$ contributions at the one-loop level.  

\begin{figure}[tb]
	\centering
    \includegraphics[scale=1]{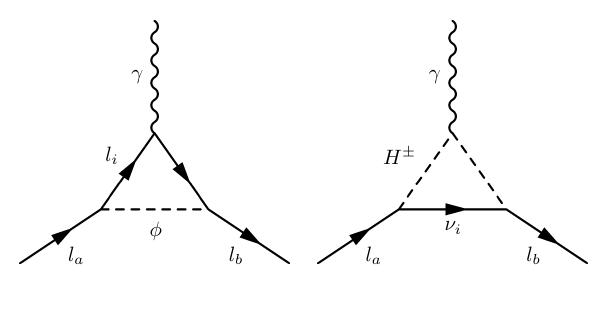}
	\caption{\emph{One-loop diagrams contributing to $\ell\rightarrow\ell^\prime\gamma$ with a neutral scalar diagram on the left and a charged scalar diagram on the right.  The indices $a,b,i$ correspond to any of the lepton flavours $e,\mu,\tau$, and we have $\phi=h,H,A$.  \label{fig:ltolpgammaOneLoop}}}
\end{figure}

Due to the similarity of the diagrams between $\ell\rightarrow\ell^\prime\gamma$ and muon $g-2$ (which is effectively $\mu\rightarrow\mu\gamma$, see figure\ \ref{fig:ltolpgammaOneLoop}), these two observables share nomenclature and contributions.  For both muon $g-2$ and $\tau\rightarrow\mu\gamma$ we can break the contributions into the same three groups: one-loop, $A^{(1)}_{ij L,R}$; two-loop fermionic, $A^{(2,f)}_{ij L,R}$; and two-loop bosonic, $A^{(2,b)}_{ij L,R}$, contributions, so that the observables can be written as
\begin{align} 
	\label{eqn:GTHDMgm2}
	\Delta a^{\mathrm{GTHDM}}_\mu &= m_\mu^2 (A^{(1)}_{\mu\mu L} + A^{(1)}_{\mu\mu R} + A^{(2,f)}_{\mu\mu} + A^{(2,b)}_{\mu\mu}), \\
	\label{eqn:GTHDMtaumugamma}
	\frac{{\rm BR}(\tau\rightarrow\mu\gamma)}{{\rm BR}(\tau\rightarrow\mu\bar{\nu}_{\mu}\nu_{\tau})} &= \frac{48\pi^{3}\alpha_{\rm{EM}}\left(|A_{\tau\mu L}|^{2}+|A_{\tau\mu R}|^{2}\right)}{G_{F}^{2}},
\end{align}
with $A_{\tau\mu L,R} = A^{(1)}_{\tau\mu L,R} + A^{(2,f)}_{\tau\mu L,R} + A^{(2,b)}_{\tau\mu L,R}$ and $\alpha_{EM}$ is the fine structure constant. All form factors $A^{(l)}_{ij L,R}$ have been appropriately renormalised by combining with the relevant counterterms, and are all calculated using masses and couplings that have been extracted from data at tree-level. Additionally, for the contributions to muon $g-2$ we must subtract off the SM contributions from the SM Higgs boson to obtain a purely BSM contribution to muon $g-2$.  

The entire one loop contribution for muon $g-2$ and $\ell\rightarrow\ell^\prime\gamma$ can be found by summing over the neutral scalars $\phi$ and lepton generations:
\begin{equation} \label{eqn:A1loop}
	A^{(1)}_{ab L,R} = \sum^{3}_{i=e,\mu,\tau} \sum_{\phi=h,H,A} \bigg(A^{(FFS)}_{ab L,R}(\phi,i) - A^{(SSF)}_{ab L,R}(H^\pm,i)\bigg),
\end{equation}
where the functions $A^{(FFS)}_{ab L,R}(\phi,i)$ and $A^{(SSF)}_{ab L,R}(\phi,i)$ involve neutral scalars ($h$,$H$,$A$) and the charged scalar $H^\pm$ respectively.  They are defined in Eqs.\ (\ref{eqn:A1loopFFS}-\ref{eqn:A1loopSSF}) in appendix \ref{sec:Loop_Functions}, and shown in figure\ \ref{fig:ltolpgammaOneLoop}.  
To obtain the BSM contributions to muon $g-2$, we must also subtract off the contribution from the SM Higgs boson to obtain a truly-BSM one-loop contribution.  

\begin{figure}[tb]
	\centering
    \includegraphics[scale=0.55]{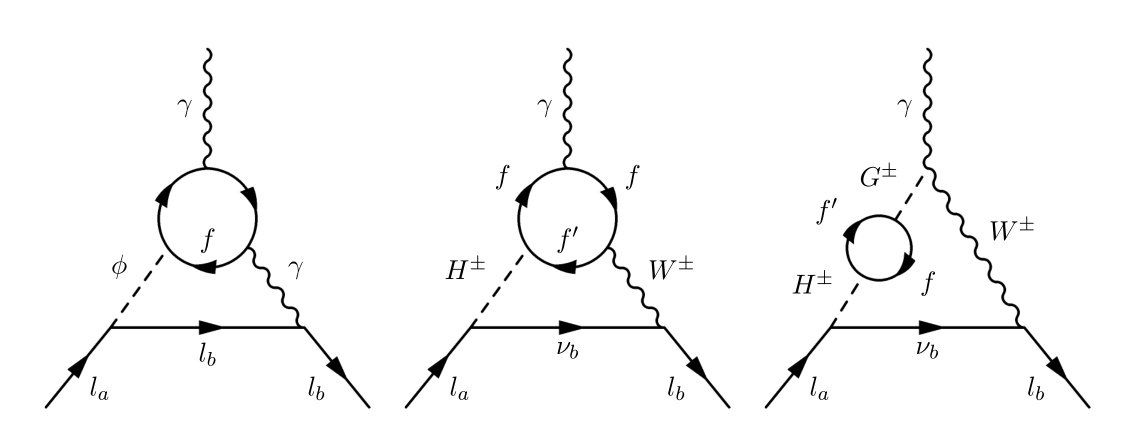}
	\caption{\emph{Two-loop fermionic Barr-Zee diagrams contributing to muon $g-2$ and $l\rightarrow l' \gamma$.  The indices $a,b$ correspond to any of the lepton flavours $e,\mu,\tau$, and $\phi=h,H,A$.  The internal photon $\gamma$ may be replaced by a $Z$ boson.    \label{fig:ltolpgammaBZFermionic}}}
\end{figure}

\begin{figure}[tb]
	\centering
    \includegraphics[scale=0.55]{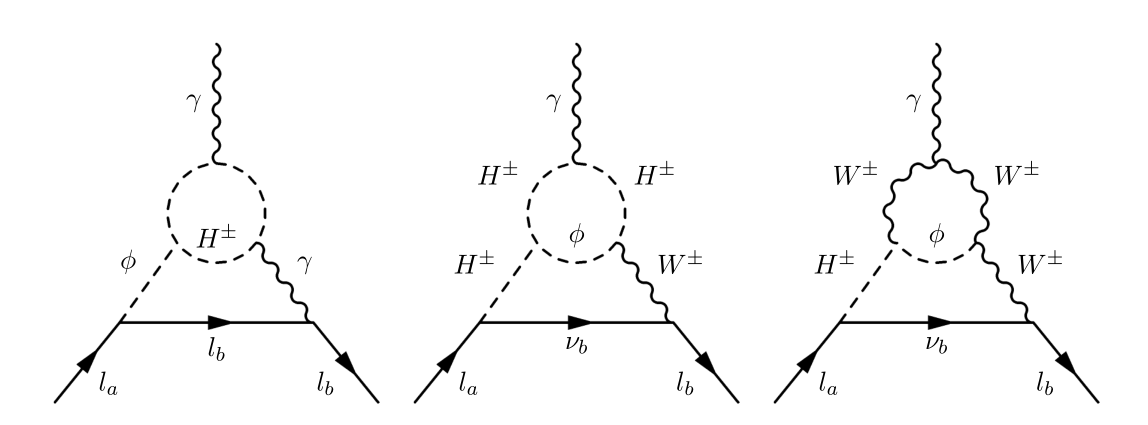}
	\caption{\emph{Two-loop bosonic Barr-Zee diagrams contributing to muon $g-2$ and $l\rightarrow l' \gamma$.  The indices $a,b$ correspond to any of the lepton flavours $e,\mu,\tau$, and we have $\phi=h,H,A$.  In the left panel, the internal photon $\gamma$ may be replaced by a $Z$ boson, and the internal $H^\pm$ with a $W^\pm$ boson.    \label{fig:ltolpgammaBZBosonic}}}
\end{figure}

At the two-loop level we consider the Barr-Zee diagrams, shown in figures~\ref{fig:ltolpgammaBZFermionic} and \ref{fig:ltolpgammaBZBosonic}.  
Just as for the one-loop contributions before, we can subdivide each of these contributions into diagrams involving charged leptons ($l^-_i$) paired with neutral bosons ($h$,$H$,$A$,$Z$,$\gamma$) and neutral leptons ($\nu_i$) paired with charged bosons ($H^\pm$,$W^\pm$).\footnote{We do not consider two-loop bosonic diagrams that are not Barr-Zee diagrams, since their maximum contributions to muon $g-2$ are relatively small \cite{Cherchiglia:2017uwv}, whereas Barr-Zee contributions have been proved to be dominant for some regions of the parameter space \cite{Omura:2015xcg}. Additionally, two-loop diagrams involving neutral bosons where both legs are Higgs bosons are suppressed by a factor $m_\mu^4$, while diagrams with both legs being either $\gamma$ or $Z$ are SM contributions, so we do not consider either, only those with both a $\phi$ and a $\gamma$ or $Z$ boson leg.  Similarly for diagrams involving charged legs of $H^\pm$,$W^\pm$, we only consider a $H^\pm$ and $W^\pm$ boson paired together, as a pair of $H^\pm$ legs lead to diagrams with suppressed contributions \cite{Ilisie:2015tra}.}  
The two-loop bosonic and fermionic diagrams involve an internal loop made of either bosons or fermions respectively.  The total fermionic two-loop contribution to muon $g-2$ is given by \cite{Cherchiglia:2016eui}
\begin{equation} \label{eqn:gm2Contribution2loopfermionic}
	A^{(2,f)}_{\mu\mu} = \sum_{f=u,d,l} \bigg(A^{(FC)}_{\mu\mu}(H^\pm,f) + \sum_{\phi=h,H,A} A^{(FN)}_{\mu\mu}(\phi,f) - A^{(FN)}_{\mu\mu}(h_{SM},f)\bigg),
\end{equation}
where the form factors are given in Eqs.~ (\ref{eqn:gm2ContributionFN}-\ref{eqn:gm2ContributionFC}) in appendix \ref{sec:Loop_Functions}. Note that only contributions from the heaviest generations of the fermions are considered, via $\Gamma_{\phi33}^{f}~(f=u,~d,~e)$.  Similarly the total bosonic two-loop contributions to muon $g-2$ are
\begin{equation} \label{eqn:gm2Contribution2loopbosonic}
	A^{(2,b)}_{\mu\mu} = \sum_{\phi=h,H} \bigg(A^{(BHN)}_{\mu\mu}(\phi) + A^{(BWN)}_{\mu\mu}(\phi) + A^{(BHC)}_{\mu\mu}(\phi) + A^{(BWC)}_{\mu\mu}(\phi) \bigg) - A^{(BWN)}_{\mu\mu}(h_{SM}),
\end{equation}
where again the bosonic two-loop functions are in  Eqs.~(\ref{eqn:gm2ContributionBHN}-\ref{eqn:gm2ContributionBWC}) in the same appendix.  Note that these contributions do not include 2-loop diagrams with an internal $Z$ boson leg, as in \cite{Ilisie:2015tra}. 

In the case of the $\tau\to\mu\gamma$ decay, the contributions from the fermionic and bosonic Barr-Zee two loop diagrams, $A^{(2,f)}_{ab L,R}$ and $A^{(2,b)}_{ab L,R}$ respectively, have the same form for each Higgs bosons and fermion or boson in the loop, and can be found in Eqs.~(\ref{eqn:ltolpgammafermionic},\ref{eqn:ltolpgammabosonic}) in appendix \ref{sec:Loop_Functions}.  

The contributions to $\tau\rightarrow3\mu$ decay can be divided up into 3 separate groups, the tree-level, dipole, and the contact contributions.  The contributions from tree-level decay are computed in \cite{Crivellin2013}.  We have found that the dipole contributions, which involve the penguin-photon diagrams of the form of $\tau\rightarrow\mu\gamma$ decays, are quite sizable compared to those at tree-level and cannot be ignored. Namely, they are given by \cite{Hou:2020itz}:
\begin{align} \label{eqn:Tauto3MuDipole}
    \mathrm{BR}(\tau\to3\mu)^{\textrm{(dipole)}} =&  \frac{\alpha_{\rm{EM}}}{3\pi} \bigg(\log{\bigg(\frac{m_\tau^2}{m_\mu^2}\bigg)}-\frac{11}{4}\bigg) \frac{\mathrm{BR}(\tau\to\mu\gamma)}{{\rm BR}(\tau\rightarrow\mu\bar{\nu}_{\mu}\nu_{\tau})}.
\end{align}

Similarly, the contact terms involving effective four-fermion interactions \cite{Kuno:1999jp} could be at first comparable to the dipole contributions. The contact contributions are given by
\begin{align} \label{eqn:Tauto3MuContact}
    \mathrm{BR}(\tau\to3\mu)^{\textrm{(contact)}} =& 
    \frac{|g_2|^2}{8} + 2|g_4|^2 + \frac{16\pi\alpha_{em}}{\sqrt{2}G_F} \textrm{Re}\bigg(g_4^*\bigg (A^{(1)}_{\tau\mu L,R} + A^{(2,f)}_{\tau\mu L,R} + A^{(2,b)}_{\tau\mu L,R}\bigg)\bigg),
\end{align}
where the coefficients $g_2$ and $g_4$ are given in appendix \ref{sec:Loop_Functions}.  

Another observable that we include is the lepton violating $h\to \tau\mu$ decay.  This is given at tree level by\footnote{We computed the contributions coming from one-loop diagrams with two charged Higgses in the loop and found them to be 7 orders of magnitude suppressed compared to the tree level. Diagrams involving a pair of heavy neutral Higgses are possible as well but even more suppressed. The GTHDM only takes into account the tree level, which relies on being close to the alignment limit but not exactly, otherwise this tree level contribution would be zero.}
\begin{align}
\mathrm{BR}(h\to\tau\mu)=\frac{3c_{\beta\alpha}^{2}m_{h}}{8\pi \Gamma_{h}}\Big(|\xi_{\mu\tau}^{l}|^{2}+|\xi_{\tau\mu}^{l}|^{2}\Big)\left(1-\frac{m_{\tau}^{2}}{m_{h}^{2}}\right)^{2}\,,
\end{align}
with the total decay width of $h$ given by $\Gamma_{h}=3.2\,{\rm MeV}$ \cite{ParticleDataGroup:2020ssz}.

Lastly, besides $g-2$ and LFV observables, experiments have also provided constraints for the LFU ratio in $\tau$ decays. This ratio is commonly known as $(g_\mu/g_e)^2$ and is given as~\cite{HernandezSanchez:2012eg,Jung:2010ik} 

\begin{eqnarray}
\left(\frac{g_{\mu}}{g_{e}}\right)^{2}=\frac{\mathrm{BR}(\tau\to\mu\bar{\nu}\nu)}{\mathrm{BR}(\tau\to e\bar{\nu}\nu)}\frac{f(m_{e}^{2}/m_{\tau}^{2})}{f(m_{\mu}^{2}/m_{\tau}^{2})}\simeq1+\sum_{i,j=\mu,\tau}\left(0.25R_{ij}^{2}-0.11R_{ii}\right),
\end{eqnarray}
where $f(x)=1-8x+8x^{3}-x^{4}-12x^{2}\,\log x$ 
and $R_{ij}$ is the BSM scalar contribution, given in the GTHDM as
\begin{eqnarray}
R_{ij}=\frac{\upsilon^{2}}{2m_{H^{\pm}}^{2}}\,\left(\xi_{\tau i}^{l}\,\xi_{j\mu}^{l}\right).\label{R scalar}
\end{eqnarray}

All of the experimental measurements and upper bounds for leptonic observables are shown in Table~\ref{tab:bound_lepton_flavour_violating}.

\begin{table}[H]
\begin{centering}
\begin{tabular}{|c|c|}
\hline 
Observable  & Experiment \tabularnewline
\hline 
\hline 
${\Delta a_\mu}$  & $2.51\pm59\times10^{-9}$ \cite{PhysRevLett.126.141801} \tabularnewline
\hline 
{\small{}$\mathrm{BR}(\tau\rightarrow\mu\gamma)$}  & $<4.4\times10^{-8}$ at 90\% C.L. \cite{Zyla:2020zbs} \tabularnewline
\hline 
{\small{}$\mathrm{BR}(\tau\rightarrow3\mu)$}  & $<2.1\times10^{-8}$ at 95\% C.L. \cite{Zyla:2020zbs} \tabularnewline
\hline 
{\small{}$\mathrm{BR}(h\rightarrow\tau\mu)$}  & $<1.5\times10^{-3}$ at 95\% C.L. \cite{CMS:2021rsq} \tabularnewline
\hline 
{\small{}$(g_\mu/g_e)$} & $1.0018\pm0.0014$ \cite{Bifani:2018zmi}\tabularnewline
\hline 
\end{tabular}
\par\end{centering}
\caption{\emph{World average measurement of $\Delta a_\mu$ and experimental bounds for the LFV decay and LFU observables considered in our analysis.\label{tab:bound_lepton_flavour_violating}}}
\end{table}

\section{Results}
\label{sec:Results}

Our main goal is to study the impact of these observables on the GTHDM parameter space and, in particular, infer the goodness-of-fit of the model in light of these anomalies.  Given the plethora of observables defined in the previous section and the large multidimensional parameter space,  it is very important to combine them in a statistically rigorous manner in a global fit.  This avoids serious shortcomings from more naive approaches like simply overlaying constraints from confidence intervals \cite{AbdusSalam:2020rdj}.

To visualize the results we will  project the high dimensional parameter space onto two-dimensional planes. To this end, the central quantity of interest is the profile likelihood,
\begin{equation}
\log\mathcal{L}_{prof}\left(\theta_{1},\theta_{2}\right)=\underset{\boldsymbol{\eta}}{\max}\log\mathcal{L}\left(\theta_{1},\theta_{2},\boldsymbol{\eta}\right) ,
\end{equation}
which is, for fixed parameters of interest $\theta_{1}$ and $\theta_{2}$, the maximum value of the log-likelihood function that can be obtained when maximizing over the remaining parameters $\boldsymbol{\eta}$. All profile likelihood figures in this study are created with \textsf{pippi}~\cite{Scott:2012qh}.

As mentioned earlier, we use here the \gambit framework for our study. The theoretical predictions of the model and the experimental likelihoods are either implemented natively in GAMBIT or from external tools interfaced with GAMBIT. In particular, the likelihoods related to $b\to s\mu^{+}\mu^{-}$ transitions are obtained from \textsf{HEPLike}, which retrieves experimental results and their correlated uncertainties from the \textsf{HEPLikeData} repository. To efficiently explore the parameter space, we employ the differential evolution sampler \textsf{Diver}, which is a self-adaptive sampler. We choose a population size of \textsf{NP} = 20000 and a convergence threshold of \textsf{convthresh} = $10^{-6}$. The data we present in this work comes from scans that took between 6 and 8 hours of running time on the Australian supercomputer \textsf{GADI} with cores varying between 1400 and 2000.

\subsection{Parameter space}

\begin{figure}[h]
\includegraphics[scale=0.5]{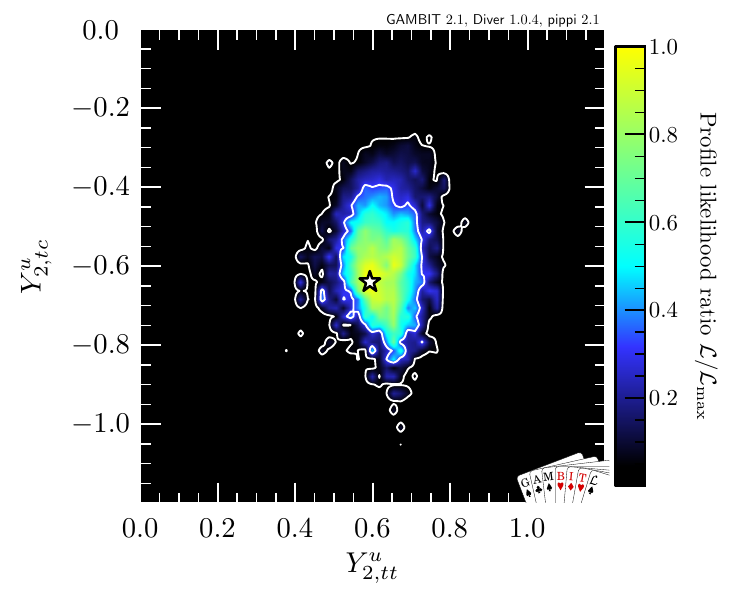}$\qquad$\includegraphics[scale=0.5]{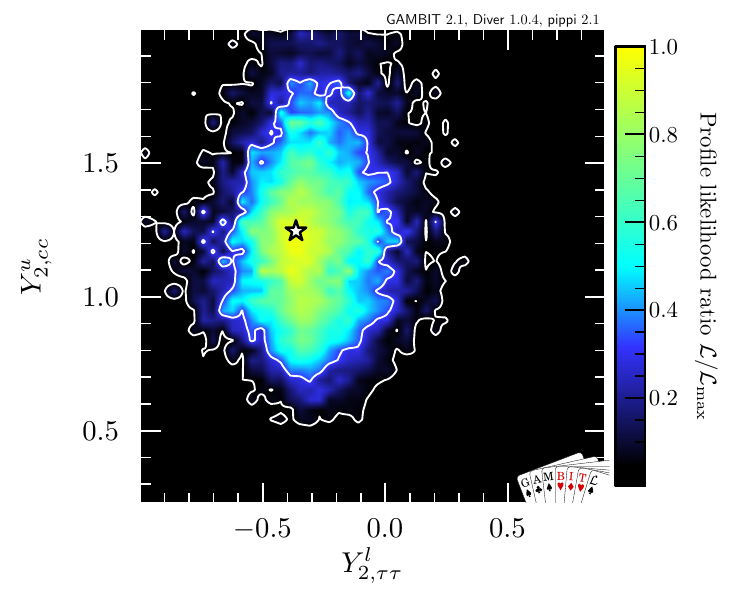}

\includegraphics[scale=0.5]{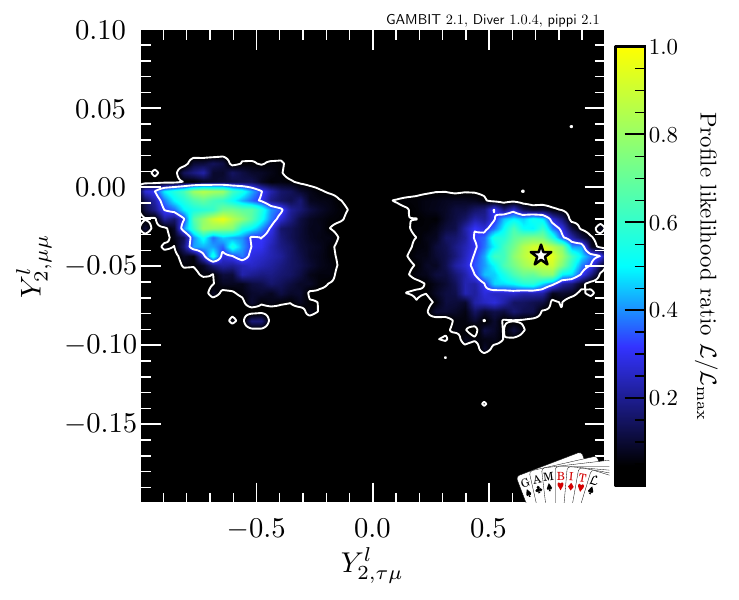}$\qquad$\includegraphics[scale=0.5]{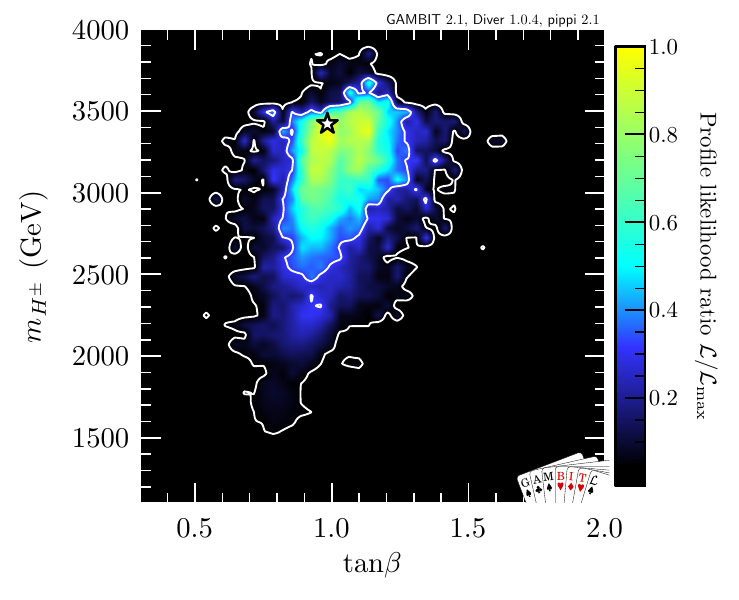}

\caption{\emph{Profile likelihood ratios $\mathcal{L}/\mathcal{L}_{max}$ for different
2D plots of the parameter space for $Y_{2,tc}^{u}\in[-2,0]$}.\label{fig:params-profiles}}
\end{figure}

We perform the parameter scans in the physical basis, i.e., where the tree-level masses of the heavy Higgses, $m_H$, $m_A$ and $m_{H^\pm}$ are taken as input. The remaining model parameters are $\tan\beta$, $m_{12}$ and the Yukawa couplings $Y_{2,ij}^{f}$ as in Eq.~(\ref{eq:Xis}). In order to avoid collider constraints, we work in the alignment limit choosing $s_{\beta-\alpha}$ close to $1$, and we select a conservative lower limit on the masses of the heavy Higgses $m_{H,A,H^\pm} \geq 500$ GeV~\footnote{From preliminary results we found that low Higgs masses are disfavoured by the contribution of various constraints and thus we do not attempt to include precise constraints on the masses from BSM Higgs searches (see e.g.\cite{Arbey:2017gmh} for a discussion of the limits on the charged Higgs mass). We leave a detailed collider study to future work.}. We also fix $m_A = m_H$ in our study, motivated by the requirement to satisfy the oblique parameter constraints which favour small mass splittings and in order to simplify the sampling of the parameter space. So as to choose reasonable priors for the Yukawa couplings, we take into account various constraints on them (or equivalently on $\xi_{ij}^f$) from previous studies. The tighter theoretical constraints come from perturbativity which requires $\left|\xi_{ij}^{f}\right|\leq\sqrt{4\pi}\sim 3.5$. On the phenomenological side, the studies in \cite{Hou:2020chc,Hou:2021sfl} have found values as large as $\xi_{tt}^{u}\sim 0.1$ and $\xi_{tc}^{u}\sim 0.32$ for masses of the heavy Higgses of order 500 GeV. With respect to the $\xi_{cc}^{u}$ coupling, it has been shown in~\cite{Iguro:2017ysu} that $\mathcal{O}(1)$ values are possible within the charged anomalies, and similar values were considered in \cite{Crivellin:2019dun} in the context of the neutral anomalies, not only for $\xi_{cc}^{u}$ but for all the Yukawa matrix elements. As for the new leptonic couplings, the results in \cite{Omura:2015xcg,Iguro:2018qzf,Crivellin:2019dun} indicate they should be $\mathcal{O}(1)$ or less in order to fit the charged anomalies. Lastly, the extra down Yukawa couplings $\xi_{ij}^{d}$ are in general expected to be $\mathcal{O}(0.1)$ \cite{Crivellin:2017upt, Iguro:2019sly} and in particular $\xi_{sb}^{d}$ is expected to be strongly constrained by $B_{s}-\overline{B}_{s}$ mixing. With all these considerations, the chosen priors on our scan parameters are
\begin{align}
\tan\beta\in[0.2,\,50] & ,\qquad m_{12}\in[-1000,\,2700]\mathrm{ GeV},\qquad m_{H^{\pm}},\,m_{A}=m_{H}\in[500,\,4000]\mathrm{ GeV},\nonumber \\
Y_{2,tt}^{u}\in[0.0,\,2.0] & ,\qquad Y_{2,cc}^{u},\,Y_{2,tc}^{u}\in[-2.0,\,2.0],\nonumber \\
Y_{2,bb}^{d}\in[-0.1\,0.1] & ,\qquad Y_{2,ss}^{d}\in[-0.2,\,0.2],\qquad Y_{2,sb}^{d}=Y_{2,bs}^{d}\in[-0.01,0.01],\nonumber \\
Y_{2,\mu\mu}^{l}\in[-0.5,0.5] & ,\qquad Y_{2,\tau\tau}^{l}\,,Y_{2,\mu\tau}^{l}=Y_{2,\tau\mu}^{l}\in[-1.0,1.0],\label{eq:Ranges}
\end{align}

The results of our scans show two degenerate regions of solutions according to the sign of $Y_{2,tc}^{u}$. We indeed verified that these regions are degenerate and the final results are unaffected by this choice, hence we select $Y_{2,tc}^{u}\in[-2,0]$ for the phenomenological analysis from now on. Namely, this degeneracy is a result of the dependency of various observables on products like $Y_{2,tc}^{u} Y_{2,ij}^{f}$ where $Y_{2,ij}^{f}$ also flips its sign.\footnote{We first found those two regions of solutions via an auxiliary scanning method based on the quadratic approximation to $\chi^{2}$ as a function of the WCs (see appendix \ref{sec:chi2_method}).}

We show in figure \ref{fig:params-profiles} different 2D planes with the most relevant parameters obtained by the scan. The values for $Y_{2,tt}^{u}$ and $Y_{2,tc}^{u}$ are displayed in the top left panel where we can observe that for the best fit point $|Y_{2,tt}^{u}|\approx|Y_{2,tc}^{u}|\approx\,0.6$. Then, in the top right panel we see a preferred value for $Y_{2,cc}^{u}\approx1.1$ (-1.1 for the positive sign solution of  $Y_{2,tc}^{u}$ from the degeneracy of solutions).  This,  along with the lepton Yukawa couplings $Y_{2,\mu\mu}^{l}$ and $Y_{2,\tau\mu}^{l}$ (bottom right panel), helps to enhance the contributions from the box diagrams in figures\ \ref{fig:a}-\ref{fig:b}.  Additionally, the LFV coupling $Y_{2,\tau\mu}^{l}$ also contributes to the $B^{+}\rightarrow K^{+}\mu^{\pm}\tau^{\mp}$ decay, requiring  $|Y_{2,\tau\mu}^{l}|\gtrsim 0.4$ in order to get $\mathrm{BR}(B^{+}\rightarrow K^{+}\mu^{\pm}\tau^{\mp})\times10^{5}<4.8$.  As for the $Y_{2,ij}^{d}$ couplings, we find $Y_{2,ss}^{d}=0.1\pm0.1$, $Y_{2,sb}^{d}=0.004\pm0.005$ and $Y_{2,bb}^{d}=0.017\pm0.005$ assuming Gaussian distributions. In particular, both $Y_{2,ss}^{d}$ and $Y_{2,sb}^{d}$ flip their signs for the positive solutions of $Y_{2,tc}^{u}$ whereas $Y_{2,bb}^{d}$ remains unaffected.

Finally, in the bottom right panel of figure \ref{fig:params-profiles} we observe that the preferred values for the charged Higgs mass are of order $3\,\mathrm{TeV}$ with $\tan\beta\approx1$. We find that the combined contribution of FCNC likelihoods fits better the data for this particular mass range. Similarly, although values of $\tan\beta$ up to 50 are possible in the GTHDM when using theoretical constraints alone, we identified that once we take into account all flavour constraints, there is a clear preference for low values, close to $\tan\beta\approx 1$, in agreement with \cite{WahabElKaffas:2007xd,Arhrib:2009hc,Branco:2011iw}. This preference can be understood as follows. The box contributions in figures \ref{fig:a}-\ref{fig:b} depend on the Green function $\mathcal{B}^{H(0)}$ in Eq.~(\ref{eq:Greenf_Box}), which for values of the charged Higgs mass $m_{H^{\pm}}<2$ TeV or $m_{H^\pm} > 4$ TeV significantly over- or undershoot, respectively, the observed value of $\Delta C_9\approx -1$ (see below).  Furthermore, the measurement of the $B_{c}$ lifetime and the BaBar collaboration $B\to D^{(\star)}\tau\overline{\nu}$ distributions, both of which depend strongly on $\tan\beta$ and $m_{H^\pm}$ through the $C_{R,L}^{cb}$ in Eq.~(\ref{semileptonicWCs}), push both $\tan\beta$ and $m_{H^{\pm}}$ to values lower than 2 and greater than $2\,\mathrm{TeV}$ respectively. In addition to this, we have also noticed a strong penalty for large $\tan\beta$ values coming from the $B_{s}\to \mu^{+}\mu^{-}$ decays, which is due to the strong $\tan\beta$ dependence on the $C_{10}$ and (pseudo) scalar WCs. Lastly the preferred masses of the other heavy Higgses, $m_H$ and $m_A$, are of the same order as $m_{H^\pm}$ as was expected because of the oblique parameter constraints. The best fit values for some relevant scan parameters can be found in table \ref{tab:WilsonCoeff}.

\subsection{Neutral and charged anomalies}

\begin{figure}
$\quad$\includegraphics[scale=0.41]{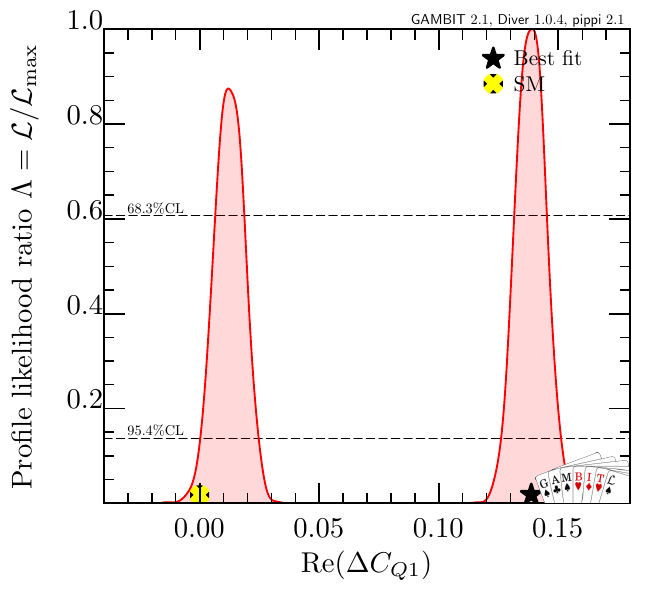}

\includegraphics[scale=0.43]{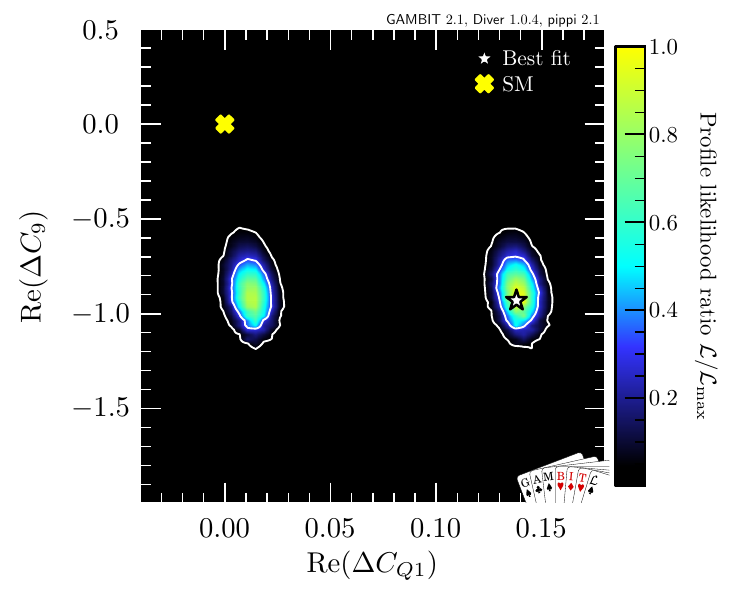}$\quad$\includegraphics[scale=0.41]{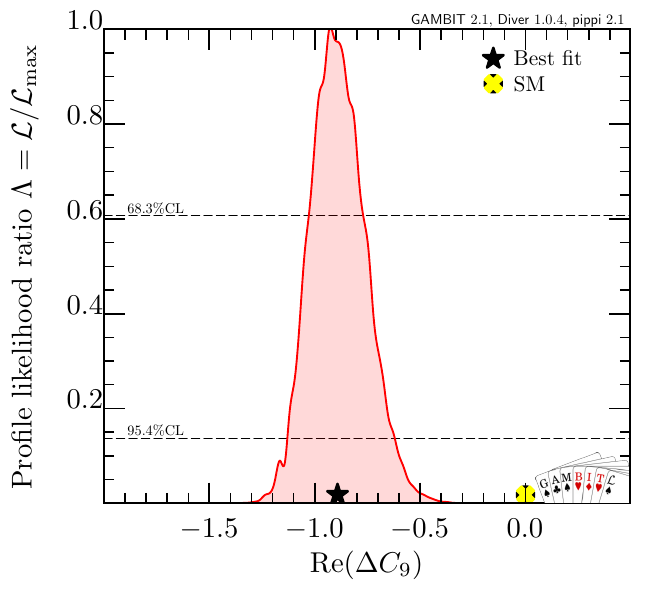}

\includegraphics[scale=0.43]{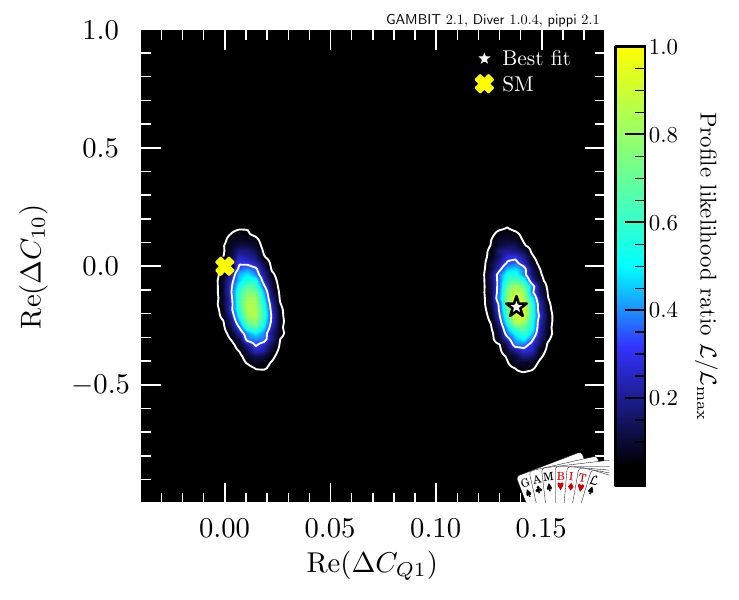}\includegraphics[scale=0.43]{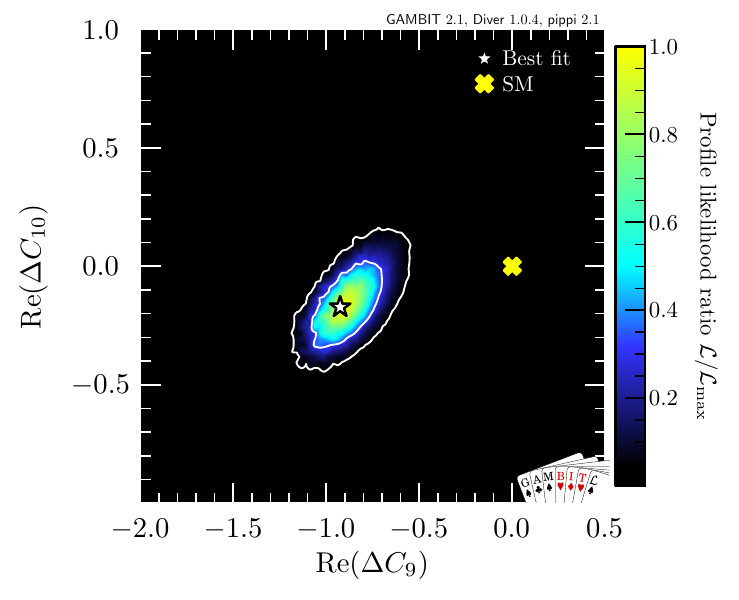}\includegraphics[scale=0.41]{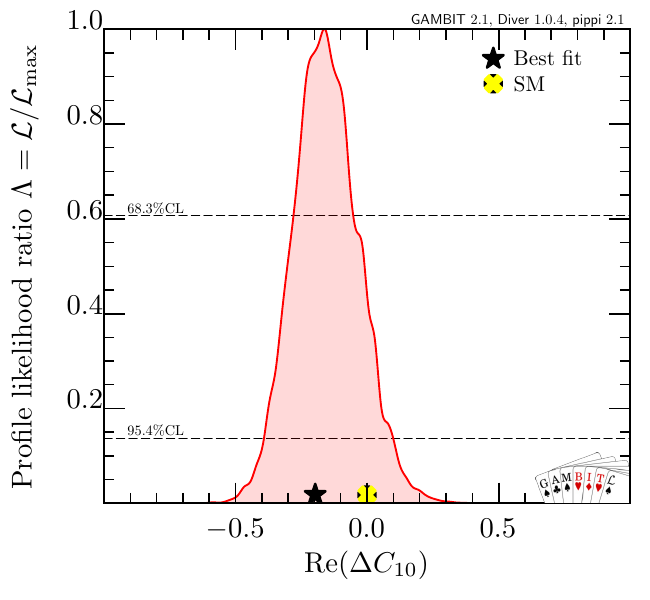}
\caption{\emph{One- and two-dimensional profile likelihoods for three of the Wilson coefficients computed from the fit.}\label{fig:WCs_1d_2d}}
\end{figure}

\begin{table}
\begin{centering}
\begin{tabular}{|c|c|}
\hline 
Parameter & Best fit \tabularnewline
\hline 
\hline 
$m_{H,A}$ & $3485$ GeV\tabularnewline
\hline 
$m_{H^\pm}$ & $3429$ GeV\tabularnewline
\hline 
$m_{12}$ & $2426$ GeV\tabularnewline
\hline 
$\tan\beta$ & $0.98$\tabularnewline
\hline 
$Y_{2,tt}^{u}$ & $0.60$\tabularnewline
\hline 
$Y_{2,cc}^{u}$ & $1.15$\tabularnewline
\hline 
$Y_{2,tc}^{u}$ & $-0.64$\tabularnewline
\hline 
$Y_{2,bb}^{d}$ & $0.017$\tabularnewline
\hline 
$Y_{2,ss}^{d}$ & $0.10$\tabularnewline
\hline
$Y_{2,sb}^{d}$ & $0.004$\tabularnewline
\hline
$Y_{2,\mu\mu}^{l}$ & $-0.04$\tabularnewline
\hline
$Y_{2,\tau\tau}^{l}$ & $-0.36$ \tabularnewline
\hline
$Y_{2,\mu\tau}^{l}$ & $0.75$ \tabularnewline
\hline
\end{tabular}\hspace{20pt}
\begin{tabular}{|c|c|}
\hline 
Wilson coefficient & Best fit \tabularnewline
\hline 
\hline 
$\mathrm{Re}(\Delta C_{Q_{1}})$ & $0.14\pm0.01$\tabularnewline
\hline 
$\mathrm{Re}(\Delta C_{2})$ & $-0.018\pm0.005$\tabularnewline
\hline 
$\mathrm{Re}(\Delta C_{7})$ & $0.002\pm0.01$\tabularnewline
\hline 
$\mathrm{Re}(\Delta C_{7}^{'})$ & $0.01\pm0.01$\tabularnewline
\hline 
$\mathrm{Re}(\Delta C_{8})$ & $0.002\pm0.015$\tabularnewline
\hline 
$\mathrm{Re}(\Delta C_{8}^{'})$ & $0.01\pm0.01$\tabularnewline
\hline 
$\mathrm{Re}(\Delta C_{9})$ & $-0.89\pm0.15$\tabularnewline
\hline 
$\mathrm{Re}(\Delta C_{10})$ & $-0.19\pm0.14$\tabularnewline
\hline 
\end{tabular}
\par\end{centering}
\caption{\emph{Best fit values for the scan parameters (left) and WCs for
$b\rightarrow s\mu^{+}\mu^{-}$ transitions (right). We show only $\mathrm{Re}(\Delta C_{Q_{1}})$
given that at tree level and in the alignment limit $\mathrm{Re}(\Delta C_{Q_{1}})=\mathrm{Re}(\Delta C_{Q_{2}})$
and $m_{s}/m_{b}\,\mathrm{Re}(\Delta C_{Q_{1}})=\mathrm{Re}(\Delta C_{Q_{1}}^{'})=-\mathrm{Re}(\Delta C_{Q_{2}}^{'})$. The uncertainties on the WCs were computed with \textsf{GAMBIT} assuming a symmetric Gaussian distribution from the resulting one-dimensional profile likelihoods. We do not display the $\mathrm{Re}(\Delta C_{9,10}^{'})$ WCs either which we find to be suppressed by a factor of $m_{b}/m_{t}$ compared to their non prime counterparts.} \label{tab:WilsonCoeff}}
\end{table}
In table \ref{tab:WilsonCoeff} we show the best fit values for the parameters from the scans (\textit{left}) and the muon specific WCs evaluated at the best fit point (\textit{right}), where in particular, $\Delta C_{9}$ is consistent with the value obtained by model independent fits at the 1$\sigma$ level. In this sense, the neutral anomalies can indeed be explained in the GTHDM  as shown in figure\ \ref{fig:WCs_1d_2d}. Furthermore, coming from the quadratic dependence in the branching ratio $\mathrm{BR}(B_{s}\rightarrow\mu^{+}\mu^{-})$, we can see two regions of solutions for the scalar WC $\Delta C_{Q_{1}}$, one of them containing the SM prediction within 2$\sigma$. In addition, we ran a complementary scan invalidating points for $|\Delta C_{Q_{1}}|>0.1$ and found that the corresponding region of solutions gives an equally good fit to the data, i.e., the preference over the second region of solutions is completely arbitrary.

\begin{figure}[htb]

\includegraphics[scale=0.5]{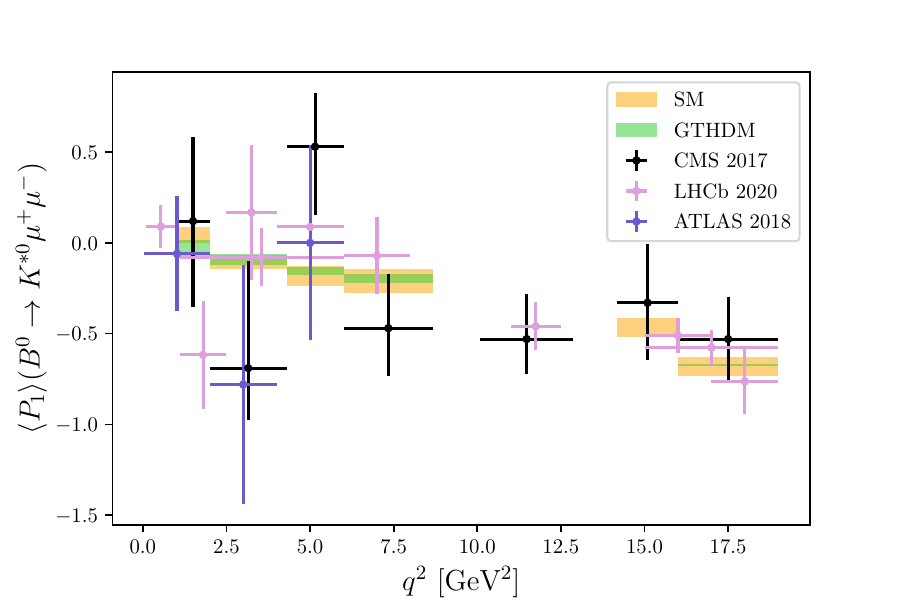}$\qquad$\includegraphics[scale=0.5]{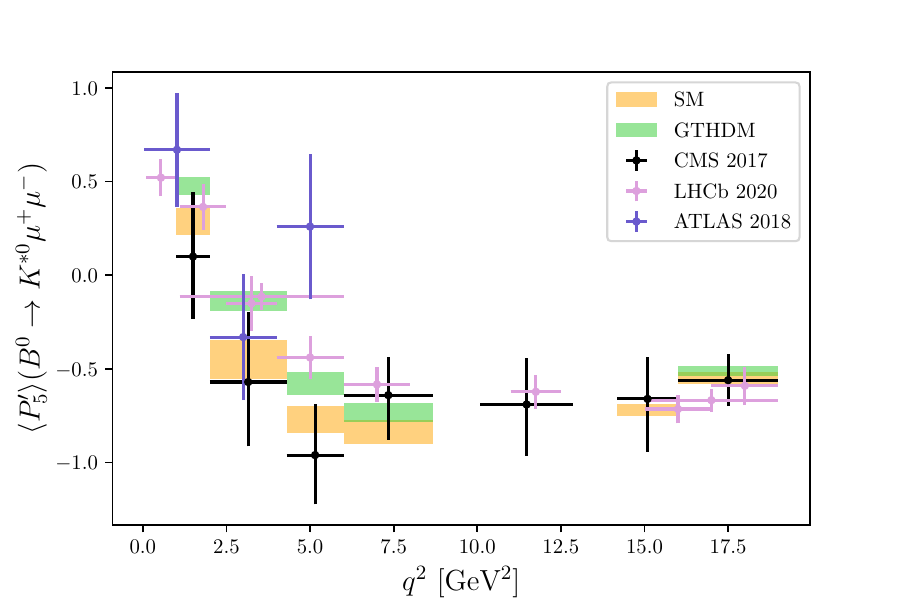}

\caption{\emph{Predicted distributions for \textit{Left}: $\left\langle P_{1}\right\rangle$ and \textit{Right}: $\left\langle P_{5}^{\prime}\right\rangle$ compared to the CMS 2017  \cite{CMS:2017ivg}, ATLAS 2018 \cite{ATLAS:2018gqc} and LHCb 2020 \cite{LHCb:2020lmf} data. The theoretical uncertainties using \textsf{GAMBIT} have been computed assuming a symmetric Gaussian distribution for the resulting one-dimensional profile likelihoods for each one of the bins. The theory predictions close to the $J/\psi(1S)$ and $\psi(2S)$ narrow charmonium resonances are vetoed from all our plots.} \label{fig:P1-P5p}}
\end{figure}

\begin{figure}[htb]
\includegraphics[scale=0.5]{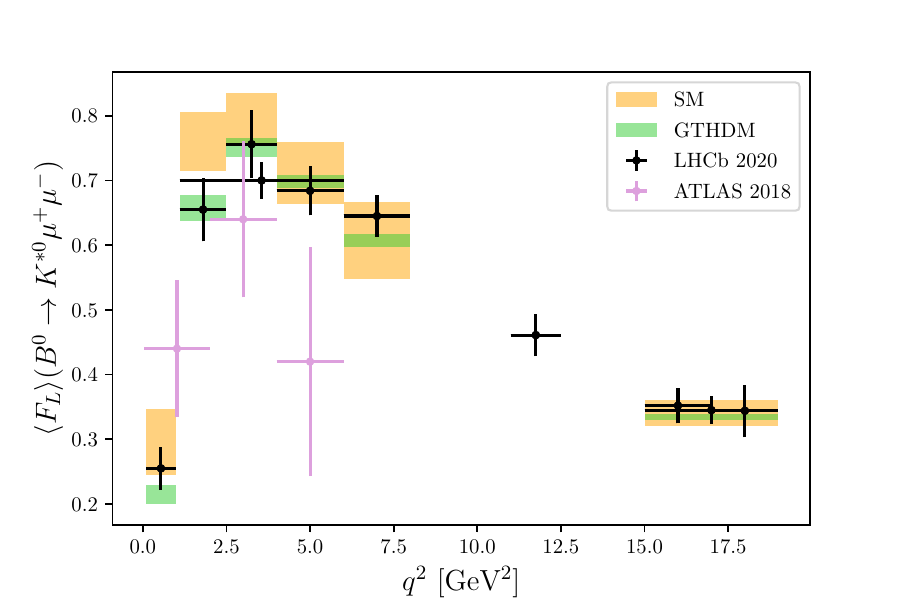}$\qquad$\includegraphics[scale=0.5]{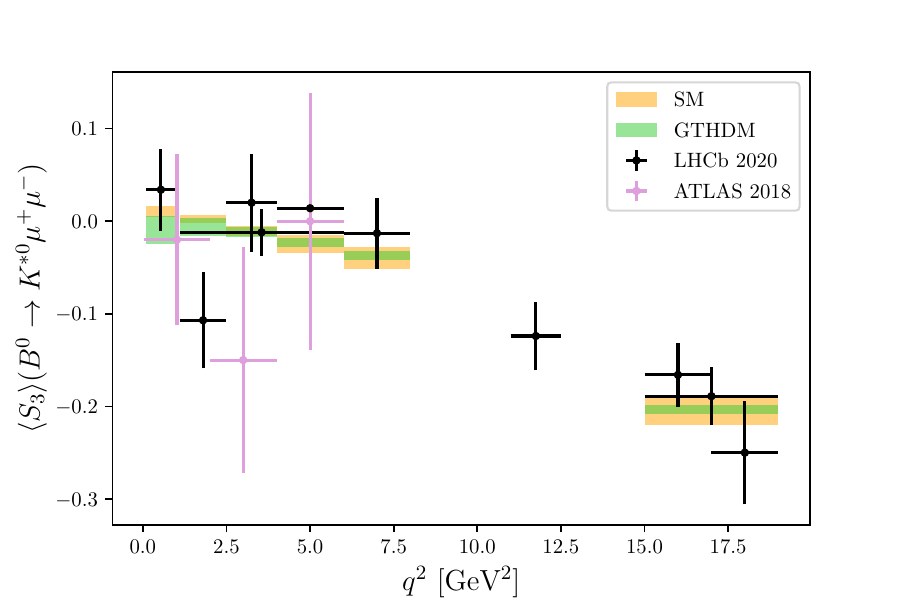}

\includegraphics[scale=0.5]{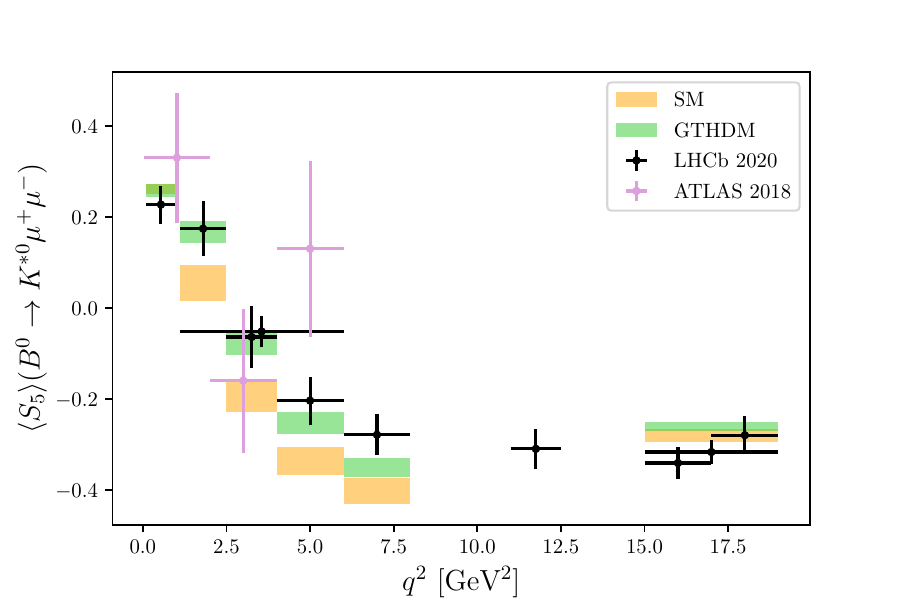}$\qquad$\includegraphics[scale=0.5]{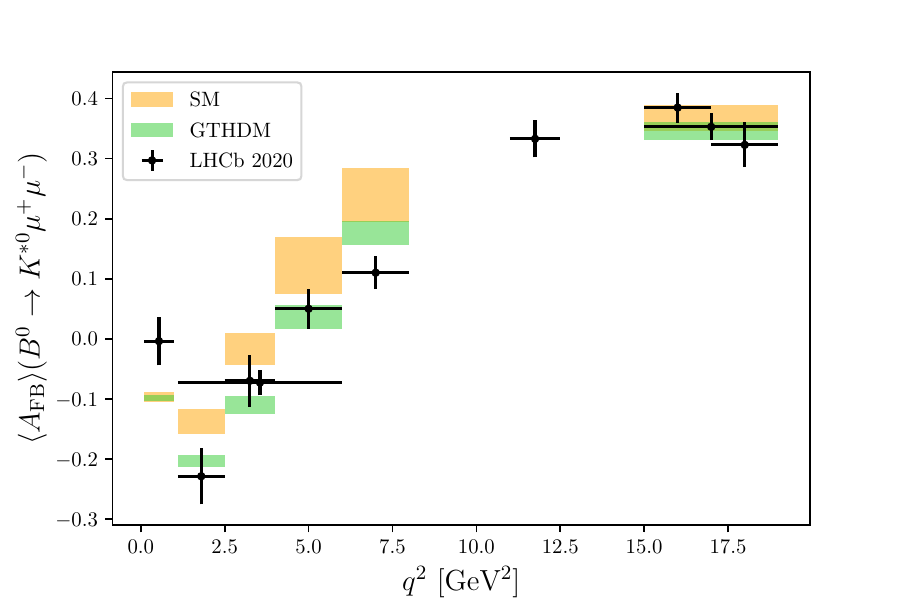}

\caption{\emph{Predicted distributions for the form factor dependent observables in the $S_{i}$ basis using both the ATLAS 2018 \cite{ATLAS:2018gqc} and the LHCb 2020 \cite{LHCb:2020lmf} data.}\label{fig:Si-obs}}
\end{figure}

In order to better understand the contribution of the GTHDM to the various rates and angular observables, we display various plots comparing both the SM and the GTHDM predictions along the experimental data. For the angular observables  $\left\langle P_{1}\right\rangle$ and $\left\langle P_{5}^{\prime}\right\rangle$ defined in Eqs.(\ref{eq:P1P2}) and (\ref{eq:P5p}), we show in figure\ \ref{fig:P1-P5p} their predictions compared to the CMS 2017 \cite{CMS:2017ivg}, ATLAS 2018 \cite{ATLAS:2018gqc} and LHCb 2020 \cite{LHCb:2020lmf} data. For $\left\langle P_{1}\right\rangle$ (figure\ \ref{fig:P1-P5p} \textit{left}) the GTHDM distribution is rather indistinguishable from the SM one, except in the $[1,\,2]$ $\mathrm{GeV}^{2}$ bin close to the photon pole and sensitive to $C_{7}^{(\prime)}$. The situation is different for $\left\langle P_{5}^{\prime}\right\rangle$ (figure\ \ref{fig:P1-P5p} \textit{right}) in which the GTHDM prediction fits the LHCb 2020 data better, particularly in the $C_{7}^{(\prime)}$- $C_{9}^{(\prime)}$ interference region ($1<q^2<6\,\textrm{GeV}^2$). We also provide in figure\ \ref{fig:Si-obs} predictions for the angular observables in the $S_{i}$ basis using the same LHCb 2020 measurements and also the ATLAS 2018 \cite{ATLAS:2018gqc} data. We can see that the GTHDM fits better the LHCb data \cite{LHCb:2020lmf} in the large recoil region than the SM by 2$\sigma$. We also note that neither the SM or the GTHDM can explain the central values (with larger uncertainties) from the ATLAS 2018 data.

\begin{figure}[h]

\includegraphics[scale=0.5]{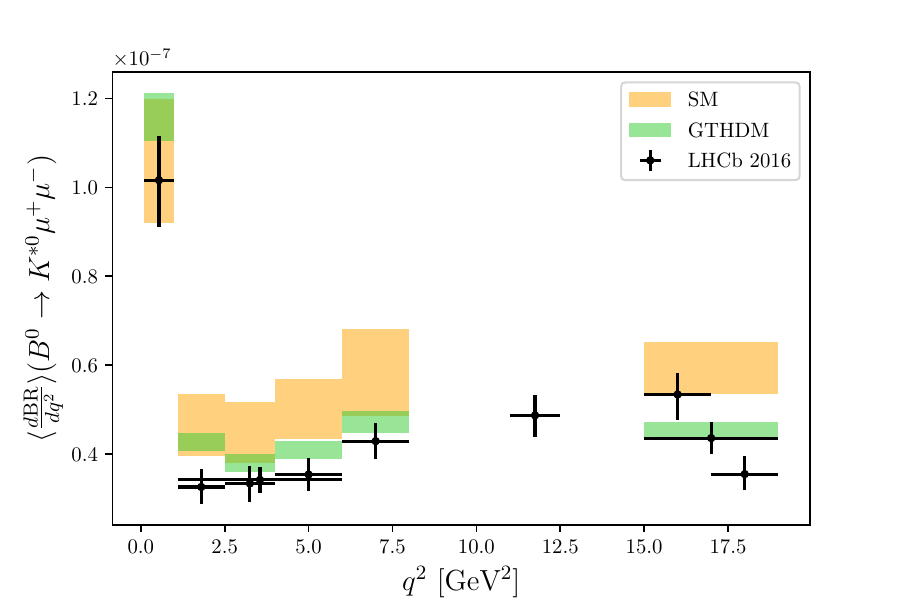}$\qquad$\includegraphics[scale=0.5]{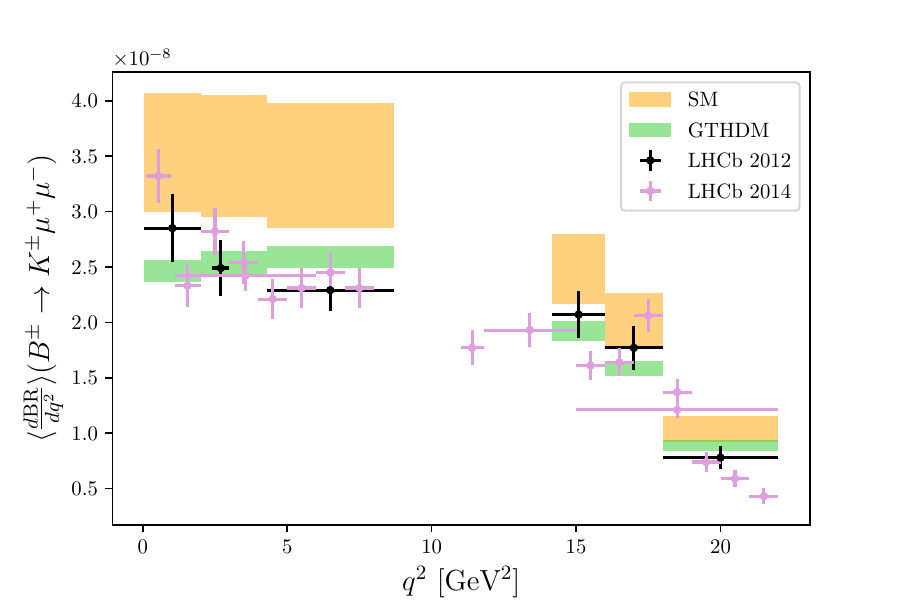}

\caption{\emph{\textit{Left}: Differential branching ratio for $\frac{d\mathrm{BR}}{dq^{2}}(B^{0}\rightarrow\,K^{*0}\mu^{+}\mu^{-})$ with the LHCb 2016 data \cite{LHCb:2016ykl}. \textit{Right}: $\frac{d\mathrm{BR}}{dq^{2}}(B^{+}\rightarrow\,K^{+}\mu^{+}\mu^{-})$ compared to the LHCb 2012 and 2014 measurements \cite{LHCb:2012juf,LHCb:2014cxe}.} \label{fig:bsmumu_predictions}}
\end{figure}

\begin{figure}[h]

\includegraphics[scale=0.5]{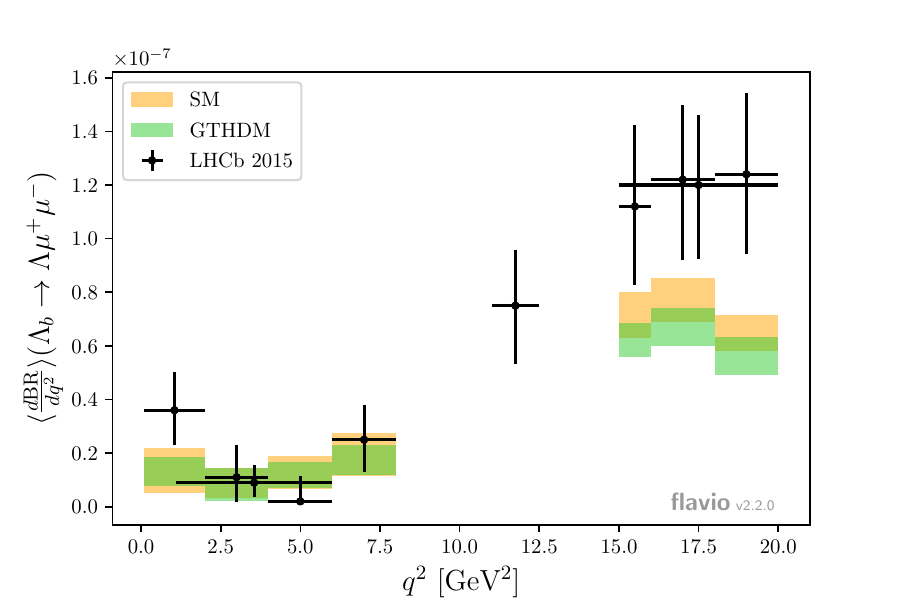}$\qquad$\includegraphics[scale=0.5]{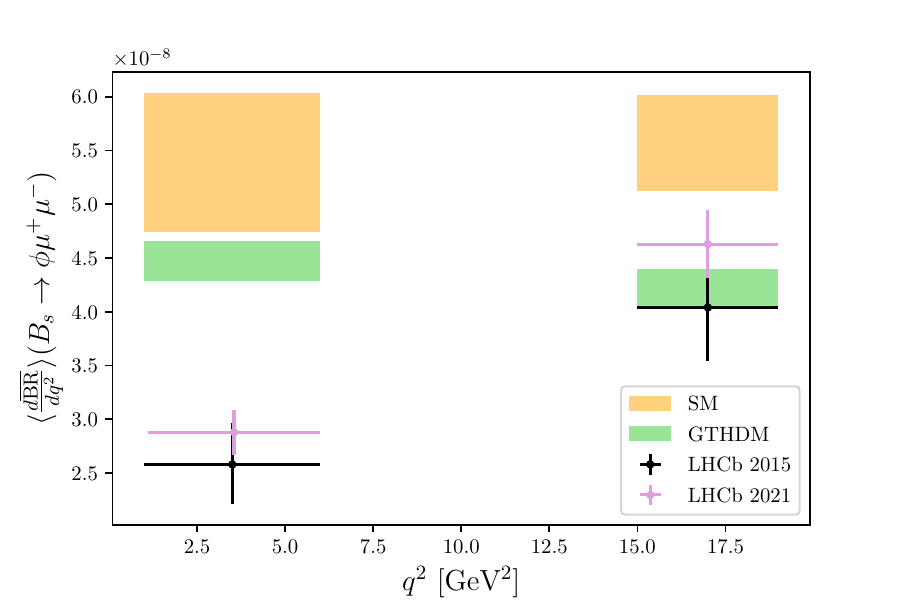}

\caption{\emph{\textit{Left}: Differential branching ratio $\frac{d\mathrm{BR}}{dq^{2}}(\Lambda_{b}\rightarrow\Lambda\mu^{+}\mu^{-})$ obtained with} \texttt{flavio} \emph{\cite{Straub:2018kue} compared to the LHCb 2015 \cite{LHCb:2015tgy} data. \textit{Right}: $\frac{d\mathrm{BR}}{dq^{2}}(B_{s}\rightarrow\phi\mu^{+}\mu^{-})$ compared to the LHCb 2015  and 2021 data \cite{LHCb:2015wdu,LHCb:2021zwz}.} \label{fig:Lambdab_Lmumu}}
\end{figure}

As for the measured branching ratios of $B^{0}\rightarrow K^{0*}\mu^{+}\mu^{-}$ and $B^{+}\rightarrow K^{+}\mu^{+}\mu^{-}$, in figure\ \ref{fig:bsmumu_predictions} we show the SM and GTHDM predictions using the LHCb results \cite{LHCb:2016ykl,LHCb:2012juf,LHCb:2014cxe}, where we can see again how the GTHDM fits better the data compared to the SM, specially in the region above the open charm threshold, sensitive to both $C_{9}^{(\prime)}$ and $ C_{10}^{(\prime)}$. In contrast, the performance of the model is worse than the SM (figure\ \ref{fig:Lambdab_Lmumu} \textit{left}) in the low recoil region of the differential branching ratio $\frac{d\mathrm{BR}}{dq^{2}}(\Lambda_{b}\rightarrow\Lambda\mu^{+}\mu^{-})$ when comparing to the LHCb 2015 \cite{LHCb:2015tgy} data. As pointed out in \cite{Bhom:2020lmk}, the decays of the $\Lambda_b$ baryon, such as $\Lambda_{b}\rightarrow\Lambda\mu^{+}\mu^{-}$ have much larger uncertainties than those of the corresponding meson decays. However, once more experimental data is available, recent \cite{Detmold:2016pkz} and future developments of lattice calculations would eventually make this decay providing similar constraints as other $b\to s\mu^+\mu^-$ transitions. Finally, the results for the $\frac{d\mathrm{BR}}{dq^{2}}(B_{s}\rightarrow\phi\mu^{+}\mu^{-})$ distribution are shown in figure \ref{fig:Lambdab_Lmumu} \textit{right}. The large recoil region of the experimental data deviates from both the SM and GTHDM predictions by approximately 3$\sigma$, and for the low recoil bin the GTHDM performs slightly better than the SM by approximately 1$\sigma$.

\begin{figure}[htb]

\includegraphics[scale=0.5]{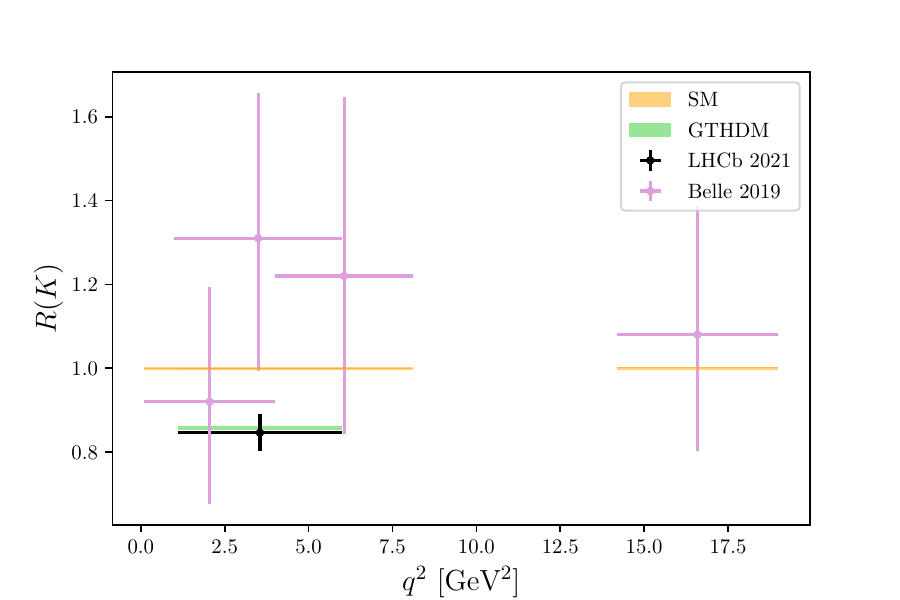}$\qquad$\includegraphics[scale=0.5]{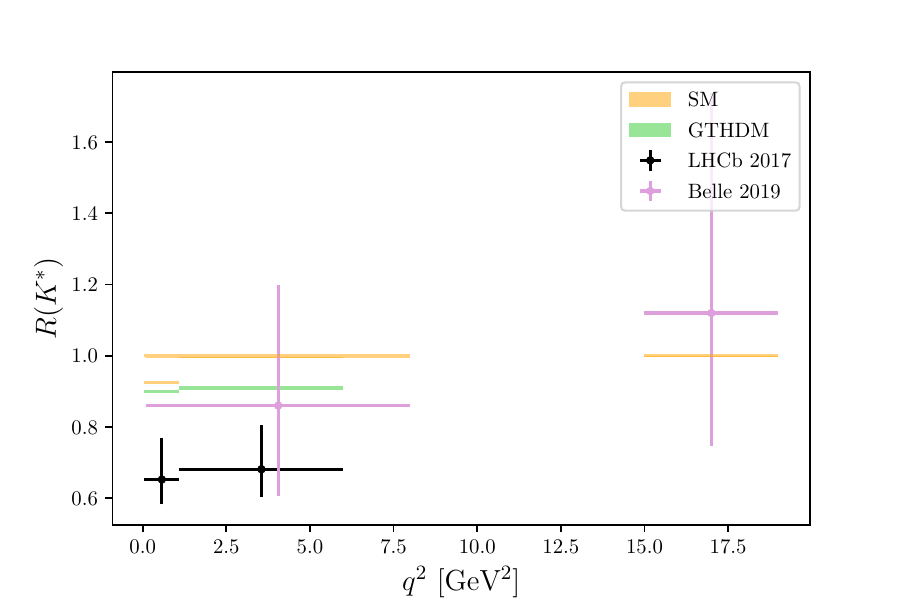}

\caption{\emph{$R(K^{(*)})$ theoretical ratios compared to both the LHCb \cite{LHCb:2017avl,LHCb:2021trn} and Belle data \cite{Belle:2019oag,BELLE:2019xld}.} \label{fig:RK-RKstar}}
\end{figure}

Last but not least important observables related to the $b\to s\mu^{+}\mu^{-}$ transitions are the ratios $R(K^{(*)})$. Despite being only three bins in total \cite{LHCb:2017avl,LHCb:2019hip,LHCb:2021trn}, these measurements have been intensively studied as they provide evidence for LFU violation. We include in our fit the latest LHCb collaboration data for the $R(K^{*})$ and $R(K)$ ratios from 2021 \cite{LHCb:2021trn} and 2017 \cite{LHCb:2017avl} respectively and obtain the plots in figure\ \ref{fig:RK-RKstar}, where we compare also to the Belle 2019 experiment data \cite{Belle:2019oag,BELLE:2019xld}. The effect from the fit on the $R(K^{(*)})$ ratios is significant, explaining the LHCb 2021 measurement of $R(K)$ at the 1$\sigma$ level.

\begin{figure}[htb]
\begin{centering}
\includegraphics[scale=0.60]{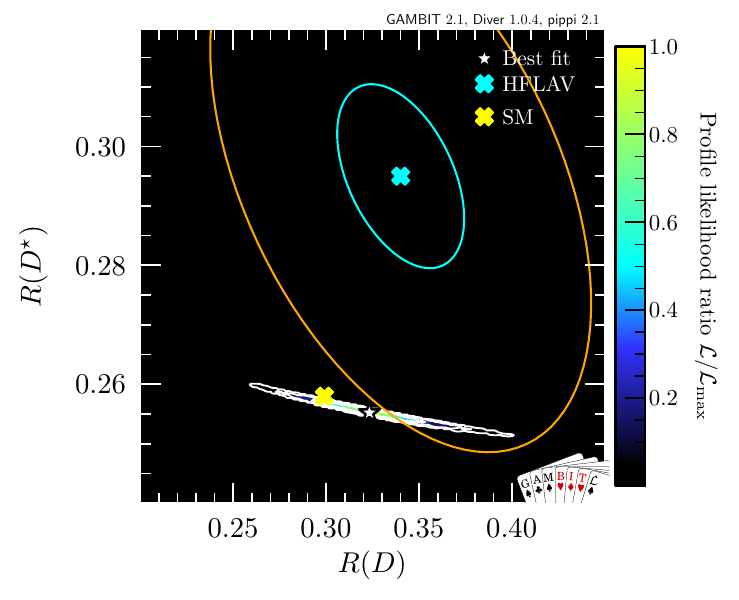}
\par\end{centering}
\caption{\emph{$R(D^{*})$ versus $R(D)$ correlated ratios. The cyan and orange lines are the 1$\sigma$ and 3$\sigma$ deviations from the HFLAV average respectively.}\label{fig:RD_RDstar}}
\end{figure}

The next interesting results are related with the charged anomalies, in particular we find that the $R(D^{(*)})$ ratio can (can not) be explained at the $1\sigma$ level with the GTHDM, a result in agreement with the phenomenological analysis of \cite{Iguro:2017ysu}. We furthermore corroborate that the constraint coming from the $B_{c}$ lifetime makes it very difficult to fit  $R(D^{*})$ and $R(D)$ simultaneously. In figure \ref{fig:RD_RDstar} we show the preferred values by the profile likelihood. We see a slightly better performance of the GTHDM compared to the SM with respect to the HFLAV average. Regarding the $d\Gamma(B\to D^{(\star)}\tau\overline{\nu})/(\Gamma dq^{2})$ distributions measured by BaBar \cite{Lees:2013uzd}, we find that the GTHDM prediction is indistinguishable from the SM, in agreement with  \cite{Martinez:2018ynq}. We find furthermore that the longitudinal polarisation $F_{L}(D^{*})$ is strongly correlated with $R(D^{*})$ and the model is not able to explain the Belle measurement, giving a best fit value of $0.458\pm0.006$.

\subsection{Anomalous $(g-2)_{\mu}$}

\begin{figure}[htb]

\begin{centering}
\includegraphics[scale=0.60]{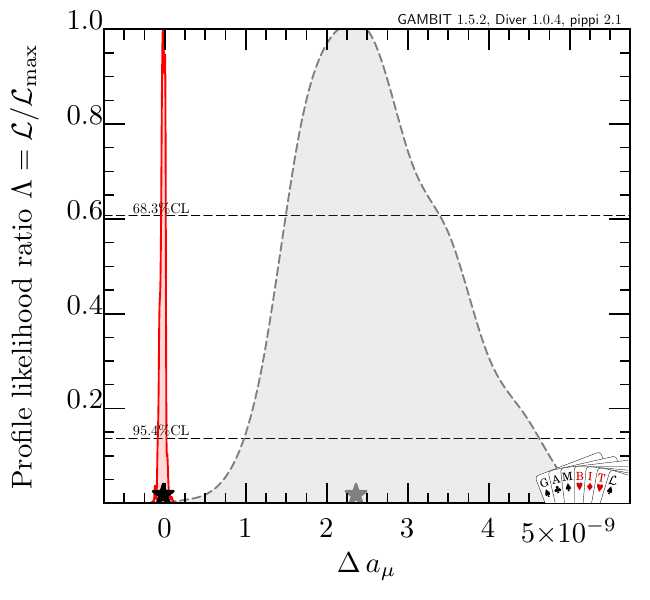}
\par\end{centering}

\caption{\emph{One-dimensional profile likelihood for $\Delta a_{\mu}$. The solid red line shows the result from the fit using all likelihoods and observables defined in this study. The dashed gray line is obtained using all but the \textsf{HEPLike} likelihoods instead.}\label{Delta_a_mu}}
\end{figure}

With regards to the anomalous magnetic moment of the muon, $(g-2)_{\mu}$, we find that a simultaneous explanation using all the likelihoods defined before is not possible (solid red line in figure\ \ref{Delta_a_mu}). However, when doing a fit to all other observables except the neutral anomalies, i.e., without using the \textsf{HEPLike} likelihoods, the model is able to explain the measured $\Delta a_{\mu}$ by Fermilab at the 1$\sigma$ level (dashed gray line in figure\ \ref{Delta_a_mu}). Furthermore, when evaluating the performance of the \textsf{HEPLike} likelihoods for the best fit value, we find a SM-like behavior with all NP WCs close to zero, except for those scalar WCs that enter in $\mathrm{BR}(B_{s}\rightarrow\mu^{+}\mu^{-})$.

\subsection{Projections for future and planned experiments}

Although a detailed collider analysis is beyond the scope of the present work, we have included as pure observables the branching ratio for $t\to c\,h$ and $h\to b\,s$ \footnote{We are not aware of current bounds for the $h\to b\,s$ branching ratio so we did not define an associated likelihood function for it.} at tree level. These tree level branching ratios in the GTHDM are suppressed as $c_{\beta\alpha}^{2}|\xi_{tc(bs)}^{u(d)}|^{2}$, respectively, so that in the alignment limit they will be exactly zero. In order to study the effects of this fined tuned suppression, we have ran a second scan with $s_{\beta\alpha}\in[0.9999,\,1]$ and we found the branching ratio of  $t\to c\,h$ decays are of order $10^{-11}-10^{-7}$, which although are outside future searches sensitivities, they are larger than the SM loop prediction ($\sim10^{-15})$ and well below the current experimental upper bound obtained by the ATLAS collaboration \cite{ATLAS:2018jqi}

\begin{align}
\mathrm{BR}(t\to c\,h)<1.1\cdot 10^{-3}\,.
\end{align}

\begin{figure}[htb]

\begin{centering}
\includegraphics[scale=0.60]{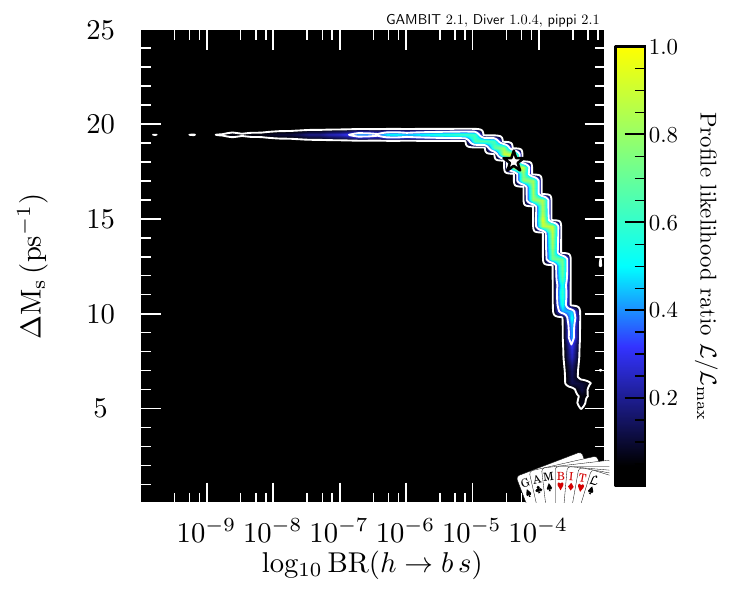}
\par\end{centering}

\caption{\emph{Profile likelihood contours in the $\Delta M_{s}$-$\mathrm{BR}(h\to b\,s)$ plane obtained with a scan using $s_{\beta\alpha}\in[0.9999,\,1]$. The observed correlation is expected from Eq.(4.18) in \cite{Herrero-Garcia:2019mcy}.} \label{fig:quark-likelihoods}}
\end{figure}

Concerning the $\mathrm{BR}(h\to b\,s)$ observable, it was shown in \cite{Herrero-Garcia:2019mcy} that it is related to tree level $B_{s}-\overline{B}_{s}$ oscillations which are not only proportional to $c_{\beta\alpha}^{2}$ but also to pseudoscalar contributions independent of the scalar CP-even mixing. Hence, in figure\ \ref{fig:quark-likelihoods} we see that 
$h\to b\,s$ is not as constrained as $t\to c\,h$ with values ranging from $10^{-7}$ up to $10^{-3}$ at the 1$\sigma$ level, which may be within range of the ILC~\cite{Barducci:2017ioq}.

\begin{figure}[htb]

\begin{centering}
\includegraphics[scale=0.65]{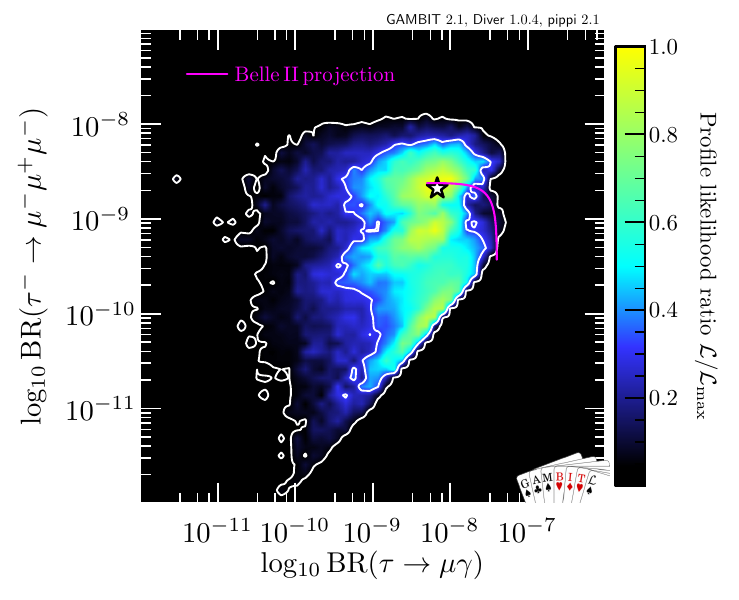}
\par\end{centering}
\caption{\emph{$\mathrm{BR}(\tau\to 3\mu)$ versus $\mathrm{BR}(\tau\to \mu\gamma)$. The magenta solid line is the combined Belle II experiment future sensitivity  obtained for both observables using a one-sided Gaussian upper limit likelihood function at 90$\%$C.L.} \label{fig:leptonic-likelihoods}}
\end{figure}

Regarding LFV searches, we show in figure\ \ref{fig:leptonic-likelihoods} the profile likelihood for the $\tau\rightarrow3\mu$ and $\tau\to\mu\gamma$ branching ratios. We see that the best fit value for the $\tau\rightarrow3\mu$ decay is well within the projected sensitivity in the Belle II experiment \cite{Belle-II:2018jsg} with a discovery potential for $\mathrm{BR}(\tau\rightarrow3\mu)\sim 10^{-9}$. Regarding the $\tau\to\mu\gamma$ decay, we find that with the projected future sensitivity, the GTHDM prediction could be confirmed with values for the branching ratio varying from $10^{-9}$ up to $10^{-8}$. As mentioned earlier, the $\tau\rightarrow3\mu$ decay receives contributions in the GTHDM from all tree, dipole and contact terms, in such a way that a possible detection in the $\tau\to\mu\gamma$ channel will not necessarily imply a strong constraint for $\mathrm{BR}(\tau\rightarrow3\mu)$.

With respect to $h\to \tau\mu$, with the model best fit point values, we computed the branching ratio $\mathrm{BR}(h\to \tau\mu)$ obtaining values from  $10^{-2}$ down to $10^{-6}$ which are within the future sensitivity at the HL-LHC, reaching the 0.05$\%$ limit \cite{Hou:2020tgl}. 

Finally, as for the $B_{s}\rightarrow\tau^{+}\tau^{-}$ decay, we find  values of at most  $\mathrm{BR}(B_{s}\rightarrow\tau^{+}\tau^{-})\sim\mathcal{O}(10^{-6})$ with our best fit point, which is one order of magnitude higher than the SM prediction, but out of reach of the future sensitivity in both the HL-LHC and the Belle-II experiments with limits at $\mathcal{O}(10^{-4})$
\cite{LHCb:2018roe,Belle-II:2018jsg}. Regarding the branching ratio 
$\mathrm{BR}(B^{+}\rightarrow K^{+}\tau^{+}\tau^{-})$, as in the $B_{s}\rightarrow\tau^{+}\tau^{-}$ decay, the predicted branching ratio $\mathrm{BR}(B^{+}\rightarrow K^{+}\tau^{+}\tau^{-})$ is of order $10^{-7}-10^{-6}$, out of reach for Belle-II projections at $2\times10^{-5}$.

\section{Conclusions and Outlook}
\label{sec:Conclusions}

We presented a frequentist inspired likelihood analysis for the GTHDM including the charged anomalies, $b\to s\mu^{+}\mu^{-}$ transitions and the anomalous magnetic moment of the muon along with other flavour observables.  
The analysis was carried out using the open source global fitting framework \textsf{GAMBIT}. 
We computed the GTHDM WCs and validated them obtaining full agreement with the one loop calculations reported in the literature after the different notation factors were taken into account.
As expected, we found that the GTHDM can explain the neutral anomalies at the $1\sigma$ level. 
Additionally, we also confirmed that the model is able to fit the current experimental values of the $R(D)$ ratio at the $1\sigma$ level, but it can not accommodate the $D^{*}$ charmed meson observables $R(D^{*})$ and $F_{L}(D^{*})$. 
Furthermore, we inspected the fitted values for the angular observables in $b\to s\mu^{+}\mu^{-}$ transitions, obtaining in general a better performance with the GTHDM in comparison to the SM.

Then, based on the obtained best fit values of the model parameters and their 1$\sigma$ and 2$\sigma$ C.L. regions, we made predictions impacting directly in the future collider observables $\mathrm{BR}(t\to ch)$, $\mathrm{BR}(h\to bs)$, $\mathrm{BR}(h\to \tau\mu)$, $\mathrm{BR}(B_{s}\rightarrow\tau^{+}\tau^{-})$, $\mathrm{BR}(B^{+}\rightarrow K^{+}\tau^{+}\tau^{-})$ and the flavour violating decays of the $\tau$ lepton,  $\mathrm{BR}(\tau\rightarrow3\mu)$ and $\mathrm{BR}(\tau\to\mu\gamma)$. We find that the model predicts values of $\mathrm{BR}(t\to ch)$, $\mathrm{BR}(B_{s}\rightarrow\tau^{+}\tau^{-})$ and $\mathrm{BR}(B^{+}\rightarrow K^{+}\tau^{+}\tau^{-})$ that are out of reach of future experiments, but its predictions for $\mathrm{BR}(h\to bs)$ and $\mathrm{BR}(h\to \tau\mu)$ are within the future sensitivity of the HL-LHC or the ILC.  We also find that the predictions for the $\tau\rightarrow3\mu$ and $\tau\to\mu\gamma$ decays are well within the projected limits of the Belle II experiment. In summary, the next generation of particle colliders will have the sensitivity to probe, discover or exclude large parts of the parameter space of the GTHDM, and thus it serves as a further motivation for the development of higher energy and higher intensity particle colliders.

We can envision many avenues of future investigation using the tools and techniques developed for this work. The complete parameter space of the GTHDM is enormous, and thus for this study we have only focused on a subset of CP-conserving interactions between second and third generation fermions. The inclusion of the first generation in the Yukawa textures would introduce additional interactions and decay channels, possibly improving the fit to various of the flavour anomalies, while at the same time introducing new relevant constraints, such as rare kaon decays, e.g. from the NA62 experiment, and LFV muon decays, e.g. from the Mu2e experiment. CP-violation in kaon and $B$-meson decays would also become important constraints in case of complex Yukawa textures. Modifications of the GTHDM may also lead to improved fits to some flavour observables, for instance it has been shown that with the addition of right-handed neutrinos the model can better accommodate the neutral anomalies. Lastly, in this study we have not included detailed collider constraints from, e.g., searches for heavy Higgs bosons at colliders. Such a detailed study is a clear follow up from this work and it will showcase the complementarity of flavour and collider searches to constrain models of new physics that tools such as GAMBIT can explore.

Finally, in view of the latest experimental measurement made by the Fermilab Muon $g-2$ Collaboration, we performed a simultaneous fit to $\Delta a_{\mu}$ constrained by the charged anomalies finding solutions at the $1\sigma$ level. Once the neutral anomalies are included, however, a simultaneous explanation is unfeasible.  A detailed study looking for a simultaneous explanation of both $g-2$ and the neutral anomalies in the GTHDM will be presented in a follow-up work.

\acknowledgments

We thank Martin White, Filip Rajec and the rest of the \gambit community for their suggestions and advice.  We would also like to thank Dominik St\"ockinger and Hyejung St\"ockinger-Kim for their help and guidance on the dominant contributions to muon $g-2$.  C.S. thanks Ulrik Egede for useful comments on the future sensitivity of the HL-LHC, and Peter Stangl for discussions about correlations in FCNC observables. F.M. thanks Grégoire Uhlrich for his help in cross-checking the Wilson coefficients with \texttt{MARTY}. The work of C.S. was supported  by the Monash Graduate Scholarship (MGS) and the Monash International Tuition Scholarship (MITS). T.E.G. is supported by DFG Emmy Noether Grant No. KA 4662/1-1. The research placement of D.J. for this work was supported by the Australian Government Research Training Program (RTP) Scholarship and the Deutscher Akademischer Austauschdienst (DAAD) One-Year Research Grant. The work of P.A.\ was supported by the Australian Research Council Future Fellowship grant FT160100274.  The work of P.A., C.B. and T.E.G. was also supported with the Australian Research Council Discovery Project grant DP180102209. The work of C.B. was supported by the Australian Research Council through the ARC Centre of Excellence for Particle Physics at the Tera-scale CE110001104. This project was also undertaken with the assistance of resources and services from the National Computational Infrastructure, which is supported by the Australian Government.  We thank Astronomy Australia Limited for financial support of computing resources, and the Astronomy Supercomputer Time Allocation Committee for its generous grant of computing time.


\appendix

\section{Gauge dependent term}
\label{sec:gauge-term}
\noindent The box diagrams in figures\ \ref{fig:c}-\ref{fig:d} will be gauge dependent, in the Feynman gauge and with $H=m_{H^{\pm}}^{2}/m_{W}^{2}$ we get,
\begin{equation}
\mathcal{B}^{H(0)}_{\textrm{mix}}(t,\,H)=\frac{\left[G_{1}(t,\,H)+V_{ts}^{*}G_{2}(t,\,H)\right]}{(H-1)^{2}\sqrt{H}(t-1)^{2}\sqrt{t}(Ht-1)^{2}},
\end{equation}
where
\begin{align}
G_{1}(t,\,H)= & \,V_{tb}V_{cs}^{*}\,H\,\text{\ensuremath{\text{\ensuremath{\xi}}_{ct}^{u}}}(t-1)tB_{1}(t,\,H),\nonumber \\
G_{2}(t,\,H)= & \,\text{\ensuremath{V_{tb}}}\left[H\,\text{\ensuremath{\xi}}_{tt}^{u}(t-1)tB_{1}(t,\,H)-\text{\ensuremath{\xi}}_{tt}^{u*}(Ht-1)B_{2}(t,\,H)\right]\\
 & -\text{\ensuremath{V_{cb}}}\,\text{\ensuremath{\xi}}_{ct}^{u*}(Ht-1)B_{2}(t,\,H),
\end{align}
with
\begin{align}
B_{1}(t,\,H)= & \,(Ht-1)^{2}\log H-(H-1)\left[(t-1)(Ht-1)+(H-1)t\log(Ht)\right],\\
B_{2}(t,\,H)= & \, (t-1)^{2}\left[H((H-1)H-1)t-(H-1)^{2}\right]\log H-(H-1)B_{3}(t,\,H),\\
B_{3}(t,\,H)= & -\left[H^{2}(t-1)^{2}t+H((t-2)(t-1)t-1)+(t-1)t+1\right]\log t\nonumber \\
 & +(t-1)\left[-Ht^{2}+(t-1)(H(Ht+t-1)+1)\log(Ht)+t\right].
\end{align}

\section{Loop Functions and Vertex Couplings} \label{sec:Loop_Functions}

The one-loop contributions can be separated into fermion-fermion-scalar (FFS) and scalar-scalar-fermion (SSF) diagrams shown in the diagrams shown in figure\ \ref{fig:ltolpgammaOneLoop}.  As seen in these diagrams, we can have any one of the SM charged leptons or neutrinos paired with any neutral Higgs boson $\phi=h,H,A$ or the charged Higgs boson $H^\pm$ respectively.  The contributions from each of these diagrams with a scalar-fermion pair from a lepton of generation $a$ to a lepton of generation $b$ are shown below:
\begin{align}
	A^{(FFS)}_{ab L}(\phi,i) =& 
	\frac{1}{16 \pi^2 m_\phi^2} \bigg(
	\frac{\Gamma^{l*}_{\phi ib}\Gamma^{l}_{\phi ia}}{24} E\begin{pmatrix}\frac{m_{l_i}^2}{m_\phi^2}\end{pmatrix} 
	+ \frac{m_{l_b}}{m_{l_a}} \frac{\Gamma^{l*}_{\phi ai}\Gamma^{l}_{\phi bi}}{24} E\begin{pmatrix}\frac{m_{l_i}^2}{m_\phi^2}\end{pmatrix} \nonumber\\
	\label{eqn:A1loopFFS}
	&+ \frac{m_{l_i}}{m_{l_a}} \frac{\Gamma^{l}_{\phi ib}\Gamma^{l*}_{\phi ai}}{3}  F\begin{pmatrix}\frac{m_{l_i}^2}{m_\phi^2}\end{pmatrix} \bigg), \\
	\label{eqn:A1loopSSF}
	A^{(SSF)}_{ab L}(H^\pm,i) =& 
	\frac{1}{16 \pi^2 m_{H^\pm}^2} \frac{\Gamma^{l*}_{H^\pm ib}\Gamma^{l}_{H^\pm ia}}{24} B\begin{pmatrix}\frac{m_{\nu_i}^2}{m_{H^\pm}^2}\end{pmatrix},
\end{align}
\begin{equation} \label{eq:GammafCba}
\Gamma_{H^\pm ba}^{f} \equiv\begin{cases}
-V_{ca}^{*}\xi_{cb}^{f} & \textrm{if }f=u,\\
V_{bc}\xi_{ca}^{f} & \textrm{if }f=d,\\
\xi_{ba}^{f} & \textrm{if }f=l, \\
\end{cases}
\end{equation}
where $A_{ab R} = A_{ab L} (\Gamma^{\phi}_{ij} \rightarrow \Gamma^{\phi*}_{ji})$.  

Additionally, to get the BSM contributions to muon $g-2$, one must subtract of the SM contribution.  
This contribution is obtained from Eq.~(\ref{eqn:A1loopFFS}), $A^{(FFS)}_{\mu\mu L,R}(h_{SM},\mu)$, by using a mass of $m_{h_{SM}} = 125.09$ GeV, and a $\Gamma^f_{h_{SM}}$ coupling in Eq.~(\ref{eq:Gammafhba}) with $c_{\beta\alpha}=0$ to obtain a SM-like coupling.  For muon $g-2$, the dominant BSM contributions at the one-loop level come from the chirality flipping term involves an internal $\tau$ lepton, $m_\mu m_\tau/(48 \pi^2 m_\phi^2) \Gamma^{\ell}_{\phi \tau\mu} \Gamma^{l*}_{\phi \mu\tau}  F\begin{pmatrix}m_\tau^2/m_\phi^2\end{pmatrix}$, with an enhancement of $m_\tau^2/m_\mu^2$.  The coupling $\xi_{\tau\mu}^{\ell}$ should be nonzero to get the chirality flip enhancement from the internal $\tau$ lepton in figure\ \ref{fig:ltolpgammaOneLoop}. 

The GTHDM contributions to muon $g-2$ from Barr-Zee diagrams with a fermionic loop are given by \cite{Cherchiglia:2016eui}
\begin{align}
	A^{(FN)}_{\mu\mu}(\phi,f) =& \frac{\alpha_{EM}^2 Q_f N^f_c}{4\pi^2 m_W^2 s_W^2} \frac{v^2}{m_\mu^2} \Gamma^{f}_{\phi 33}\Gamma^{l}_{\phi 22} \bigg(Q_f \frac{m_f^2}{m_\phi^2} {\cal F}_\phi\begin{pmatrix}m_\phi,m_f\end{pmatrix} \nonumber\\
	\label{eqn:gm2ContributionFN}
	&- \frac{g^l_v g^f_v}{4 s_W^2 c_W^2} \frac{m_f^2}{m_\phi^2-m_Z^2} \bigg({\cal F}_\phi\begin{pmatrix}m_\phi,m_f\end{pmatrix} - {\cal F}_\phi\begin{pmatrix}m_Z,m_f\end{pmatrix}\bigg)\bigg), \\
	A^{(FC)}_{\mu\mu}(H^\pm,f) =& -\frac{\alpha_{EM}^2 N^f_c v^2}{32 \pi^2 m_W^2 s_W^4} \frac{v^2}{m_\mu^2} \frac{m_f^2}{m_{H^\pm}^2-m_W^2} \Gamma^{f}_{A 33}\Gamma^{l}_{A 22} \nonumber\\
	\label{eqn:gm2ContributionFC}
	& \bigg({\cal F}_{H^\pm}^f\begin{pmatrix}m_{H^\pm}\end{pmatrix} - {\cal F}_{W^\pm}^f\begin{pmatrix}m_W\end{pmatrix}\bigg), 
\end{align}
where $c_{W}^{2}=\cos^{2}\theta_{W}$, $g^f_v=T_{3f}-2Q_f s_W^2$, and $T_{3f}$ denotes the isospin of the loop fermion ($T_{3f} = (1/2,-1/2,-1/2)$ for $(u, d, l)$).  The contributions $A^{(FN)}_{\mu\mu}(\phi,f)$ corresponds to the left diagram of figure \ref{fig:ltolpgammaBZFermionic}, while $A^{(FC)}_{\mu\mu}(H^\pm,f)$ corresponds to the other two.  

The loop function ${\cal F}_\phi$ used to calculate the two-loop neutral fermionic Barr-Zee contributions to muon g-2 are defined as \cite{Cherchiglia:2016eui}: 
\begin{equation}\label{eqn:gm2functionFSA}
    {\cal F}(M,m) = \begin{cases} 
    -2 + \log\begin{pmatrix}\frac{M^2}{m^2}\end{pmatrix} - \frac{M^2-2 m^2}{M^2} \frac{\Phi(M,m,m)}{M^2-4 m^2}, & \phi=h,H, \\ 
    -\frac{\Phi(M,m,m)}{M^2-4 m^2}, & \phi=A, \end{cases}
\end{equation}
and the loop functions ${\cal F}_{H^\pm}^f,{\cal F}_{W^\pm}^f$ for the charged fermionic Barr-Zee contributions are defined as
\begin{align}
    \label{eqn:gm2functionFCl}
    {\cal F}_{H^\pm}^l(M) =& x_l + x_l (x_l-1) \begin{pmatrix}\mathrm{Li}_2(1-1/x_l) -\frac{\pi^2}{6}\end{pmatrix} + \begin{pmatrix}x_l -\frac{1}{2}\end{pmatrix} \log(x_l), \\
    \nonumber
    \label{eqn:gm2functionFCd}
    {\cal F}_{H^\pm}^d(M) =& -(x_u-x_d) + \begin{pmatrix}\frac{\overline{c}}{y} - \frac{c(x_u-x_d)}{y}\end{pmatrix} \Phi(\sqrt{x_d},\sqrt{x_u},1) \\\nonumber
    &+ c \begin{pmatrix} \mathrm{Li}_2(1-x_d/x_u) - \frac{1}{2} \log(x_u) \log(x_d/x_u) \end{pmatrix} \\
    &+ (s+x_d)\log(x_d) + (s-x_u)\log(x_u), \\
    \nonumber
    \label{eqn:gm2functionFCu}
    {\cal F}_{H^\pm}^d(M) =& -(x_u - x_d) + \begin{pmatrix}\frac{\overline{c}}{y} - \frac{c(x_u-x_d)}{y}\end{pmatrix} \Phi(\sqrt{x_d},\sqrt{x_u},1) \\\nonumber
    &+ c \begin{pmatrix} \mathrm{Li}_2(1-x_d/x_u) - \frac{1}{2} \log(x_u) \log(x_d/x_u) \end{pmatrix} \\\nonumber
    &+ (s+x_d)\log(x_d) + (s-x_u)\log(x_u) - \frac{4}{3} \frac{x_u-x_d-1}{y}  \Phi(\sqrt{x_d},\sqrt{x_u},1) \\
    &- \frac{\log(x_d)^2 - \log(x_u)^2}{3}, \\
  c =& (x_u - x_d)^2 - Q_u x_u + Q_d x_d, \\
  \overline{c} =& (x_u - Q_u) x_u - (x_d + Q_d) x_d, \\
  y =& (x_u - x_d)^2 - 2 (x_u + x_d) + 1, \\
  s =& (Q_u + Q_d)/4,
\end{align}
where $x_f=m_f^2/M^2$, and ${\cal F}_{W^\pm}^f={\cal F}_{H^\pm}^f(m_{H^\pm} \rightarrow m_W)$. The function $\Phi$ needed for the above loop functions is from \cite{DAVYDYCHEV1993123}:
\begin{align}
    \nonumber
    \label{eqn:gm2functionPhi}
    \Phi(m_1,m_2,m_3) =& \frac{\lambda}{2} \bigg( 2 \log(a_{+}) \log(a_{-}) - \log\begin{pmatrix}\frac{m_1^2}{m_3^2}\end{pmatrix} \log\begin{pmatrix}\frac{m_2^2}{m_3^2}\end{pmatrix} \\
    &- 2 \mathrm{Li}_2(a_{+}) - 2 \mathrm{Li}_2(a_{-}) +\frac{\pi^2}{3} \bigg), \\
    \lambda =& \sqrt{m_1^4+m_2^4+m_3^4-2 m_1^2 m_2^2-2 m_2^2 m_3^2-2 m_3^2 m_1^2} \\
    a_{\pm} =& \frac{m_3^2\pm m_1^2\mp m_2^2-\lambda}{2 m_3^2}, 
\end{align}
where the masses have been ordered so that $m_1<m_2<m_3$.  

The contributions to muon $g-2$ from Barr-Zee diagrams with a bosonic loop are given by \cite{Ilisie:2015tra}
\begin{align}
	\label{eqn:gm2ContributionBHN}
	A^{(BHN)}_{\mu\mu}(\phi) =& \frac{\alpha_{EM}}{8\pi^3 m_\phi^2} \frac{v}{m_\mu} {\rm Re}(\Gamma^{l}_{\phi \mu\mu}) \lambda_{\phi H^\pm H^\mp} {\cal A}\begin{pmatrix}\frac{m_{H^\pm}^2}{m_\phi^2}\end{pmatrix}, \\
	\label{eqn:gm2ContributionBWN}
	A^{(BWN)}_{\mu\mu}(\phi) =& \frac{\alpha_{EM}}{8\pi^3 v m_\mu} {\rm Re}(\Gamma^{l}_{\phi \mu\mu}) g_{\phi W^\pm W^\mp} {\cal B}\begin{pmatrix}\frac{m_W^2}{m_\phi^2}\end{pmatrix}, \\
	\nonumber
	A^{(BHC)}_{\mu\mu}(\phi) =& \frac{\alpha_{EM} {\rm Re}(\Gamma^{l*}_{H^\pm \mu\mu} \lambda_{\phi H^\pm W^\mp})}{64 \pi^3 s_w^2 (m_{H^\pm}^2-m_W^2)} \frac{v}{m_\mu} \lambda_{\phi H^\pm H^\mp} \int^1_0 dx \quad x^2(x-1) \\
	\label{eqn:gm2ContributionBHC}
	&\bigg({\cal G}\begin{pmatrix}1,\frac{m_\phi^2}{m_{H^\pm}^2},x\end{pmatrix} - {\cal G}\begin{pmatrix}\frac{m_{H^\pm}^2}{m_W^2},\frac{m_\phi^2}{m_W^2},x\end{pmatrix}\bigg), \\
	\nonumber
	A^{(BWC)}_{\mu\mu}(\phi) =& \frac{\alpha_{EM} {\rm Re}(\Gamma^{l*}_{H^\pm \mu\mu} \lambda_{\phi H^\pm W^\mp})}{64 \pi^3 s_w^2 v m_\mu (m_{H^\pm}^2-m_W^2)} g_{\phi W^\pm W^\mp} \int^1_0 dx \quad x^2 ((m_{H^\pm}^2+m_W^2-m_{\phi}^2)(1-x)-4m_W^2)) \\
	\label{eqn:gm2ContributionBWC}
	& \bigg({\cal G}\begin{pmatrix}1,\frac{m_\phi^2}{m_{H^\pm}^2},x\end{pmatrix} - {\cal G}\begin{pmatrix}\frac{m_{H^\pm}^2}{m_W^2},\frac{m_\phi^2}{m_W^2},x\end{pmatrix}\bigg),
\end{align}     
where $A^{(BHN)}_{\mu\mu}(\phi)$, $A^{(BHC)}_{\mu\mu}(\phi)$, and $A^{(BWC)}_{\mu\mu}(\phi)$ correspond to figure\ \ref{fig:ltolpgammaBZBosonic}, and $A^{(BHC)}_{\mu\mu}(\phi)$ to the left panel of figure\ \ref{fig:ltolpgammaBZBosonic} with $H^\pm$ replaced by $W^\pm$.  

The couplings $g_{\phi W^\pm W^\mp}$, $\lambda_{\phi H^\pm W^\mp}$, and $\lambda_{\phi H^\pm H^\mp}$ (between $\phi=h,H,A$ and $H^\pm H^\mp$, $H^\pm W^\mp$, $W^\pm W^\mp$) can be found by mixing the gauge states in the Lagrangian in Eq.~(\ref{eq:yuk2d}) according to Eqs.\ (\ref{eqn:HiggsMixingMatrix},\ref{eqn:GenericMixingMatrix}).  Reading off the coefficient of the $\phi-W^\pm-W^\mp$ term, where $\phi=h,H,A$, we obtain
\begin{align}\label{eqn:BosonicCouplingsphiWW}
    g_{\phi W^\pm W^\mp} = & \begin{cases} \frac{(2 m_{H^\pm}^2-m_h^2) * \cos(\alpha-3 \beta) * \sin(2 \beta) + \cos(\alpha+\beta) * ((3 m_H^2 + 2 m_{H^\pm}^2)*\sin(2 \beta)-8 m_{12}^2)}{8 v^2 \cos(\beta)^2 \sin(\beta)^2}, & \phi = h, \\ 
    \frac{(2 m_{H^\pm}^2-m_H^2) * \sin(\alpha-3 \beta) + (3 m_H^2 + 2 m_{H^\pm}^2 - 4 m_{12}^2/(\sin(\beta) \cos(\beta))) * \sin(\alpha+\beta)}{2 v^2 \sin(2 \beta)}, & \phi = H, \\
    0, & \phi = A. \\ \end{cases}
\end{align}
Similarly we can read off the coefficients of the terms involving $\phi-H^\pm-H^\mp$ and $\phi-H^\pm-W^\mp$ terms: 
\begin{align} \label{eqn:BosonicCouplingsphiHH}
    \lambda_{\phi H^\pm H^\mp} = & \begin{cases} \sin(\alpha-\beta), & \phi = h, \\  \cos(\alpha-\beta), & \phi = H \\ 0, & \phi = A, \end{cases}  \\
     \label{eqn:BosonicCouplingsphiHW}
    \lambda_{\phi H^\pm W^\mp} = & \begin{cases} \cos(\alpha-\beta), & \phi = h, \\ -\sin(\alpha-\beta), & \phi = H, \\ -i, & \phi = A. \end{cases}
\end{align}

The loop functions used for the two-loop muon $g-2$ bosonic Barr-Zee contributions come from \cite{Ilisie:2015tra} and are defined as one dimensional integrals:
\begin{align}
    \label{eqn:gm2functionA}
    {\cal A}(z) =& \frac{1}{2} \int_{0}^{1} dx \frac{x(x-1)}{z-x(1-x)} \log\begin{pmatrix}\frac{z}{x(1-x)}\end{pmatrix}, \\
    \label{eqn:gm2functionB}
    {\cal B}(z) =& \frac{1}{2} \int_{0}^{1} dx \frac{x*z*(3x(4x-1)+10)-x(1-x)}{z-x(1-x)} \log\begin{pmatrix}\frac{z}{x(1-x)}\end{pmatrix}, \\
    \label{eqn:gm2functionGtilde}
    {\cal G}(z_a,z_b,x) =& \frac{1}{x(1-x)-z_a*x-z_b*(1-x)}\log\begin{pmatrix}\frac{z_a*x+z_b*(1-x)}{x(1-x)}\end{pmatrix}.
\end{align}
Similarly, the Barr-Zee fermionic and bosonic contributions to $l \rightarrow l' \gamma$ are given by
\begin{align}
    A_{abL}^{{\rm (2, f)}} =& -\sum_{\phi=h,H,A} \sum_{f=t,b,\tau}\frac{N^f_c Q_f \alpha_{EM}}{8\pi^{3}}\frac{\Gamma_{\phi\;ab}^{l*}}{m_{l_a} m_f}\nonumber \\
    & \left[\frac{(1-4s_{W}^{2})g^f_v}{8 s_{W}^{2}c_{W}^{2}} \left\{ {\rm Re}(\Gamma_{\phi\;33}^{f})\tilde{F}_{H}\begin{pmatrix}\frac{m_f^2}{m_\phi^2},\frac{m_\phi^2}{m_Z^2}\end{pmatrix}-i{\rm Im}(\Gamma_{\phi\;33}^{f})\tilde{F}_{A}\begin{pmatrix}\frac{m_f^2}{m_\phi^2},\frac{m_\phi^2}{m_Z^2}\end{pmatrix}\right\} \right.\nonumber \\
    \label{eqn:ltolpgammafermionic}
    &\left.+Q_f\left\{ {\rm Re}(\Gamma_{\phi\;33}^{f})F_{H}\begin{pmatrix}\frac{m_f^2}{m_\phi^2}\end{pmatrix}-i{\rm Im}(\Gamma_{\phi\;33}^{f})F_{A}\begin{pmatrix}\frac{m_f^2}{m_\phi^2}\end{pmatrix}\right\} \right], \\
    A_{abL}^{{\rm (2, b)}} =& 
    \sum_{\phi=h,H}\frac{\alpha_{EM}}{16\pi^{3}}\frac{g_{\phi W^\pm W^\mp}\Gamma_{\phi\;ab}^{l*}}{m_{l_a}v} \bigg[3F_{H}\begin{pmatrix}\frac{m_W^2}{m_\phi^2}\end{pmatrix}+\frac{23}{4}F_{A}\begin{pmatrix}\frac{m_W^2}{m_\phi^2}\end{pmatrix}+\frac{3}{4}G\begin{pmatrix}\frac{m_W^2}{m_\phi^2}\end{pmatrix}\nonumber \\
    &+\frac{m_\phi^2}{2m_W^2}\left\{ F_{H}\begin{pmatrix}\frac{m_W^2}{m_\phi^2}\end{pmatrix}-F_{A}\begin{pmatrix}\frac{m_W^2}{m_\phi^2}\end{pmatrix}\right\}+\frac{1-4s_{W}^{2}}{8s_{W}^{2}}\left\{\frac{3}{2}\left\{F_A\begin{pmatrix}\frac{m_W^2}{m_\phi^2}\end{pmatrix}+G\begin{pmatrix}\frac{m_W^2}{m_\phi^2}\end{pmatrix}\right\}\right.\nonumber \\
    &\left.+ \left(5-t_{W}^{2}+(1-t_{W}^{2})\frac{m_\phi^2}{2m_W^2}\right)\tilde{F}_{H}\begin{pmatrix}\frac{m_W^2}{m_\phi^2},\frac{m_W^2}{m_Z^2}\end{pmatrix}\right.\nonumber \\
    \label{eqn:ltolpgammabosonic}
    & \left.+\left(7-3t_{W}^{2}-(1-t_W^2)\frac{m_\phi^2}{2m_W^2}\right)\tilde{F}_{A}\begin{pmatrix}\frac{m_W^2}{m_\phi^2},\frac{m_W^2}{m_Z^2}\end{pmatrix} \right\} \bigg], \\
    \nonumber
    A_{R}^{{\rm (2, f,b)}}=& A_{L}^{{\rm (2, f,b)}}(\Gamma_{\phi\;\tau\mu}^{l*}\rightarrow\Gamma_{\phi\;\mu\tau}^{l},~i\rightarrow-i).  
\end{align}
These do include the $Z$ boson contributions as per \cite{Omura:2015xcg}.  The coupling $g_{\phi W^\pm W^\mp}$ is defined in Eq.~(\ref{eqn:BosonicCouplingsphiWW}), and $t_W^2=\tan^2\theta_W$.  

Finally, the loop functions $F_{H,~A}$, $G$ and $\tilde{F}_{H,~A}$ used for $l\rightarrow l'\gamma$ flavour-violating processes are defined as
\begin{align}
    \label{eqn:gm2functionfH}
    F_{H}(z) =& \frac{z}{2}\int_{0}^{1}dx\frac{1-2x(1-x)}{x(1-x)-z}\log\frac{x(1-x)}{z},\\
    \label{eqn:gm2functionfA}
    F_{A}(z) =& \frac{z}{2}\int_{0}^{1}dx\frac{1}{x(1-x)-z}\log\frac{x(1-x)}{z},\\
    \label{eqn:gm2functionG}
    G(z) =& -\frac{z}{2}\int_{0}^{1}dx\frac{1}{x(1-x)-z}\left[1-\frac{z}{x(1-x)-z}\log\frac{x(1-x)}{z}\right],\\
    \label{eqn:gm2functionftildeH}
    \tilde{F}_{H}(x,y) =& \frac{xF_{H}(y)-yF_{H}(x)}{x-y},\\
    \label{eqn:gm2functionftildeA}
    \tilde{F}_{A}(x,y) =& \frac{xF_{A}(y)-yF_{A}(x)}{x-y}.
\end{align}

The one-loop contributions to $\tau\rightarrow3\mu$ depend on the coefficients $A_{\tau\mu L,R}$ and $g_i$ given below \cite{Kuno:1999jp}:
\begin{align}
    \nonumber
    {\cal L}_{\tau\rightarrow 3\mu} =& -\frac{e\, m_\mu}{2} A_{\tau\mu R} (\bar{\tau}_R \sigma^{\mu\nu} \mu_L) F_{\mu\nu} - \frac{e\, m_\mu}{2} A_{\tau\mu L} (\bar{\tau}_L \sigma^{\mu\nu} \mu_R) F_{\mu\nu} \\
    \nonumber
    & -\frac{4G_F}{\sqrt{2}} \bigg[g_1 (\bar{\mu}_R\mu_L)^\dagger(\bar{\tau}_R\mu_L)^\dagger + g_2 (\bar{\mu}_L\mu_R)^\dagger(\bar{\tau}_L\mu_R)^\dagger \\
    \nonumber
    & + g_3 (\bar{\mu}_R\gamma_\mu\mu_R)^\dagger(\bar{\tau}_R\gamma^\mu\mu_R)^\dagger + g_4 (\bar{\mu}_L\gamma_\mu\mu_L)^\dagger(\bar{\tau}_L\gamma^\mu\mu_L)^\dagger\\
    \label{eqn:tauto3muLagrangian}
    & + g_5 (\bar{\mu}_L\gamma_\mu\mu_L)^\dagger(\bar{\tau}_R\gamma^\mu\mu_R)^\dagger + g_6 (\bar{\mu}_R\gamma_\mu\mu_R)^\dagger(\bar{\tau}_L\gamma^\mu\mu_L)^\dagger + h.c.\bigg]. 
\end{align}
In the GTHDM, only $g_2$ and $g_4$ receive contributions:
\begin{align}
    g_2 =& \frac{i\,m_\mu^2}{192 \sqrt{2}\, \pi^2\, G_{F}\, m_{H^\pm}^4}\xi^l_{\tau\mu}(|\xi^l_{\mu\mu}|^2+|\xi^l_{\tau\mu}|^2)(\xi^l_{\mu\mu}+\xi^l_{\tau\tau}), \\
    g_4 =& \frac{-i}{128 \sqrt{2}\, \pi^2\, G_{F}\, m_{H^\pm}^2} \xi^l_{\tau\mu}(|\xi^l_{\mu\mu}|^2+|\xi^l_{\tau\mu}|^2)(\xi^l_{\mu\mu}+\xi^l_{\tau\tau}).
\end{align}

\section{Auxiliary scanning method}\label{sec:chi2_method}

The two regions of solutions for $Y_{2,tc}^{u}$ were expected already when applying the quadratic approximation to the $\chi^{2}$ function defined in \cite{Capdevila:2018jhy} for a fit to the $b\rightarrow s\mu^{+}\mu^{-}$ observables solely. Explicitly, the likelihood function is approximated by 
\begin{equation}
\log\mathcal{L}=-\frac{\chi^{2}}{2},\quad\chi^{2}(\mathbf{C})\approx\chi_{min}^{2}+\frac{1}{2}\left(\mathbf{C}-\mathbf{C}_{\mathrm{bf}}\right)^{T}\mathrm{Cov}^{-1}\left(\mathbf{C}-\mathbf{C}_{\mathrm{bf}}\right),
\end{equation}
where $\mathbf{C}=\{\Delta C_{7},\Delta C_{9},\Delta C_{10},\Delta C_{7}^{'},\Delta C_{9}^{'},\Delta C_{10}^{'}\}$ are the WCs used as parameters to be fitted, and $\mathrm{Cov}^{-1}$ is the covariance matrix or Hessian obtained using the \texttt{minuit} and \texttt{flavio} packages,
\begin{equation}
\mathrm{Cov}^{-1}=\left(\begin{array}{cccccc}
5594.96 & -128.83 & 0.1604 & -1156.88 & -0.0139 & -0.0146\\
-128.83 & 44.89 & -10.11 & -102.95 & -7.153 & -14.66\\
0.1604 & -10.11 & 34.81 & -90.76 & -6.29 & -12.91\\
-1156.88 & -102.95 & -90.76 & 3613.3 & -64.07 & -131.44\\
-0.0139 & -7.153 & -6.29 & -64.07 & 17.34 & -0.037\\
-0.0146 & -14.66 & -12.91 & -131.44 & -0.037 & 72.17
\end{array}\right) ,
\end{equation}
which encodes a fit using the likelihoods from \cite{Bhom:2020lmk} (excepting the associated likelihoods for the Belle experiment measurements not available in \texttt{flavio}). After obtaining the Hessian, a random generator in \texttt{Mathematica} is requested to find points inside the ellipsoid defined by $\Delta\chi^{2}\leq\sigma_{2d}(1)$
and $\Delta\chi^{2}\leq\sigma_{2d}(2)$ for 2 degrees of freedom and
boundaries defined by the values of the parameter space in Eq.~(\ref{eq:Ranges}). With this auxiliary method, we were able to help the  \textsf{Diver} sampler to scan over different regions of the parameter space.

\bibliographystyle{JHEP}
\bibliography{GTHDM}
\end{document}